# Straintronics: Manipulating the Magnetization of Magnetostrictive Nanomagnets with Strain for Energy-Efficient Applications

Supriyo Bandyopadhyay[1,*], Jayasimha Atulasimha[2] and Anjan Barman[3]

[1]Department of Electrical and Computer Engineering, Virginia Commonwealth University, Richmond, VA 23284, USA

[2]Department of Mechanical and Nuclear Engineering, Virginia Commonwealth University, Richmond, VA 23284, USA

[3]Department of Condensed Matter Physics and Material Sciences, S. N. Bose National Center for Basic Sciences, Block JD, Sector III, Salt Lake, Kolkata 700 106, India

## ABSTRACT

The desire to perform information processing, computation, communication, signal generation and related tasks, while dissipating as little energy as possible, has inspired many ideas and paradigms. One of the most powerful among them is the notion of using magnetostrictive nanomagnets as the primitive units of the hardware platforms and manipulating their magnetizations with electrically generated static or time varying mechanical strain to elicit myriad functionalities. This approach has two advantages. First, information can be retained in the devices after powering off since the nanomagnets are non-volatile unlike charge-based devices such as transistors. Second, the energy expended to perform a given task is exceptionally low since it takes very little energy to alter magnetization states with strain. This field is now known as "straintronics", in analogy with electronics, spintronics, valleytronics, etc. We review the recent advances and trends in straintronics, including digital information processing (logic), information storage (memory), domain wall devices operated with strain, control of skyrmions with strain, non-Boolean computing and machine learning with straintronics, signal generation (microwave sources) and communication (ultra-miniaturized acoustic and electromagnetic antennas) implemented with strained nanomagnets, hybrid straintronics-magnonics, and interaction between phonons and magnons in straintronic systems. We identify key challenges and opportunities, and lay out pathways to advance this field to the point where it might become a mainstream technology for energy-efficient systems.

---

[*] Corresponding author. Email: sbandy@vcu.edu

TABLE OF CONTENTS

# I.     INTRODUCTION: INFORMATION AND ENERGY

One of the most numerous living species on this planet are ants. There are fewer than 10 billion humans inhabiting the earth, but there are more than 1 trillion ants. That is an astronomical number, but it fades into insignificance compared to the number of stars in the universe. Even that fades into insignificance compared to the bits of information that is sensed, produced, processed, stored or communicated in our society every year. Whether it is uploading one's vacation photos on Facebook, reaching out to a loved one in What's App, forecasting tomorrow's weather, decoding the human genome, or inciting a rebellion in Twitter, an enormous amount of information is processed annually and the number of digital bits that are used up to encode that information would easily exceed the number of stars in the visible universe.

Information, it turns out, is "physical" [1] and therefore there is always some energy cost associated with every digital bit of information. Today, roughly 10% of the energy produced in the United States is consumed by its information processing infrastructure. A data center can require as much energy as the city of Athens in Greece and its individual carbon footprint can be comparable to that of a nation like Malaysia. Ukraine recently announced a plan to build a data center for mining cryptocurrency data next to a nuclear power plant because the power requirement is anticipated to be 2-3 GW [2]! Such is the energy demand of information processing. Even humans as computers are energy-hungry and roughly 20% of the calories consumed by a human is used by the brain to "think". It therefore behooves us to seek increasingly energy efficient devices and hardware for information processing. This has impelled device physicists and engineers to seek out unusual approaches to manipulate digital bits of information with an eye to remaining as energy-frugal as possible.

Conventional hardware platforms for computing, information processing and information communication are built with electronics that leverage electric charge-based devices, e.g. transistors, to carry out computational tasks. However, charge-based electronics can be energy-hungry and it is unlikely that transistors will ever evolve to the point when they dissipate sub-aJ of energy to process one bit of information, as we show later in this review. This, and the fact that transistors are "volatile" and cannot retain information once powered off, has motivated engineers to seek out alternates such as magnetic devices where information is encoded not in the charge degree of freedom of electrons (as in transistors), but in the spin degree of freedom. Magnetic devices are not a panacea – they have a few shortcomings as well – but they are "non-volatile" and can be more energy-efficient than transistors in some circumstances if, *and only if*, their states are switched with certain energy-efficient mechanisms. Of course, not all switching mechanisms are energy-efficient, but one that is particularly energy-efficient is "straintronics" which switches the magnetization of a magnetostrictive nanomagnet (with two stable magnetization states) from one stable state to the other using electrically generated mechanical strain. The two magnetization states of the nanomagnet encode the binary bits 0 and 1, allowing the nanomagnet to act as the basic *binary switch* which is the primitive constituent of all digital hardware (a transistor is also essentially a binary switch). The energy dissipated in the switching action can be as low as ~1 aJ, which is about two orders of magnitude lower than the energy dissipated in switching a state-of-the-art transistor used in today's most advanced computing hardware. That is why straintronics has attracted attention and motivates this review.

Before we proceed further, one word of caution may be in order. The term "straintronics" has multiple connotations and hence it is important that we clarify what we refer to when we use the term. We use the term to describe specifically the rapidly burgeoning field of altering the magnetization state of magnetostrictive nanomagnets with mechanical strain, which is generated in myriad ways, but mostly electrically or optically. This has applications in both digital and analog devices and systems, and is attractive because of its energy efficiency. The field witnessed rapid growth over the last fifteen years as more device applications became apparent. We will not delve into the history of the field, nor will we discuss much of the physics, since this review focuses on applications in information processing and communication. Therefore, we will concentrate mostly on such constructs as binary switches for computer logic and memory, neurons and synapses for neuromorphic computing, belief networks for Bayesian computing and some analog applications such as microwave generators and extreme sub-wavelength antennas for embedded applications. While there may be other applications of straintronics, they are outside the purview of this review. Finally, we will also discuss the rich physics of interaction between acoustic waves (phonons) and spin waves (magnons) since it sheds light on the intricacies of manipulating magnetic states with time-varying strain. This is a burgeoning field with exciting possibilities. We call it "hybrid straintronics-magnonics".

## II. BINARY SWITCH IN DIGITAL INFORMATION PROCESSORS

The hardware for *digital* information processing is always built around a basic *binary switch*. The switch has two states – ON and OFF – which encode the binary bits 0 and 1. The electronic switch that has ruled the roost for the last seven decades in digital information processors, and the one used to benchmark all digital switches, is the celebrated "metal-oxide-semiconductor-field-effect-transistor" (MOSFET) and its various avatars such as the "fin field effect transistor" (FINFET), "tunnel-field-effect-transistor" (TFET), "negative capacitance transistor" (n-CFET), etc. We start with a discussion of the transistor switch because of its overwhelming dominance in digital information processors.

The conductance of the transistor has two states – high conductance (ON-state) and low conductance (OFF-state) – that encode the binary bits 0 and 1. This device has been scaled relentlessly and aggressively to the point that transistors with 5-nm gate length are currently on the anvil. The energy dissipation incurred to switch a transistor between its two stable conductance states (ON and OFF) has progressively decreased with time while the switching speed has increased to the point where switching delays less than 100 ps are now routine. Few technologies have been as successful as the transistor. This is a device that has been amazingly successful with unprecedented staying power. So, it is natural to ask why anyone would be concerned about its potential demise. Opinion is divided on this issue, but there are two inherent shortcomings of the transistor that are its Achilles' heel and they behoove us to look at "transistor-alternatives".

Both shortcomings are related to the fact that the transistor is a *charge-based device*. Its two conductance states – high and low – which are used to encode the binary bits 0 and 1 are delineated by the amount of charge resident within the transistor. The high conductance state is attained when charge carriers (electrons in an n-channel MOSFET or holes in a p-channel MOSFET) flood into the transistor's channel region and establish a conducting path between the source and the drain contacts. When these charge carriers are expelled from the channel, the conducting path is disrupted and the transistor switches off, thereby reaching the low conductance state. This is shown in Fig. 1. Therefore, the two states are demarcated by the amount of electrical charge stored in the channel. A larger amount of channel charge $Q_1$ represents the ON-state and a smaller amount of charge $Q_2$ represents the OFF-state. Since charge is a *scalar* quantity and has only "magnitude" and no "direction", as long as we wish to encode binary bits in charge, we have no other option but do so using two different magnitudes (or amounts) of charge, such as $Q_1$ and $Q_2$.

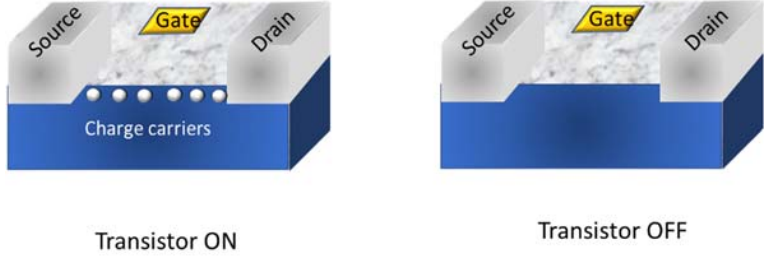

Fig. 1: Basic working principle of a MOSFET transistor. The transistor is on when charges reside in the channel and off when charges are expelled from the channel.

It is obvious that every time we switch a charge based device, we must move charge into or out of the device to change the amount of charge from $Q_1$ to $Q_2$, or vice versa, thereby causing the flow of a (time-averaged) current $I$ given by

$$I = |Q_1 - Q_2|/\Delta t = \Delta Q/\Delta t, \tag{1}$$

where $\Delta t$ is the amount of time it takes to change the channel charge from $Q_1$ to $Q_2$, or vice versa. This will then cause energy dissipation of the amount

$$E_d = I^2 R \Delta t = (\Delta Q/\Delta t) I R \Delta t = \Delta Q I R = \Delta Q \Delta V. \tag{2}$$

Here, $R$ is the resistance in the path of the current and $\Delta V = IR$. We can think of $\Delta V$ as the amount of voltage needed to change the charge in the channel by the amount $\Delta Q$. Note that the energy dissipation given in Equation (2) is *not* independent of the switching time, because $\Delta V$ depends on the switching time for a fixed $\Delta Q$ and $R$ $(\Delta V = \Delta Q R/\Delta t)$. We can actually write the energy dissipation in Equation (2) also as $E_d = (\Delta Q)^2 R/\Delta t$, which clearly shows that we will dissipate *more* energy if we switch *faster* (smaller $\Delta t$). Therefore, a more meaningful quantity to benchmark a device may be the energy-delay product which is $E_d \Delta t = (\Delta Q)^2 R$. We cannot reduce this quantity arbitrarily by reducing $\Delta Q$, since a sufficiently large $\Delta Q$ is needed to distinguish between bits 0 and 1, especially when operating in a noisy environment. We cannot reduce $R$ arbitrarily either, since that would require us to increase the cross-section of the current path at the cost of a large device footprint. Therefore, there is very likely a lower bound on the energy that will have to be dissipated as long as we use a charge-based device like the transistor.

We can try to estimate this lower bound by considering a modern day MOSFET (or FINFET). The Intel® Core™ i7-6700K processor released in 2015 uses 14-nm scale FINFETs, operates with a power supply voltage of 1.2 V and clock frequency of 4 GHz, while dissipating 91 W of power. It has roughly 1.75 billion transistors, which dissipate the bulk of the power, and about 10% of them switch at any given time during the chip's operation (i.e. the so-called "activity level" is 10%). We can therefore estimate the average energy dissipation per transistor as

$$E_d^{\text{FINFET}} = P_d/(Naf) = 91/(1.75 \times 10^9 \times 0.1 \times 4 \times 10^9) = 130 \text{ aJ}, \tag{3}$$

where $P_d$ is the power dissipation, $N$ is the number of transistors in the chip, $a$ is the activity level and $f$ is the clock frequency.

From Equation (2), the amount of charge that is moved in the channel of the FINFET to switch it on and off is roughly

$$\Delta Q = E_d / \Delta V = 130 \times 10^{-18} / 1.2 = 1.08 \times 10^{-16} \text{ Coulombs}, \qquad (4)$$

which is the charge carried by a mere 673 electrons or holes. Obviously, the number of charge carriers moved during switching must greatly exceed the number that can spontaneously appear in the channel due to noise and thermal fluctuations. The latter is the charge fluctuation in the transistor's "gate" and is given by [3]

$$\Delta Q|_{fluctuation} = \sqrt{C_g kT}, \qquad (5)$$

where $C_g$ is the gate capacitance, $k$ is the Boltzmann constant and $T$ is the absolute temperature. The gate capacitance for the FINFET structure can be estimated roughly as (this includes contributions of line capacitance, etc.)

$$C_g = \Delta Q / \Delta V = 1.08 \times 10^{-16} / 1.2 = 90 \text{ aF}, \qquad (6)$$

which makes $\Delta Q|_{fluctuation} = 6.45 \times 10^{-19}$ Coulombs at room temperature, and that is the charge of only ~4 charge carriers.

The minimum amount of $\Delta Q$ that we will need to move through the switch in order to be able to distinguish between the bits and switch reliably should be considerably larger than $\Delta Q|_{fluctuation}$ and let us say that it is $\Delta Q_{min} = \xi \Delta Q|_{fluctuation}$, where we can interpret $\xi$ ($\xi \gg 1$) as a measure of the reliability of switching. In that case, the minimum energy that we must dissipate to switch a FINFEET/MOSFET type device will be [4]

$$E_d^{min} = \Delta Q_{min} \Delta V_{min} = (\Delta Q_{min})^2 / C_g = \xi^2 (\Delta Q|_{fluctuation})^2 / C_g = \xi^2 kT. \qquad (7)$$

Equation (7) is very instructive. It tells us that the minimum energy that we must dissipate is determined by the minimum reliability that we are able to tolerate. There is a trade-off between energy dissipation and reliability; we can buy energy efficiency at the cost of reliability and vice versa. If we wish to be even "minimally reliable", we would perhaps want $\xi > 10$, and hence the minimum energy dissipation that we will have to live with may be ~100 kT at room temperature (or 0.4 aJ), if we use the MOSFET or any of its clones.

The second shortcoming of a charge-based device is that it is "volatile", meaning that if we turn off the power, information stored in the transistor (i.e. whether it was on or off) will be lost quickly since the stored charge will leak out rapidly. This is the primary reason why most computing architectures are of the von-Neumann type which consists of a processor, a memory and a switch that communicates between the processor and memory (see Fig. 2). The processor is made of fast but volatile elements while the memory has slow but non-volatile elements. The instruction sets are stored in the memory and are fetched to the processor via the switch when a program is executed. This is an inefficient approach since the back-and-forth communication between the processor and memory slows down the program execution (it is, in fact, responsible for the boot delay in a computer). If the processor could be made of non-volatile elements as well, then the instructions sets could have been stored in the processors in-situ which would eliminate the need for the switch and a partition between processor and memory. In fact, this is the driving force behind non-von-Neumann architectures and "processing in memory" (PIM) and "computing in memory" (CIM) approaches [5].

Clearly, transistor type volatile devices will not be ideal for non-von-Neumann architectures. "Non-volatile switches", which can retain memory of their states after the power has been switched off, will be preferable. There are number of non-volatile switches, all with switching delays of 1 ns or less, that have attracted attention. Table I lists them and compares switching energy dissipations (the list is not exhaustive).

**Table I: Switching times and energy dissipation of some non-volatile switches**

| Type of switch | Switching delay | Switching energy dissipation |
| --- | --- | --- |
| Memristors [6] | Sub-nanosecond | ~3 pJ |
| Phase change memories [7] | ~1 ns | 0.1 – 1 pJ |
| Strain switched nanomagnets [8] | Sub-nansceond | 1-10 aJ |

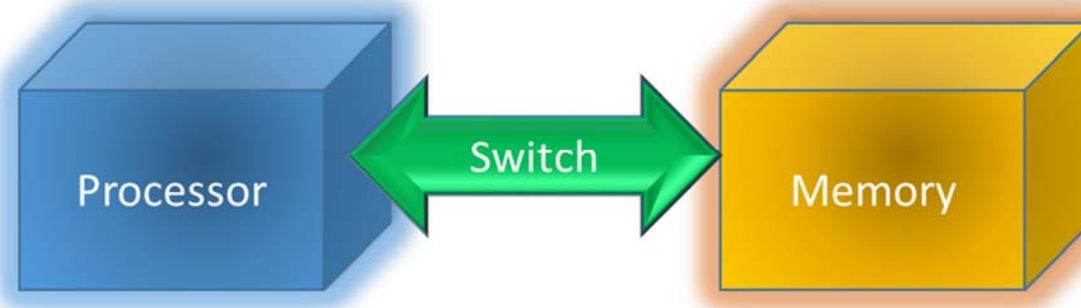

Fig. 2: Basic von-Neumann architecture.

Table 1 shows that there are a number of non-volatile switches that are capable of switching in sub-nanosecond and could be potential replacements for the transistor. However, in terms of energy dissipation, only one is comparable to or better than a transistor that currently dissipates about 100 aJ to switch. The winner is the magnetostrictive nanomagnet whose magnetization can be switched with electrically generated mechanical strain. This is the basis of "straintronics" which is the topic of this review.

### III. STRAINTRONIC NANOMAGNETIC SWITCHES AS BINARY SWITCHES

The field of "straintronics" refers to the technology of digital and analog computing, signal processing, signal generation, communication, etc. implemented with *strain-switched nanomagnets*. The first step is to fashion a nanomagnet into a bistable element whose magnetization can have only two stable orientations (states) that can be harnessed to encode the bit 0 and the bit 1. One way to accomplish this is to choose particular shapes of the nanomagnets, which will ensure that their magnetizations are bistable, meaning that they can point in only two (mutually anti-parallel) directions which can encode the binary bits 0 and 1.

Fig. 3(a) shows a nanoscale ferromagnet shaped like a thin elliptical disk. If its volume is small enough (but not so small that it behaves like a super-paramagnet, instead of a ferromagnet, at the operating temperature) and the eccentricity of the ellipse is large enough, then this entity behaves as a *single-domain nanomagnet* whose magnetization can point only along the major axis – either pointing to the right or to the left – and these two magnetization orientations encode the binary bits 0 and 1. Such a nanomagnet is said to possess in-plane (magnetic) anisotropy (IPA). If the thickness of the nanomagnet is small enough, then the magnetization can point perpendicular to the surface owing to surface anisotropy, either up or down, as shown in Fig. 3(b). Such a nanomagnet is said to possess perpendicular magnetic anisotropy (PMA). PMA nanomagnets have certain advantages over IPA nanomagnets in some applications (not all) because they are relatively insensitive to imperfections such as edge roughness and they are more scalable in size, i.e. their lateral dimensions can be made smaller without causing them to lose their ferromagnetism. However, in this article, we will discuss mostly IPA nanomagnets because their physics is often easier to elucidate.

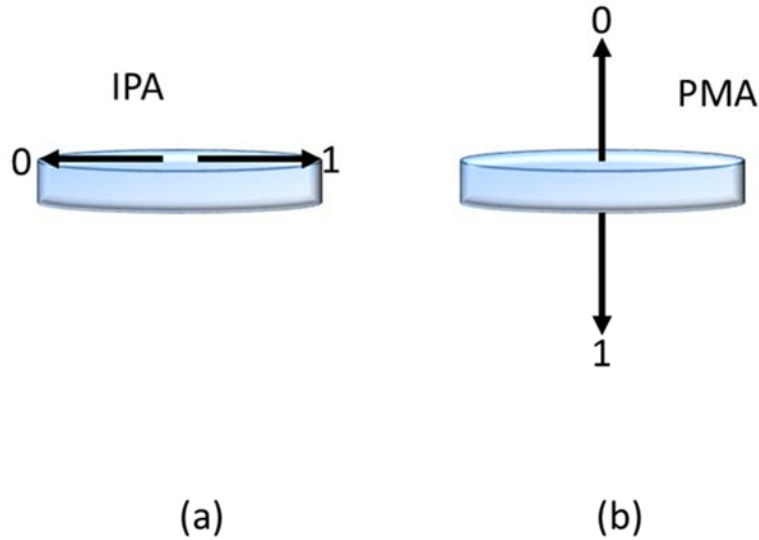

Fig. 3: A single-domain nanomagnet shaped like an elliptical disk. (a) In-plane magnetic anisotropy nanomagnet where the magnetization can point along the major axis, either pointing to the left or to the right. (b) Perpendicular magnetic anisotropy where the magnetization can point only perpendicular to the surface, either pointing up or pointing down.

There may be ~$10^4$ electron spins in a nanomagnet of the type shown in Fig. 3, but because of exchange interaction among them, they will always all *point in the same direction*. If we flip the magnetization from right to left (or up to down), and vice versa, *all* the $10^4$ spins will rotate together in unison, thus acting like one giant classical spin [9]. Unlike in a charge-based device like the MOSFET, where $N$ different charge carriers in the channel will act independently like $N$ different degrees of freedom, here all the $N$ spins act collectively like a *single* degree of freedom. This can reduce energy dissipation during the switching action dramatically [10]. Note that there is no need for any current to flow through the nanomagnet in order to flip its magnetization, and hence there is no unavoidable dissipation associated with current flow. Of course, there will be some dissipation associated with the flipping action and that may or may not involve some current flow external to the nanomagnet, but whether that dissipation is larger or smaller than the dissipation incurred in a comparable charge-based device (with inevitable current flow through the device) depends on how the flipping is accomplished. Some flipping mechanisms are energy-efficient and some are not. In general, mechanisms that rely on passing a spin current through the nanomagnet, such as spin-transfer-torque (STT) [11] and spin-orbit-torque (SOT) [12], are not particularly energy efficient, while those that rely on voltage- or electric-field control of magnetization such as voltage-controlled-magnetic-anisotropy (VCMA) [13] and straintronics, which we discuss next, tend to be more energy-efficient.

As mentioned in Section 1, the term "straintronics" refers to the science and technology of switching the magnetization of a *magnetostrictive* nanomagnet (e. g. Co, Ni, FeGa, Terfenol-D) using electrically generated mechanical strain. The idea is to place an elliptical magnetostrictive nanomagnet in elastic contact with an underlying (poled) piezoelectric film by delineating the nanomagnet on top of the film. The elastic contact allows highly efficient strain transfer from the piezoelectric to the magnetostrictive nanomagnet, as long as the piezoelectric layer is much thicker than the magnetostrictive layer. Such a system makes up a "two-phase multiferroic".

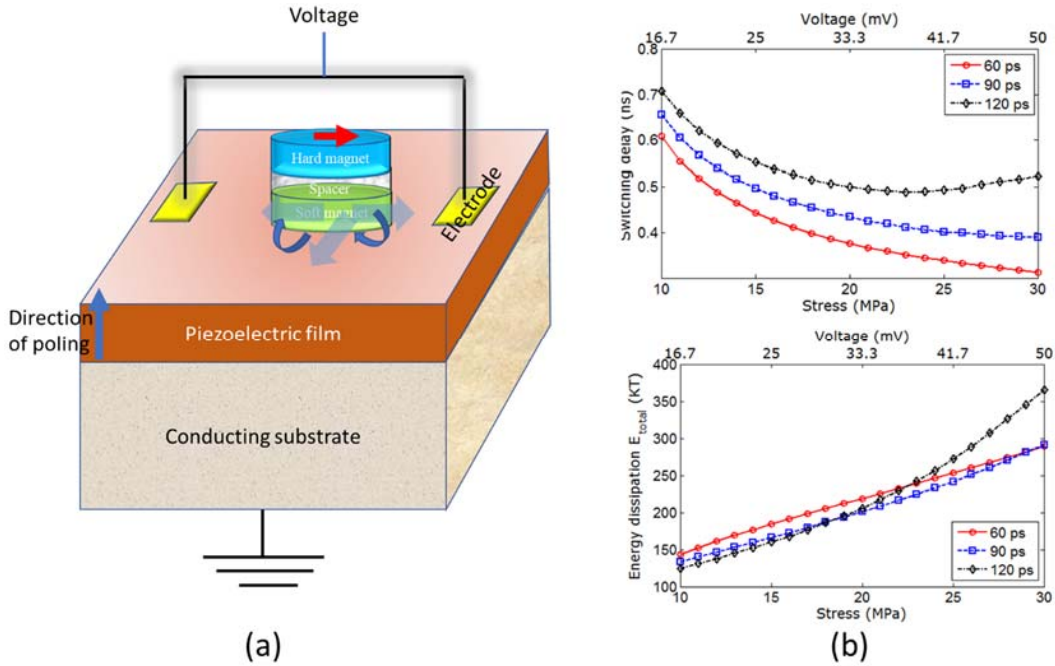

Fig. 4: (a) Straintronic switching: flipping the magnetization of a magnetostrictive elliptical nanomagnet, elastically coupled to an underlying poled piezoelectric film, with a precisely timed strain pulse generated with the applied voltage. (b) Calculated switching delay and energy dissipation as a function of the stress (and the corresponding gate voltage needed to generate the required stress) for three different stress ramp times (60, 90, 120 ps). Reproduced from [17] with permission of the American Institute of Physics.

A schematic to elucidate the switching action is shown in Fig. 4. Application of a voltage over (not across) the piezoelectric film with electrodes delineated on the film generates biaxial strain in it [14, 15], which is transferred fully or partially to the nanomagnet. If the polarity of the voltage is such that the electric field is in the direction opposite to that of the poling, as shown in Fig. 4(a), then compressive strain will be generated along the major axis of the elliptical nanomagnet and tensile strain along the minor axis. If the voltage polarity is reversed, then the signs of the strains will reverse as well. If the magnetostriction of the nanomagnet is positive (FeGa, Terfenol-D), then the former scenario will cause the magnetization to rotate away from the major axis (or the so-called easy axis of the nanomagnet) towards the minor axis (or the hard axis) while the latter scenario will not cause any rotation. On the other hand, if the magnetostriction is negative (Co, Ni), then the latter scenario will make the magnetization rotate towards the minor axis, while the former scenario will not cause rotation.

Thus, by choosing the appropriate voltage polarity (depending on the sign of the magnetostriction), we can rotate the magnetization from a stable direction along the easy axis by up to $90^0$ and make it align along the hard axis. This mechanism allows a maximum of $90^0$ rotation and not the full $180^0$ rotation needed to flip the magnetization and bring about a complete magnetic reversal. Therefore, this strategy does not allow one to write either bit 0 or 1 deterministically. In fact, after the voltage (stress) is withdrawn, the magnetization will inevitably return to assume an orientation along the easy (major) axis, but with equal probability of pointing to the right or left. Therefore, this allows for writing a desired bit into the nanomagnet with only 50% likelihood.

It might appear that we can encode the binary bits 0 and 1 in two mutually *perpendicular* directions as opposed to antiparallel directions. Say, we encode bit 0 in the magnetization state aligned along the easy axis and bit 1 in the magnetization state aligned along the hard axis. In this case, to store bit 1, we must keep the stress (voltage) on all the time. If we turn the voltage off, the magnetization will return to the easy axis and the bit 1 will be replaced by bit 0. Therefore, this makes for a *volatile* memory where information is lost if the power is turned off. Volatile memory is undesirable since the stored bit will have to be continuously refreshed and the refresh cycles consume an enormous amount of energy.

There are many ways out of this conundrum. One possibility is that *as soon as* the $90^0$ rotation is completed and the magnetization aligns along the minor axis, we withdraw the stress. In that case, the magnetization will continue to rotate further (beyond $90^0$) under an inertial torque and complete $180^0$ rotation or full reversal and stop [16]. This is depicted in Fig. 4 (a). The energy dissipated to flip the magnetization with this scheme can be exceptionally low (< 1 aJ) while the switching delay can be sub-ns as shown in the plots in Fig. 4(b) [17]. There are, however, two disadvantages of this approach. The first is that one would have to know precisely at what juncture the $90^0$ rotation is completed and withdraw the stress exactly at that juncture. This is impossible to do at room temperature because thermal noise will introduce a sizable spread in the time it takes to complete the $90^0$ rotation. Hence, we will often fail to withdraw the stress at the correct time and this will cause a large switching error probability. The second disadvantage is that this is a "toggle" approach unlike approaches like STT or SOT which are non-toggle approaches. Say, we wish to orient the magnetization to the right along the easy axis, i.e. to write a particular bit into the nanomagnet. We must first read the stored bit, i.e. determine in which direction the magnetization is pointing. If it is already pointing in the desired direction, we will do nothing. Otherwise, we will flip the magnetization as just described and this will write the desired bit. A memory cell based on this kind of writing scheme is called "toggle memory" since all we can do is toggle the magnetization. As a result, it is always necessary to *read the previously stored bit first* and then take (or not take) the action to toggle.

There is a non-toggle version of straintronic memory as well [18], but that needs the use of an in-plane magnetic field. The idea there is to apply the magnetic field (of the right strength) along the minor axis of the elliptical soft layer which will bring the two stable orientations out of the major axis and make them point in two directions that are mutually perpendicular as shown in Fig. 5. In that case, if the magnetostriction of the soft layer is positive, then applying tensile stress along one of the two directions will orient the magnetization along that direction, while compressive stress will orient the magnetization along the other direction. The opposite will be true if the magnetostriction is negative. Thus, we can write the desired bit (i.e. orient the magnetization along either direction) by simply choosing the sign of the stress applied along one of the two directions, without having to read the stored bit first. This strategy also has the advantage that no precise timing of the stress cycle is needed. Hence, the write error probability (WEP) will be considerably lower at room temperature. The disadvantage is the need for the in-plane magnetic field which also has to be of the right strength to make the two stable directions mutually perpendicular. The required field strength may vary from one nanomagnet to another (because of variations in the size and shape of the nanomagnets) and this poses a challenge. Misalignment of the nanomagnet's easy axis with the magnetic field strength poses another serious challenge. Finally, if the magnetization direction is "read" with a magnetic tunnel junction (MTJ), the resistance on/off ratio of the MTJ (one orientation is "on" and the other "off") will be poor because the orientations are not antiparallel. This can be ameliorated somewhat by using two pairs of electrodes instead of one to apply the stress [19] and that can allow the angular separation between the two stable directions to exceed $90^0$, resulting in a larger on/off ratio.

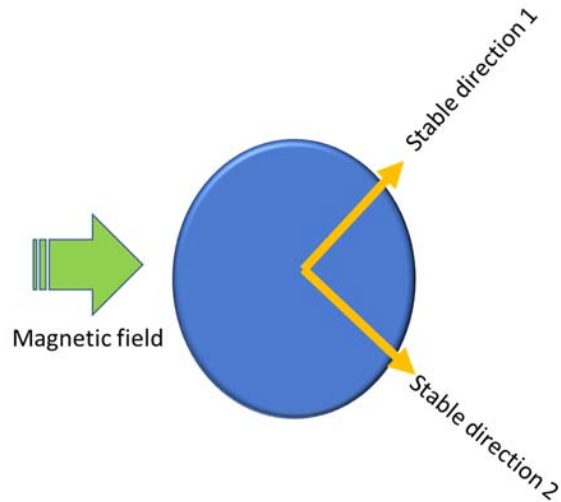

Fig. 5: A "non-toggle" straintronic writing scheme. An in-plane magnetic field of the correct strength brings the two stable magnetization directions out of the major axis an make them subtend $90^0$ angle with each other in the plane of the nanomagnet. If the magnetostriction is positive, then compressive stress along direction 1 will align the magnetization along direction 2 and tensile stress will align it along direction 1. The opposite will be true if the magnetostriction is negative. Thus, we can align along either direction by choosing the sign of the stress, without having to know the previous orientation of the magnetization.

Another strategy is to apply two uniaxial stresses in two different directions (neither of which is along the easy or hard axis of the nanomagnet) *sequentially* and that can complete the $180^0$ rotation in two steps [20]. Two antipodal gate pairs are delineated on the piezoelectric film surrounding the nanomagnet as shown in the top panel of Fig. 6. The lines joining the centers of two opposite pairs subtend an acute angle. One pair is first activated by applying a voltage to it (the members of the pair are electrically shorted together). This would generate biaxial stress in the piezoelectric, but we can approximate the effect by assuming that uniaxial stress is generated along the line joining the two activated gate pads. Such a stress will rotate the magnetization away from the major axis (easy axis) of the soft layer and roughly stabilize it in a direction perpendicular to the line joining the activated pair if the product of the stress and magnetostriction has a negative sign. The other pair is then activated (followed by deactivation of the first pair) and this rotates the magnetization further, bringing the total rotation to an angle θ, where $90^0 < θ < 180^0$. Finally when the second gate pair is deactivated, the magnetization relaxes to the *nearest* stable state along the major axis, which is opposite to the initial direction. That completes $180^0$ rotation, or complete magnetization reversal. This strategy would require four gate pads as shown in the top panel of Fig. 6, but it eliminates the requirement of having to time the stress pulse precisely, which is a very difficult proposition when thermal noise is present. It also does not require any in-plane magnetic field. This switching mechanism has been demonstrated experimentally [21] (see Fig. 6) using Co nanomagnets delineated on a piezoelectric (001) PMN-PT substrate. The downside is that the increased number of electrodes increases the device footprint.

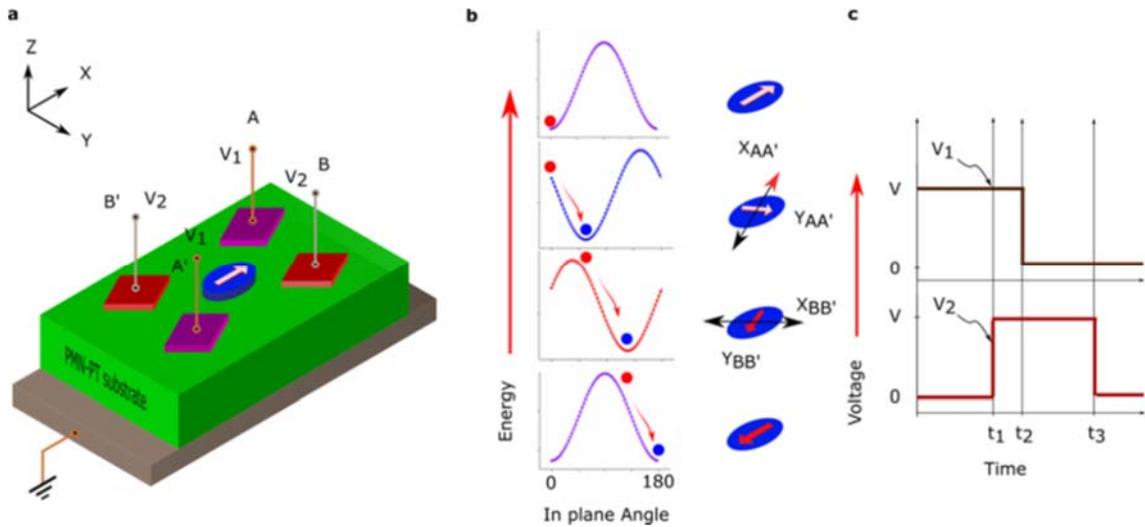

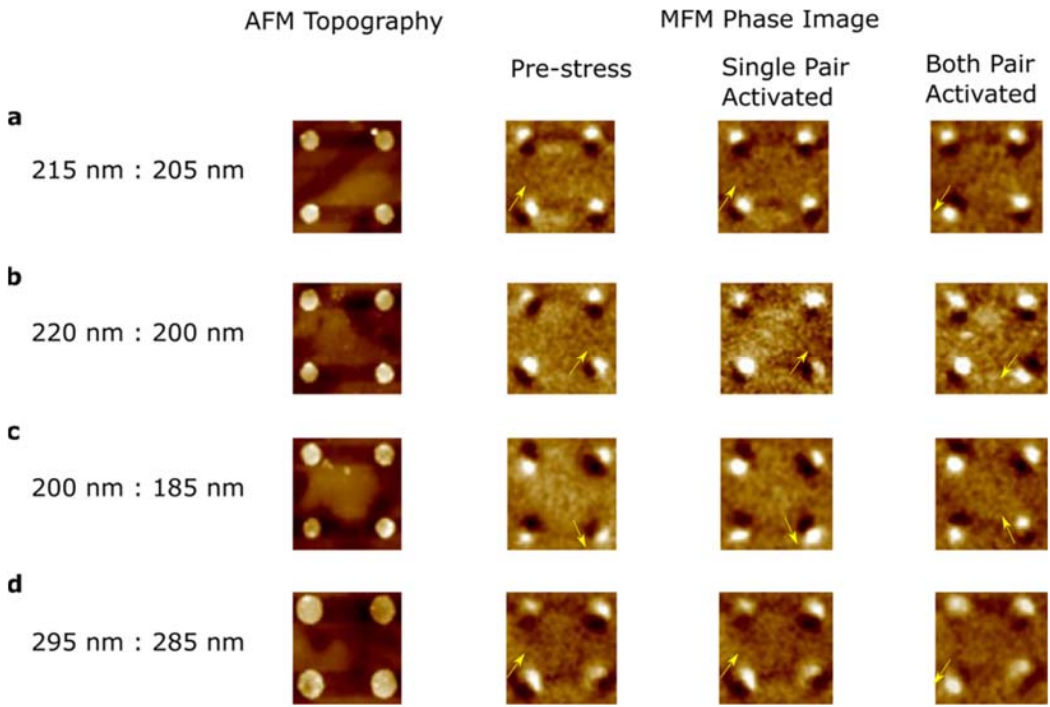

Fig. 6: (Top panel) a. Electrode placement for applying uniaxial stresses to an elliptical nanomagnet in two different directions (neither collinear with a principal axis) by sequentially activating the electrode pairs AA' and BB'. The nanomagnet is placed on a piezoelectric substrate. b. Potential energy profile of the nanomagnet as a function of the angle subtended by the magnetization with the major axis (direction of the arrow shown) - with no electrode activated, only AA' activated, BB' activated and AA' de-activated, and all electrodes de-activated. c. Timing sequence of the voltage $V_1$ applied to electrode pair AA' and $V_2$ applied to electrode pair BB'. (Bottom panel). Atomic and magnetic force micrographs of four different sets of Co nanomagnet assemblies (with different major and minor axes dimensions) on a PMN-PT substrate subjected to this stress sequence. The magnetic force micrographs are shown at three different stages of electrode activation. One out of four nanomagnets flip completely ($180^0$ rotation) upon completion of the stress cycle, showing that the switching is error-prone (25% success probability). Reproduced from [21] with permission of the American Chemical Society.

Several groups have experimentally studied the control of magnetization in magnetostrictive films deposited on piezoelectric films using voltage generated strain (i.e. a two phase multiferroic [22]), while demonstrating reversible control of nanomagnetic domains [23], repeatable reversal of perpendicular magnetization in the absence of a magnetic field in regions of a Ni film [24], and strain assisted reversal of perpendicular magnetization in Co/Ni multilayers [25]. Others have shown the use of strain control of magnetization orientation in manganite titanate [26] and lanthanum strontium manganite (LSMO) films [27], iron films [28], TbCo$_2$/FeCo multilayers [29] and strain control of magnetic properties of FeGa/NiFe multilayer films [30] as well as FeGa films [31]. Strain has been shown to reorient magnetization in Ni rings [32, 33] and Ni squares of 2 microns side [34] and the soft layer of MTJs of lateral dimensions 20 μm × 40 μm [35]. The magneto-electric effect has also been used to read the magnetization orientation in a composite multiferroic heterostructure [Ni(TbCo$_2$/FeCo)]/[Pb(Mg$_{1/3}$Nb$_{2/3}$)O$_3$]$_{1-x}$ [PbTiO$_3$]$_x$ [36].

There are many reports of demonstrated control of magnetization in nanomagnets deposited on piezoelectric substrates. For example, an electric field induced stress mediated reversible control of magnetization orientation in nanomagnets of nominal lateral dimensions 380 nm × 150 nm deposited on a 1.28 micron PZT thin film was demonstrated with the application of 1.5 V to the PZT film [36]. Furthermore, building on individual control of magnetoelectric heterostructures with localized strain to reorient the magnetization in a Ni ring of 1000 nm outer diameter, 700nm inner diameter, and 15 nm thickness, deterministic multistep reorientation of magnetization in a 400 nm Ni dot of 15 nm thickness has been reported [37].

Uniform magnetization rotation through 90º has also been demonstrated through imaging with X-ray photoemission electron microscopy (X-PEEM) and X-ray magnetic circular dichroism (XMCD) in elliptical nanomagnets of nominal lateral dimensions ~100 nm ×150 nm [38].

There are also reports of switching the resistance states of MTJs with electrically generated mechanical strain [39, 40], making this methodology mainstream for application in magnetic random access memory. In [40], the experimentally measured high/low resistance ratio, sometimes referred to as "tunneling magnetoresistance ratio" (TMR) of such MTJs exceeded 2:1 (or 100%) at room temperature, which is very respectable.

### A. Switching error in straintronic switches

One feature that stands out in Fig. 6 (which shows complete magnetic reversal under strain or straintronic switching) is that the switching is not particularly *reliable* – only one out of four nanomagnets would switch, resulting in barely 25% switching success. A possible reason for this poor statistics might have been the fact that the nanomagnets are made of Co which is weakly magnetostrictive (saturation magnetostriction ~ 60 ppm). Therefore, in a subsequent experiment, we used FeGa nanomagnets on a PMN-PT substrate to check if the statistics would improve with a material that has a higher magnetostriction (200-300 ppm). We carried out two sets of experiments with two different materials: Co and FeGa. We did not use the four electrode configuration, but instead used a different principle to ensure $180^0$ rotation of the magnetization. We fabricated pairs of dipole coupled nanomagnets, with one member of the pair more elliptical than the other. The line joining their centers is parallel to the hard axis (minor axis of the ellipse). This is shown in Fig. 7(a). The more elliptical nanomagnet has a larger shape anisotropy energy barrier and hence its magnetization state is "harder". In the ground state configuration, the two magnetizations will be mutually antiparallel because of dipole interactions.

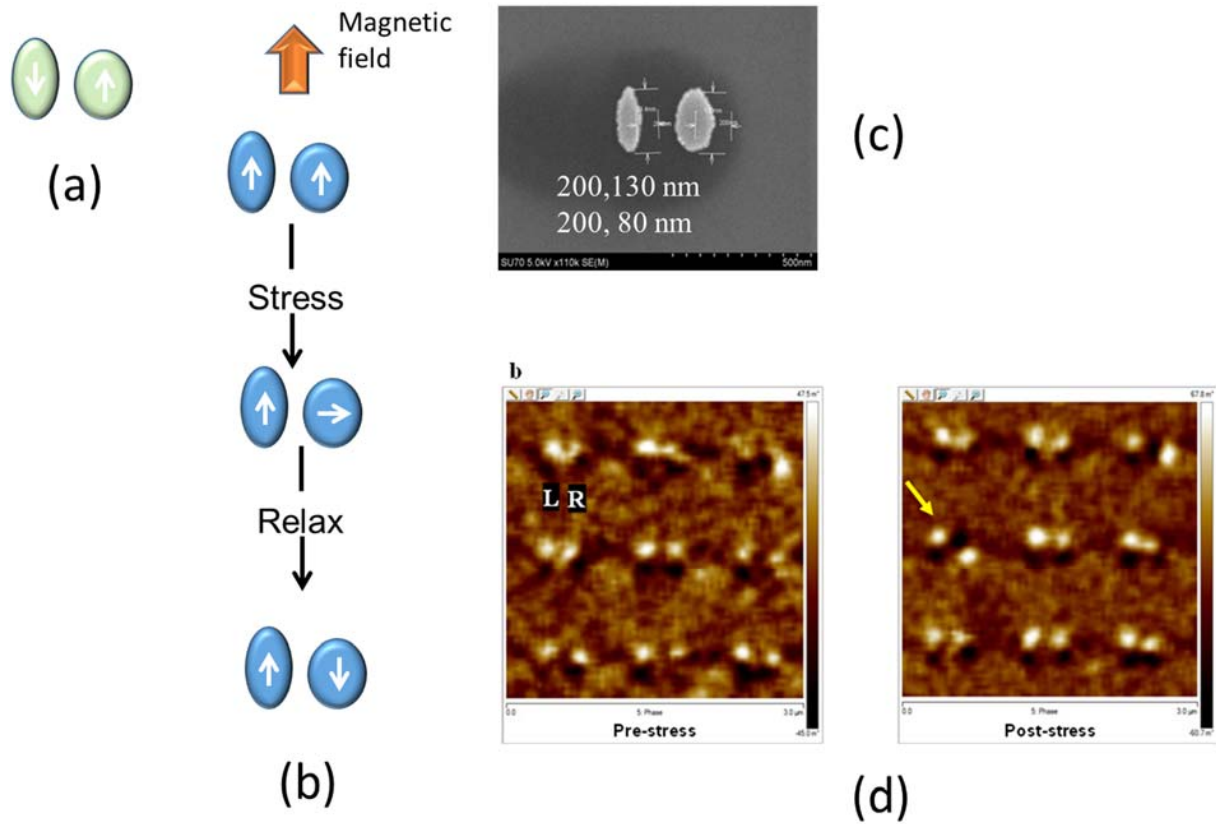

Fig. 7: (a) Two closely spaced elliptical nanomagnets, one more elliptical than the other, positioned such that the line joining their centers is parallel to the hard axis (minor axes of the ellipses). (b) A sequence of applying a global magnetic field to align the magnetizations in the same direction, followed by stress and relaxation. The magnetizations at various steps are shown. (c) Scanning electron micrograph of two Co nanomagnets fabricated on a PMN-PT substrate. The major and minor axes dimensions are stated in the figure. (d) Magnetic force micrographs of 9 pairs showing that a magnetic field aligns the magnetizations all parallel to each other (left panel). After applying stress and relaxing, one out of nine pairs assumes the expected antiparallel configuration (right panel). Reproduced from [41] with permission of the American Chemical Society.

Suppose a global magnetic field aligns the magnetizations of a pair in the same direction, placing the duo in an excited metastable state. After removal of this magnetic field, the magnetization of the softer nanomagnet may not be able to flip by itself (and assume the antiparallel ordering of the ground state) because the dipole coupling strength may not be able to overcome the shape anisotropy energy barrier in the softer nanomagnet and make its magnetization flip. A global uniaxial stress of the correct sign and sufficient strength applied along the direction of the major axes of both nanomagnets will rotate the magnetization of the softer nanomagnet by ~$90^0$ (the harder nanomagnet's magnetization is stiff and hence will barely rotate). Then, upon stress release, the softer nanomagnet will more likely (with > 50% probability) flip its magnetization to assume the antiparallel configuration because of the dipole coupling influence of its neighbor. This scenario is depicted in Fig. 7(b). We can actually think of the pair as implementing a Boolean NOT gate if we view the magnetization of the left nanomagnet as encoding the input bit and that of the right nanomagnet as implementing the output bit. The stress acts as a clock to trigger the NOT operation.

Fig. 7(c) shows a scanning electron micrograph of such a pair (Co nanomagnets fabricated on a PMN-PT substrate) where the major axis dimension of both nanomagnets is 200 nm [41]. The minor axis dimension of the softer nanomagnet is 130 nm, while that of the harder nanomagnet is 80 nm. In the left panel of Fig. 7(d), we present magnetic force micrographs showing the magnetizations of 9 such pairs after being subjected to a global magnetic field. The field orients the magnetizations of all nanomagnets in its own direction. In the right panel, we show the magnetic force micrographs after the magnetic field was removed and the nanomagnets were subjected to global stress. Note that only one out of nine pairs switched to assume the corerct antiparallel configuration [41]. We would expect the fraction of switching pairs to exceed 50%, whereas the observed fraction is ~11%.

We repeated this experiment with FeGa nanomagnets with the expectation that the switching probability will increase because of the five times higher magnetostriction of FeGa compared to Co [42]. The corresponding magnetic force micrographs depicting the magnetization states of elliptical FeGa dipole coupled pairs is shown in Fig. 8. The hard nanomagnet's major axis is 350 nm and minor axis is 200 nm, while the soft nanomagnet's major axis is 285 nm and the minor axis 265 nm. The center to center spacing is 330 nm and the magnet thickness is 8-9 nm.

In this case, one out of four pairs switch correctly, indicating that the switching success probability has improved to 25%, but that is still considerably less than the expected probability of at least 50%. All this shows that straintronic switching is rather error-prone and hence it may not be suitable for certain types of applications despite its excellent energy efficiency. It is particularly unsuitable for Boolean logic applications [43, 44] which demand high fidelity and reliability. Boolean logic has strict requirements on tolerable switching error probability because errors in logic circuits *propagate* throughout the chip. If the output bit of one gate is corrupted and that bit is then fed as input to another gate, then the latter's output gets corrupted as well. Thus, errors are "contagious" in logic circuits, unlike in memory circuits. If one cell in a memory array is corrupted, it does not corrupt any other cell, so that the error is 'contained'. This is why it is easy to construct error correction schemes for memory, but it is extremely difficult to do so in logic circuits where errors are dynamic and propagating.

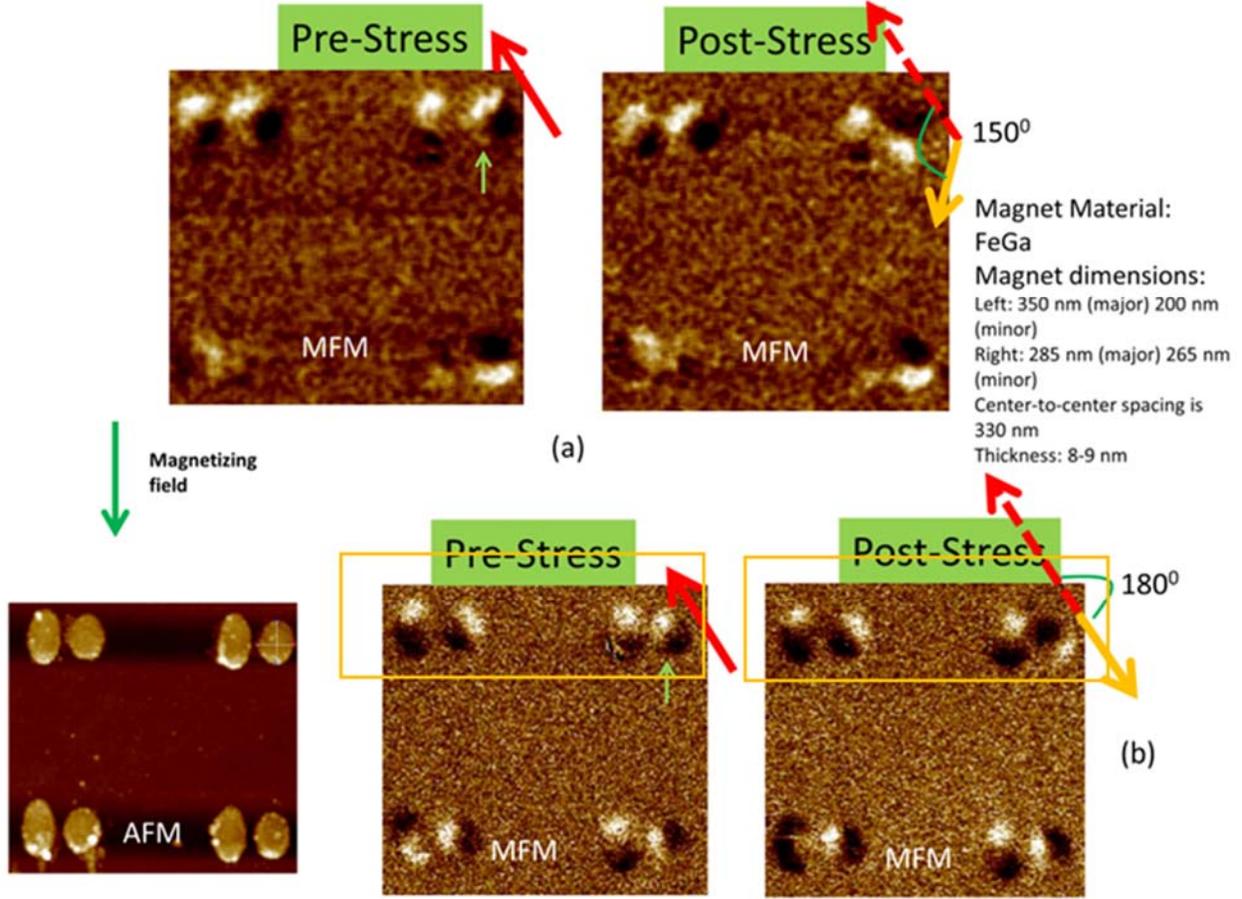

Fig. 8: Pre- and post-stress magnetic force micrographs of dipole coupled pairs of FeGa nanomagnets which are initially subjected to a global magnetic field to make their magnetizations parallel to each other. After the stress cycle, one out of four pairs switch. Reproduced from [42]. No copyright permission required.

There have been numerous theoretical simulations of magneto-dynamics in pristine (defect-free) nanomagnets in the presence of thermal noise to estimate the switching error probability when the nanomagnet is switched magnet-elastically (i.e. with strain). These simulations consider switching errors caused only by thermal noise (not defects) and are usually based on the Landau-Lifshitz-Gilbert-Langevin (or stochastic Landau-Lifshitz-Gilbert) equation [45-51]. They have shown that straintronic switching error probability can range between $10^{-3} – 10^{-8}$ at room temperature (owing to thermal noise alone) which is, of course, too high for logic where the error probability might need to be no larger than $10^{-15}$ [52].

The stochastic Landau-Lifshitz-Gilbert equation is

$$\frac{d\vec{m}(t)}{dt} = -|\gamma|\vec{m}(t) \times \left[ \vec{H}_{eff}(t) - \frac{\alpha}{|\gamma|}\left( \frac{d\vec{m}(t)}{dt} + \tau \frac{d^2\vec{m}(t)}{dt^2} \right) \right] \quad , \quad (8)$$

$$\vec{H}_{eff}(t) = \vec{H}_d + \vec{H}_c + \vec{H}_s(t) + \vec{H}_{ex}(t) + \vec{H}_{th}(t)$$

where $\vec{m}(t)$ is the time-dependent magnetization (normalized to the saturation magnetization of the nanomagnet's material), $\gamma$ is the gyromagnetic factor (a universal constant), $\alpha$ is the Gilbert damping factor (a material constant) and $\tau$ is a time-constant introduced to account for spin inertia [53]. Typically, $\tau$ is of the order 1 -100 ps. The effective magnetic field in the equation is given by

$$\vec{H}_{eff}(t) = \vec{H}_d + \vec{H}_c + \vec{H}_s(t) + \vec{H}_{ex}(t) + \vec{H}_{th}(t), \tag{9}$$

where $\vec{H}_d$ is the demagnetization field (due to the shape anisotropy arising from the elliptical shape of the nanomagnet), $\vec{H}_c$ is the field due to any magneto-crystalline anisotropy, $\vec{H}_s(t)$ is the field due to strain, $\vec{H}_{ex}(t)$ is the field due to exchange interaction between the spins (this term would be usually neglected in a single-domain nanomagnet) and $\vec{H}_{th}(t)$ is the random field due to thermal noise (white Gaussian noise). Expressions for these fields can be found in refs. [45-51]. Normally, the spin-inertia term in Equation (8) is neglected since the magnetization switching takes place over ~ 1 ns, which is much longer than $\tau$. However, it has been recently found that spin-inertia can have an effect on the switching error probability [54].

In the stochastic Landau-Lifshitz-Gilbert approach, one would solve Equations (8) and (9) numerically using different seeds for the random number generator that generates $\vec{H}_{th}(t)$. Each one generates a switching trajectory. Each trajectory is simulated until the magnetization reaches a steady-state which is either the starting state (switching failed) or the destination state (switching succeeded). The switching error probability is the fraction of trajectories that end in failure.

The switching error probabilities calculated for straintronic switching and reported in refs. [45-51] are way too high for applications in Boolean logic. The situation gets even worse when we consider real nanomagnets that have geometric defects such as surface or edge roughness [as seen in Fig. 7(c)], or structural defects such as vacancies or ridge formation at the boundaries. Localized defects (such as an isolated vacancy due to a few missing atoms) are relatively innocuous [55], but extended defects such as thickness variation along a significant fraction of the nanomagnet's surface, or along the edges, can increase the error probability by several orders of magnitude [43]. This, more than anything else, makes the viability of straintronic logic dubious.

Fortunately, there are many other information processing paradigms, very different from Boolean logic and memory, which are much more forgiving of errors. They usually involve collective computational models where the cooperative activities of many devices acting in unison elicit the computational activity and the failure of even a significant fraction of them do not impair the circuit operation. Other paradigms that leverage low-energy-barrier nanomagnets for probabilistic computing [56] are also quite resilient against structural defects in the nanomagnets since they do not depend on binary switching unlike Boolean logic and memory, but instead depend on the probability distribution of magnetization states. The probability distribution curve is relatively immune to even significant variations in nanomagnet thickness or lateral dimensions [57]. That of course does not mean that defects have no effect on performance. For example, when low-energy-barrier nanomagnets are used for binary stochastic neurons in probabilistic computing models, extended defects can still have mild deleterious effects, but they are usually somewhat benign and certainly not catastrophic [58, 59].

## IV. STRAINTRONIC MEMORY

The unacceptably large switching error rates in straintronic switches will very likely preclude any application in conventional Boolean logic devices and circuits. We will discuss this further in Section V, but here we emphasize that "memory" is much more tolerant of switching errors than "logic". There are well-known error correction protocols for memory chips, but not for logic. Hence, it is believed that straintronics may enable extremely energy-efficient memory in the long run, if we can improve error resilience to the point where the room-temperature error probability in realistic structures is $\sim 10^{-5}$ or smaller. Early voltage controlled magnetic anisotropy (VCMA) based memory reported error probabilities of that order [60] and hence error probabilities of that order are acceptable if the energy dissipation is low. Thus, while straintronic Boolean logic appears impractical because of the poor error-resilience, straintronic memory seems to be not just practical, but also attractive because of the very low energy dissipation incurred during the writing operation.

**Memory scaling issues:** While the above bodes well for the development of a purely strain switched toggle magnetic memory, there are size scaling issues that still need to be addressed. The most important consideration when it comes to "memory" is the *density* of cells (or bit density). This would mandate adopting perpendicular magnetic tunnel junctions (p-MTJs) with lateral dimensions less than 20 nm in order to be competitive with experimentally demonstrated STT-RAM of 11 nm diameter [61].

Consider such a p-MTJ with an energy barrier $E_b = K_u V$ in the soft layer, where $K_u$ is the uniaxial anisotropy energy density and $V$ is the volume of the soft layer. For a soft layer of diameter $\sim 20$ nm and thickness $\sim 1$ nm, which is typical for p-MTJs, the volume $\sim 314.2$ nm$^3$. Assume $E_b \sim 1$ eV $= 40\ kT$ (where $k$ is the Boltzmann constant and $T$ the absolute temperature, assumed to be 300 K) which ensures that the magnetization does not switch spontaneously from one stable state to another at room temperature because of thermal noise (retention time $\sim 10$ years). Thus, $K_u = 5.1 \times 10^5$ J/m$^3$.

To switch the magnetization with strain alone, we will need to overcome this energy barrier and hence need that

$$K_u \leq \frac{3}{2} \lambda_s \sigma$$

where $(3/2)\lambda_s$ is the saturation magnetostriction and $\sigma$ is the stress developed in the magnetostrictive soft magnetic layer. Even if we assume an optimistic $(3/2)\lambda_s \sim 500$ micro-strain, the stress required will be $\sim 1000$ MPa, which is impractical to apply either via direct strain transferred from an underlying piezoelectric layer or by the use of surface acoustic waves (SAW). Highly magnetostrictive materials, for example Terfenol-D [62], will not achieve a larger magnetization deflection at low stress/strain levels due to the bidirectional coupling between the magnetization and strain [63]. Hence, while toggle memory switched with strain would scale to $\sim 100$ nm lateral dimensions, there is a need to explore other strain based and hybrid switching mechanism to scale to lateral dimensions well below 50 nm. *This is currently the most serious challenge to straintronic memory.*

### A. Memory based on time varying strain (acoustic waves)

In the preceding paragraphs, we discussed switching the magnetization of a magnetostrictive nanomagnet's magnetization with static (time-invariant) strain. In this section, we will discuss the effect of time-varying strain produced by an acoustic wave launched into the piezoelectric substrate underneath the magnetostrictive naomagnets. Time-varying strain results in an equally (if not more) energy efficient modality of switching the magnetization of nanomagnets. Surprisingly, it appears to be more reliable than switching with static strain [43], although the reason behind the increased error resilience is not well understood, except that it *may* be due to the fact that multiple cycles of the strain *repeatedly* goad the nanomagnet to switch, thereby increasing the switching probability.

One way to subject a magnetostrictive nanomagnet to periodic time-varying strain is to place it on a piezoelectric substrate and then launch a surface acoustic wave (SAW) in the substrate. As the SAW passes underneath the nanomagnet, it subjects the latter to time-varying strain which will make it expand and contract periodically. The inverse magnetostriction effect (Villari effect) will then make the nanomagnet's magnetization vector rotate periodically, or precess. If the angle of precession can be made large enough, a bistable nanomagnet can switch from one stable magnetization state to the other under some circumstances, such as when there is a real or effective magnetic field, or a small spin polarized current, present to aid the switching.

The SAW can be launched with electrodes delineated on the substrate. A time-varying (ac) voltage is applied to the electrodes which causes a time varying electric field around them and that then causes a time varying strain which propagates through the substrate, resulting in an acoustic wave. The wavelength of the wave is determined by the relation $\lambda = v_{ac}/f$, where $f$ is the frequency of the time varying voltage (which is the same as the frequency of the time varying strain) and $v_{ac}$ is the acoustic wave velocity. The wave decays into the substrate with characteristic decay length on the order of the wavelength $\lambda$. Thus, if the wavelength is much smaller than the substrate thickness, then the power in the wave is confined to the surface and it is called a surface acoustic wave (SAW).

The most common type of electrode for launching a SAW is an interdigitated transducer (IDT) consisting of interwoven electrodes forming a comb-like pattern as shown later in Fig. 10. This type of electrode will typically launch the Rayleigh or the Sezawa mode of SAW [64]. Other types of electrodes can also launch a SAW, but not of the Rayleigh or Sezawa type. If we have two solid electrodes at the two edges of a piezoelectric substrate and then we apply a time varying voltage between them, the region of the substrate pinched between the two electrodes will be subjected to a time-varying electric field, which will cause a time varying strain. That will also generate an acoustic wave. If the wavelength given by the relation $\lambda = v_{ac}/f$ is much smaller than the substrate thickness, it too will be a SAW, but of course not of the Rayleigh or Sezawa mode.

### B. Spin transfer torque (STT) switching of magneto-tunneling junctions for memory

Since we are discussing magnetic random access memory (MRAM), we will introduce in this sub-section the dominant MRAM device, which is the spin-transfer-torque-random-access-memory (STT-RAM). Before doing that, we will briefly introduce the *magnetic-tunnel-junction* or MTJ since it is the device that ultimately converts the magnetization state encoding the bits 0 and 1 into an electrical conductance state (high and low) for electrical reading of a stored bit in a magnetic memory cell. This action is sometimes referred to as "spin-to-charge conversion" since the bit value encoded in the spin-based magnetic state is converted into a charge-based electrical state, namely high and low conductance states.

The MTJ is a three-layered structure consisting of a "hard" nanomagnet (with stiff magnetization that is not easily rotated), an ultrathin spacer layer, and a "soft" nanomagnet whose magnetization can be rotated by strain, or a spin-polarized current passing through it and imparting spin angular momentum to the resident electrons (Fig. 9(a)). This latter mechanism is called spin-transfer-torque (STT) [65-67]. The MTJ is shown in Fig. 9(b).

When the soft layer's magnetization is parallel to that of the hard layer, the MTJ has a lower resistance than when the two magnetizations are mutually antiparallel, as shown in Figs. 9 (c) and (d). This resistance difference is a consequence of spin-dependent tunneling through the spacer. Since we always know the magnetization orientation of the hard layer, we can simply measure the resistance of the MTJ and thus determine whether the soft layer's magnetization is pointing to the left or right (i. e. parallel or antiparallel to the known magnetization of the hard layer). Thus, the magnetization state of the soft layer (which would encode binary bit information) is read electrically, resulting in "spin-to-charge conversion".

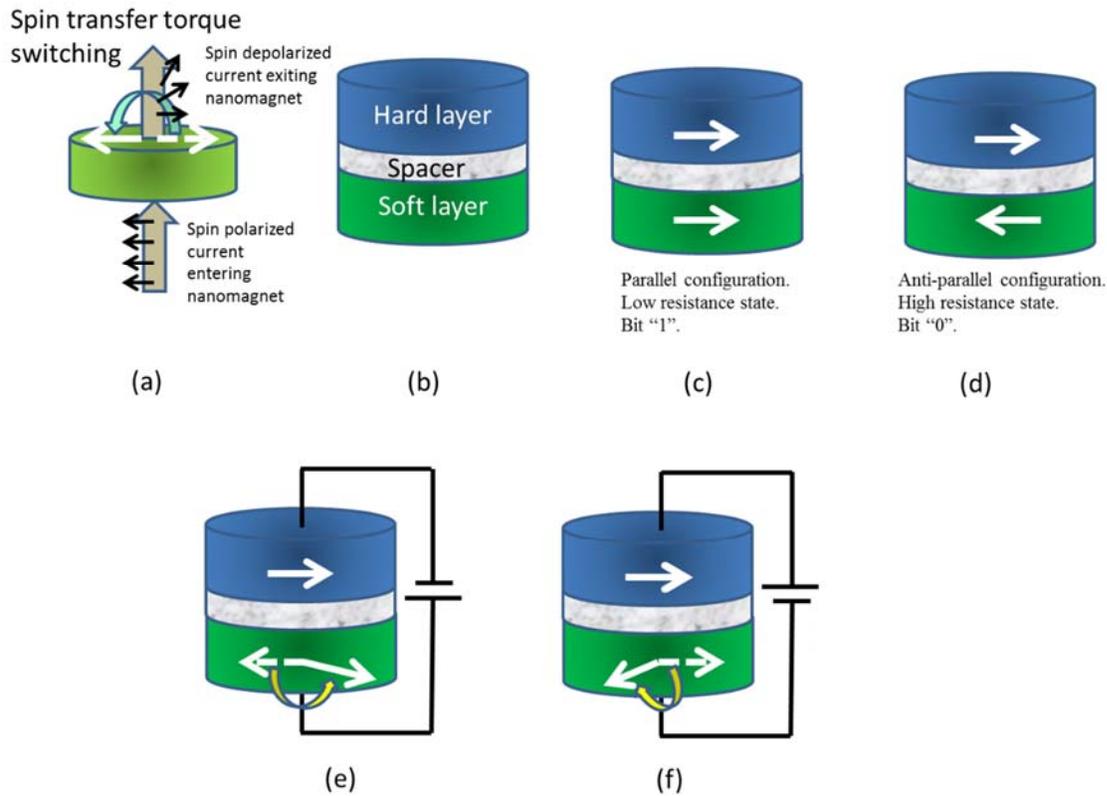

Fig. 9: (a) Spin transfer torque switching of a nanomagnet's magnetization from the right-direction (broken arrow) to the left-direction (solid arrow) with a spin polarized current in which the spins of electrons are oriented in the left-direction. (b) A magneto-tunneling junction (MTJ) structure. (c) When the hard and soft layers' magnetizations are mutually parallel, the MTJ resistance measured between the hard and soft layers is low. (d) When the hard and soft layers' magnetizations are mutually anti-parallel, the MTJ resistance is high. (e) Switching the soft layer from the anti-parallel configuration (broken arrow) to the parallel configuration (solid arrow) with the polarity of the battery shown. The hard layer acts as a spin polarizer. (f) Switching the soft layer from the parallel configuration (broken arrow) to the anti-parallel configuration (solid arrow) with the polarity of the battery reversed. The hard layer acts as a spin analyzer. Reproduced from [68] with permission of the Institute of Physics.

In addition to providing the means to "read" the magnetization state of the soft layer, the MTJ also allows one to "write" either magnetization state into the soft layer. We can align the soft layer's magnetization parallel or anti-parallel to that of hard layer by rotating the soft layer's magnetization with a spin polarized current flowing through it. This is the action of "writing" the magnetization state which then writes the bit value (0 or 1) in the resistance state of the MTJ.

In Fig. 9, we depict an MTJ built with hard and soft layers possessing IPA. The spin-polarized current flows in a direction perpendicular to the heterointerfaces by tunneling through the spacer and it is generated in the following way. If we connect the negative terminal of the battery to the hard layer and the positive terminal to the soft layer, then the hard layer will inject its majority spin electrons (spins polarized parallel to the hard layer's magnetization) into the soft layer. This constitutes a spin-polarized current injected into the soft layer. The injected spins will transfer their momenta to the spins of the resident electrons in the soft layer, which will begin to turn in the direction of the hard layer's magnetization, and ultimately the soft layer's magnetization will align along the hard layer's magnetization, thereby making the two magnetizations mutually parallel (Fig. 9 (e)). The hard layer acts as the spin polarizer and generates the spin polarized current that switches the soft layer.

If we reverse the polarity of the battery, the soft layer will inject electrons into the hard layer and generate a spin-polarized current. Electrons whose spins are aligned parallel to the hard layer's magnetization will be preferentially transmitted by the hard layer which acts as a spin analyzer or filter. Therefore, the soft layer will inject many more of those spins that are parallel to the hard layer's magnetization than those that are antiparallel. Continued injection depletes the population of the parallel spins in the soft layer, so that ultimately spins that are anti-parallel to the hard layer's magnetization become majority spins in the soft layer. This makes the soft layer's magnetization anti-parallel to that of the hard layer's (Fig. 9 (f)). Thus, we can write either bit 0 or bit 1 into the soft layer by choosing the polarity of the battery.

This method of switching magnetization with a spin polarized current is unfortunately not particularly energy-efficient because of the large magnitude of current needed to switch the magnetization of the soft layer. The switching action is likely to dissipate about $10^7$ $kT$ of energy (~1.6 pJ) to switch a single-domain nanomagnet in ~ 1 ns, even when the energy barrier $E_b$ within the magnet is only few tens of $kT$ [69]. More recent estimates bring this number down to ~100 fJ [70], which is still excessive.

Considerable amount of research has been carried out in an effort to reduce the current density, and current densities as low as 2.1 MA cm$^{-2}$ in an MTJ with a resistance-area product of 16 Ω μm$^2$ have been reported [71]. In an MTJ whose cross-sectional area is 1 μm$^2$, the power dissipated to switch would be ~ 7 mW, which is extremely high. This is why small cross-sectional areas are needed for STT switching. Attempting to reduce the switching current further by thinning the magnetic layers or the spacer layer results in dramatic reduction of the high- to low-resistance ratio, or the tunneling magnetoresistance ratio (TMR), since it is governed by spin-dependent tunneling between the magnetic layers. Typically, the energy dissipated to switch with STT is several fJ. There have been some recent efforts to reduce the energy dissipation by using spacer layers that have smaller bandgap, such as ScN, which would offer a lower tunneling resistance and hence a lower resistance-area product, but this may be counter-productive since the lower barrier to tunneling may increase the thermionic emission over the barrier. Since thermionic emission is not spin-dependent unlike tunneling, the overall effect will be to reduce the TMR even further.

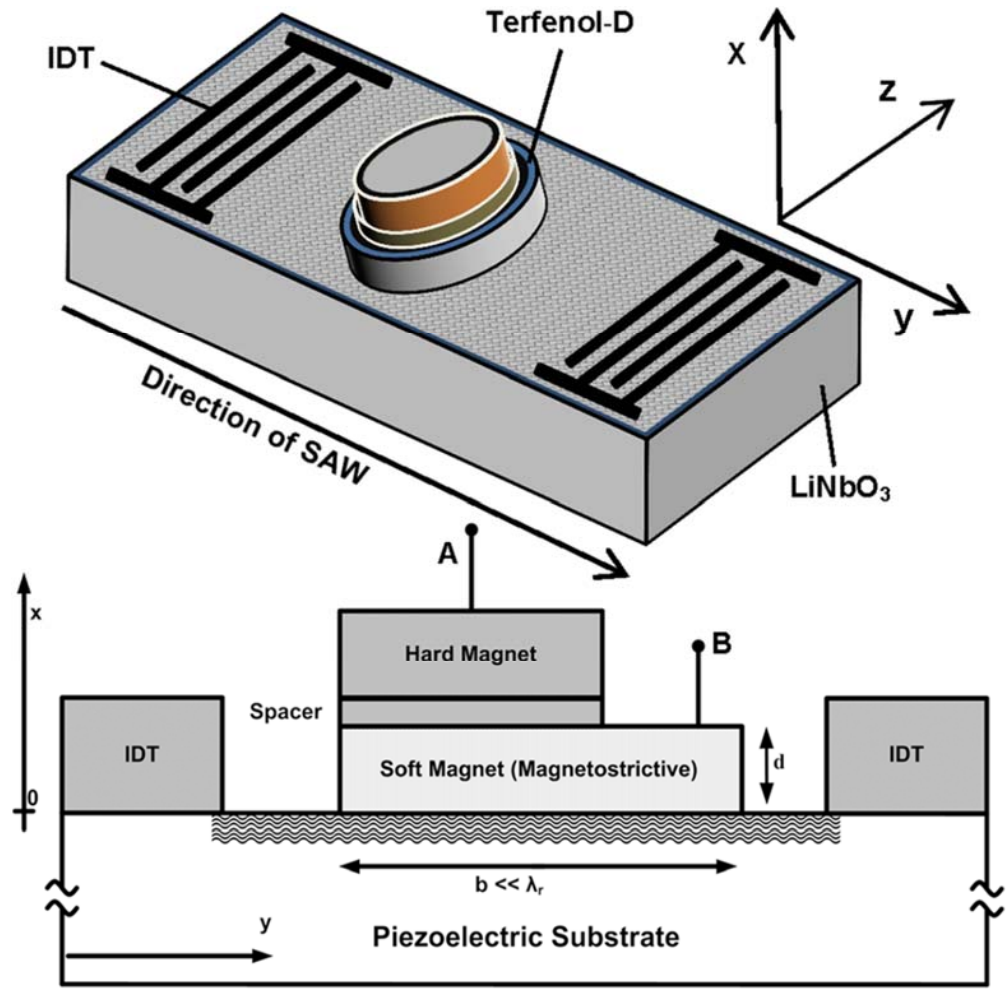

Fig. 10: (Top) Schematic illustration of the hybrid system with interdigitated transducers (IDTs) to launch the SAW and an MTJ, serving as a bit storage unit, placed between IDTs on a piezoelectric substrate. The soft layer of the MTJ is in contact with the substrate and is periodically strained by the SAW. The resistance between the terminals A and B is used to read the bit stored (we assume that both magnets are metallic). For writing, a small spin polarized current is passed between the same two terminals during the appropriate cycle of the SAW, when the magnetization rotates out of the easy axis. In this configuration, the reading and writing currents do not pass through the highly resistive piezoelectric, so the dissipation during the read/write operation is kept small. Bits are addressed for read/write using the traditional crossbar architecture. Reproduced from [74] with permission of the American Institute of Physics.

As discussed earlier, there have been proposals to replace spin-transfer-torque switching with spin-orbit-torque switching which involves passing a spin current instead of a charge current through the MTJ in order to switch the magnetization of the soft layer and thus switch the MTJ resistance. The spin current is generated by passing a charge current through a heavy metal layer (e.g. Pt, β-Ta, etc.), or a topological insulator, placed underneath the soft layer, which converts the charge current to a spin current by virtue of the giant spin Hall effect in the heavy metal layer [72] or the spin-momentum locking effect in a topological insulator [73]. The ratio of the spin current to the charge current is called the "spin Hall angle" and it is typically less than unity in the case of the heavy metal layer. However, because the charge current is passed through a *metal* layer which has a much smaller resistance than an MTJ, the power and energy dissipations are reduced because the current path is no longer though the MTJ. This also reduces damage to the MTJ caused by the flow of charge current through it. The reduction in the energy dissipation depends on the thickness of the metal layer and an analysis can be found in ref. [68] which shows how the energy dissipation depends on different geometric features. The topological insulator, on the other hand, may not have a low resistance unlike the heavy metal, but it may produce an effective spin Hall angle that exceeds unity. The salient drawback of these approaches is that they will all result in a *three-terminal* MTJ, which is unattractive for memory applications since it will have a much larger footprint than a two-terminal MTJ used in conventional spin-transfer-torque-random-access-memory (STT-RAM). On the flip side, the advantage is the reduced energy dissipation and the physical separation of the read and write paths, which avoids "read-disturb" (corrupting the stored bit during the reading operation) and reduces damage to the MTJ since the charge current path is not through the MTJ but though a different layer. This improves the memory's endurance.

### C. Hybrid switching methodology (strain and spin transfer torque)

We proposed a different approach where the memory remains two-terminal. Our approach is a *bimodal* switching mechanism in the sense that two different switching mechanisms are pressed into service at the same time to reduce the energy dissipation [74]. In our hybrid approach, we use a magnetostrictive soft layer placed atop a piezoelectric substrate. A surface acoustic wave (SAW) is launched in the substrate and flows underneath the soft layer, straining it periodically. At the same time, we synchronously pass a charge current pulse through the MTJ during the appropriate cycle of the SAW to generate spin transfer torque and drive the magnetization of the soft layer to the desired orientation. The SAW rotates the magnetization by $90^0$ during the cycle when the product of the magnetostriction and strain is negative. If during that cycle, a charge current is introduced to produce spin transfer torque (STT), then a complete $180^0$ rotation can be achieved with *reduced charge current* since the SAW lends a helping hand to the STT. In fact, SAW does the "heavy lifting" and since it is much more energy efficient than STT, the overall energy dissipation is reduced, perhaps by an order of magnitude [74]. Two conditions however must be fulfilled for reliability: (1) the probability of switching the magnetization of a magnet to the desired orientation (writing of bits) must be ~100% at room temperature when the STT charge current is injected, and (2) the probability of unintentionally switching the magnet due to the SAW alone is ~0% at room temperature when no STT current is injected. This will ensure that bits are written reliably in the target memory cells and data already stored in other cells are not corrupted. We showed in ref. [74] that both conditions can be fulfilled with proper design. The structure for this bimodal switching is shown in Fig. 10. A very small amount of energy is required to generate the global SAW that acts on all MTJs on the wafer, and when that energy is amortized over all the MTJs, the energy cost per MTJ is miniscule. We found that this approach can reduce the write energy dissipation in a memory cell by approximately an order of magnitude. Further reduction may be possible with design optimization.

Periodic switching of magnetization between the hard and the easy axis of 40 μm × 10 μm × 10 nm Co bars sputtered on $LiNbO_3$ has been shown [75]. Other authors have studied acoustically induced switching in thin films [76] including focusing surface acoustic waves (SAW) to switch a specific spot in an iron-gallium film [77]. The influence of frequency and wavevector have also been studied [78]. Several proposals suggest a complete 180° rotation with an appropriately timed acoustic pulse [79]. Stroboscopic X-ray techniques have been used to study strain waves and magnetization at the nanoscale [80].

Excitation of spin wave modes in GaMnAs layers by a picosecond strain pulse [81] as well as magnetization dynamics in GaMnAs [82] and GaMn(As,P) [83] have been demonstrated. In in-plane magnetized systems, surface acoustic waves have been utilized to drive ferromagnetic resonance in thin Ni films [84, 85]. Resonant effects have also been studied by spatial mapping of focused SAWs [86]. There are theoretical studies of the possibility of complete magnetization reversal in a nanomagnet subjected to acoustic wave pulses [87]. Interestingly, for high frequency excitation of extremely small nanomagnets, the Einstein De Haas effect seems to dominate as has been proposed [87] and experimentally demonstrated [88, 89].

### D.  Straintronic Magnetic Tunnel Junctions

There have been reports of switching the conductance state of a magnetic tunnel junction with static strain alone, in a configuration very similar to that shown in Fig. 4(a). Strain induced switching of an MTJ was first demonstrated by Li, et al. [90] in large area devices. Later, Zhao, et al. demonstrated (static) strain induced switching of the conductance of a µm-scale MTJ with a room temperature tunneling magneto-resistance ratio (TMR) exceeding 100% [91]. They also demonstrated that strain modulated the coercivity of the soft layer.

Fig. 11 shows the structure of the straintronic MTJ, the strain distribution around the MTJ in the piezoelectric film when a gate voltage is applied across the piezoelectric layer and the magneto-resistance traces obtained at different gate voltages. Fig. 12 shows the simulated magnetization distributions within the soft layer at two different gates voltages and also the resistance switching as a function of the gate voltage. The ratio of the OFF-to-ON resistance (which is essentially the TMR) exceeds 2:1 at room temperature. The applied gate voltage in this experiment was large because the piezoelectric layer had a large thickness of 0.5 mm. Reducing the layer thickness to 100 nm would reduce the gate voltage to 16 mV, making the switching extremely energy-efficient.

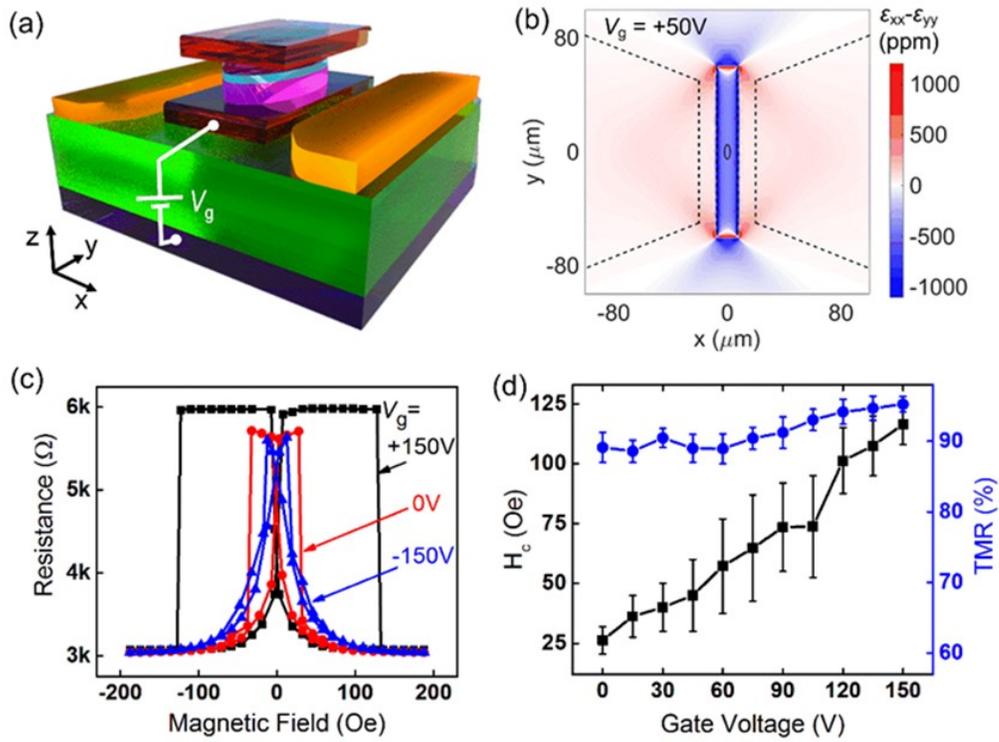

Fig. 11: (a) Schematic of a straintronic magnto-tunneling junction. A voltage $V_g$ is applied across the piezoelectric layer shown in green. (b) The in-plane anisotropic strain $\varepsilon_{xx} - \varepsilon_{yy}$ profile generated in the piezoelectric layer upon application of a gate voltage $V_g = +50\,\text{V}$. The solid line ellipse at the center denotes the MTJ pillar, and the dashed lines denote the positions of electrodes and side gates shown in (a). This result is generated with COMSOL Multiphysics software. (c) Magnetoresistance traces measured under different gate voltages $V_g$. (d) Variation of the switching (magnetic) field [squares] and tunneling magnetoresistance ratio (TMR) [circles] of the MTJ as a function of $V_g$. Reproduced from [91] with permission of the American Institute of Physics.

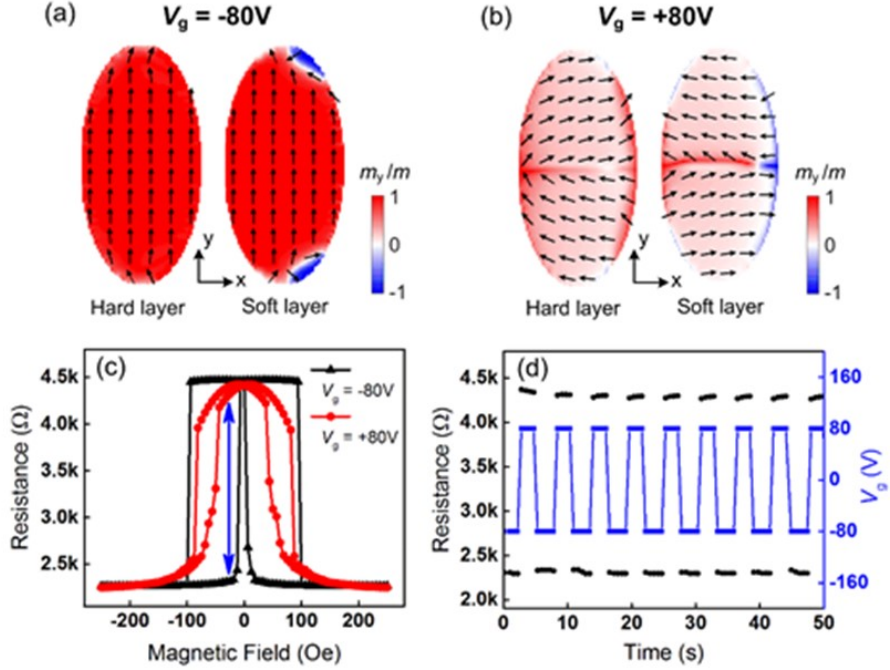

Fig. 12: (a)-(b) Micromagnetic simulation results showing the magnetization configurations of the hard and soft layers of the MTJ in Fig. 11, after application of gate voltage (a) $V_g = -80$ V and (b) $V_g = +80$ V. A small bias field of 30 Oe is applied along the major axis to overcome any effect of dipole interaction. The dimension of the MTJ is minor axis = 3 µm and major axis = 6 µm. Black arrows indicate the direction of magnetic moments. (c) Measured magnetoresistance loops for $V_g = -80$ V and $V_g = +80$ V. The blue arrow indicates the switchable high- and low-resistance states. (d) Toggling of the MTJ between high- and low-resistance states with application of ±80 V gate voltage pulsing. A small bias magnetic field of 30 Oe is applied along the +*y*-axis (refer to Fig. 11a) to overcome the dipole interaction between the two magnetic layers. Reproduced from [91] with permission of the American Institute of Physics.

### E. Mixed-mode magnetic tunnel junctions: Surface acoustic wave and spin transfer torque

One way to address the scaling problem of straintronic memory, alluded to in Section IV.B, and still achieve an order of magnitude reduction in the energy dissipation over STT-RAM that needs 100fJ/bit even at 11 nm lateral dimension p-MTJs [61], is to use a combination of resonant surface acoustic waves (r-SAWs) and spin-transfer-torque (STT) as discussed in Ref [92]. The key idea here is that the magnetization dynamics in the magnetostrictive nanomagnet is resonant with the SAW that drives the magnetization to build over few tens of cycles and eventually precess in a cone with a deflection of ~45° from the perpendicular direction. This reduces the STT current required to switch the magnetization direction without increasing the STT application time or reducing the switching probability in the presence of room temperature thermal noise. Thus, the lateral dimensions can be downscaled aggressively and yet one can use low levels of stress/SAW amplitude and moderate magnetostriction by leveraging SAW-FMR.

Fig. 13 illustrates the idea: when the SAW provides an effective AC magnetic field, which is at the ferromagnetic resonance (FMR) frequency, it drives the magnetization through large angles as the energy pumped due to strain (magnetostrictive coupling) extends over many cycles. In magnets with in-plane anisotropy, Fig. 14(a) shows the manner in which such resonant SAW drives the magnetization of a perpendicular magnetic tunnel junction (p-MTJ) to precess in a cone with large deflection. This reduces the STT current need to switch the magnetization compared to the case where the STT switching is not assisted by resonant SAW as shown in Fig. 14(b).

By incorporating inhomogeneity through lateral anisotropy variation [93], it was shown that magnetization precession in different grains can be significantly incoherent with room-temperature thermal noise. Interestingly, the precession in different grains are found to be in phase, even though the precession amplitude (angle of deflection from the perpendicular direction) varies across grains of different anisotropy as illustrated in Fig. 15. This large "mean" deflection in the presence of thermal noise and inhomogeneity improves the efficacy of SAW-assisted STT devices as the STT effective field is a function of sin θ, where θ is the angle between the fixed-layer and the free-layer magnetizations. In summary, this simulation study showed that high mean deflection angle due to acoustically induced FMR can complement the STT switching by reducing the STT current significantly in practical devices; even though the applied stress induced change in anisotropy is much lower than the total anisotropy barrier.

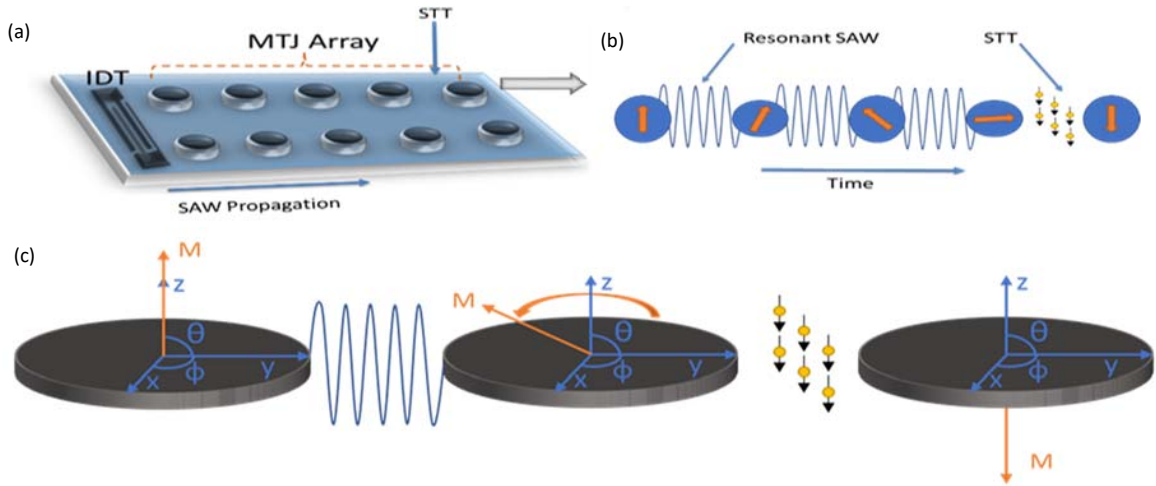

Fig. 13: (a) MTJ array switched with resonant SAW and STT (b) Magnetization dynamics with resonant SAW + STT switching of in-plane magnetization (c) Magnetization dynamics with resonant SAW + STT switching of out-of-plane magnetization. Reproduced from [92] with permission of the American Institute of Physics.

While SAW driven FMR [94] was previously reported in Ni films, applying a field to tune the FMR to the acoustic frequency has been demonstrated more recently in Ni films [95]. It has been shown that the power absorption in acoustically driven FMR (ADFMR) scales exponentially with the length of the magnetic element along the SAW propagation direction and it is consistent over a range of input power values (>65 dB) [96]. More recently, a detailed study of the optimization of acoustically driven FMR (ADFMR) devices [97] has been reported and use of resonant acoustic pulses to switching the magnetization in GaMnAs between two stable states [98] has been demonstrated. These works suggest that SAW-FMR or ADFMR is likely to become a viable way to switch scaled nanomagnets. Furthermore, materials such as YIG [99] and Rare Earth substituted YIG [100] offer low magnetostriction but their extremely small damping allow deflections to build over many cycles and reach large values as dissipation is limited. Thus, there is potentially a large material parameter space to achieve at least an order of magnitude energy saving with similar error rates. Additionally, small saturation magnetization in ferrimagnets would allow for lower STT write current which would be synergistic with assistance from SAW to enable over an order of magnitude reduction in energy dissipation compared to the existing 100 fJ/bit in STT-RAM devices. This could enable the scalability required to be competitive with current CMOS memory implementations while having the added advantage of non-volatility.

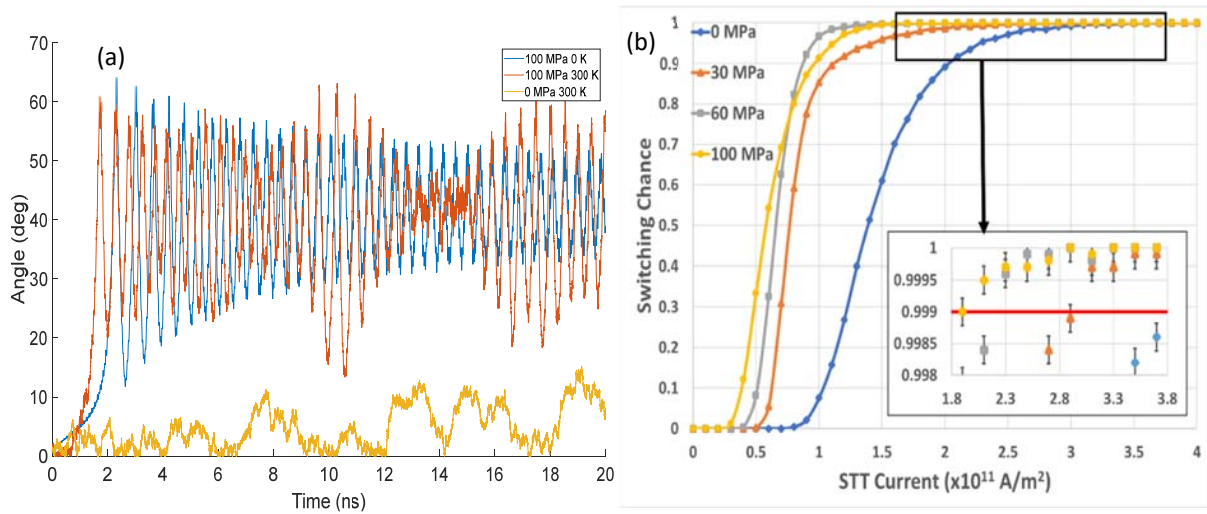

**Fig. 14:** Out-of-plane magnetization dynamics simulations with (a) comparison of resonant SAW with and without thermal noise to purely thermal noise and (b) Switching probability vs. STT current density at three different SAW magnitudes as well as for no SAW applied. Reproduced from [92] with permission of the American Institute of Physics.

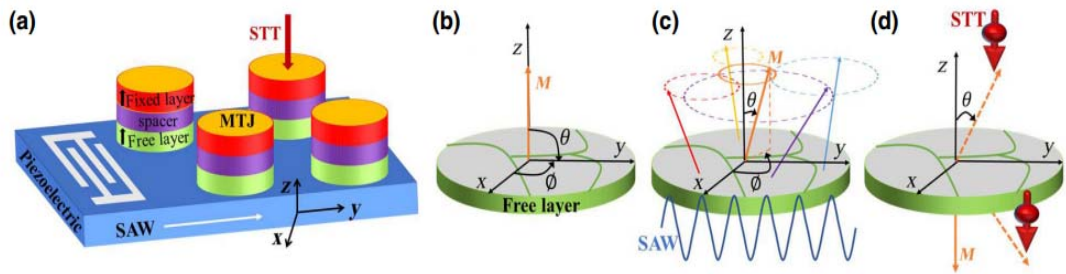

**Fig. 15:** MTJ arrays and SAW electrode over piezoelectric substrate. (b) Initial magnetization state of the inhomogeneous (i.e., granular) free layer. (c) Application of SAW induces different angle precession and the resulting incoherency reduces the net magnetization, M. (d) Final magnetization state after application of STT current. Reproduced from [93] with permission of the American Physical Society

## V. STRAINTRONIC BOOLEAN LOGIC: EXPERIMENTS

In Section III.A, we mentioned that straintronic nanomagnetic switches are not suitable for Boolean logic since the error probabilities are unacceptably high. While this is certainly true, in this sub-section, we briefly describe experimental efforts undertaken in our labs to implement straintronic Boolean logic with both static and time-varying strain. We do this for the sake of comprehensiveness. Curiously, time varying strain, in the form of SAW, results in lower switching error probability and hence more reliability. While we have not confirmed the cause for this difference, it may very well be the fact that the effect of SAW builds up over many cycles and hence the probability of switching correctly under SAW exceeds that under static strain.

The simplest Boolean logic gate for Boolean computing is the inverter or NOT gate. It is a single input-single output gate in which the output bit is always the logic complement of the input bit. We discussed such a system in Section III.A and Fig. 7 showed a nanomagnetic implementation of a NOT gate where the NOT operation was triggered with static strain. The gate however turned out to be disappointingly error-prone since only one out of nine gates operated correctly.

The obvious question is why is the statistics so poor that only 1 out of 9 pairs responds? There are many possible reasons for this: e.g. Co is only weakly magnetostrictive, there are pinning sites within the nanomagnets due to defects which prevent rotation, etc. However, when the experiment was repeated with time-varying stress generated by a SAW (see Fig. 16 for the arrangement), the statistics improved. While static stress switches 1 out of 9, time-varying stress switched 4 out of 4 pairs as shown in the magnetic force micrograph in Fig. 17 [101]. It appears that repeated cycles of stressing coaxes the nanomagnets to respond better, but this remains to be investigated further before it can be confirmed.

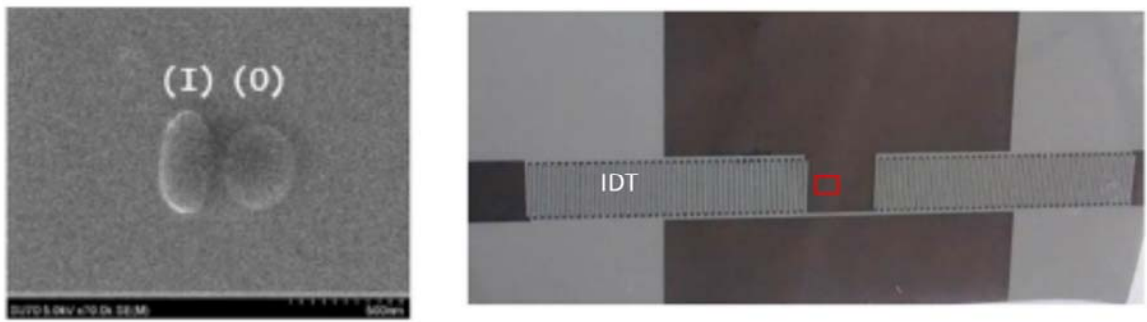

Fig. 16: The left panel shows scanning electron micrograph of a pair of nanomagnets acting as a NOT gate. The more elliptical nanomagnet hosts the input bit (I) and the other the output bit (O). The right panel shows the delay line with the interdigitated transducers (IDT) for launching the surface acoustic wave. The red square houses the nanomagnet pairs. Reproduced from [101] with permission of the American Institute of Physics.

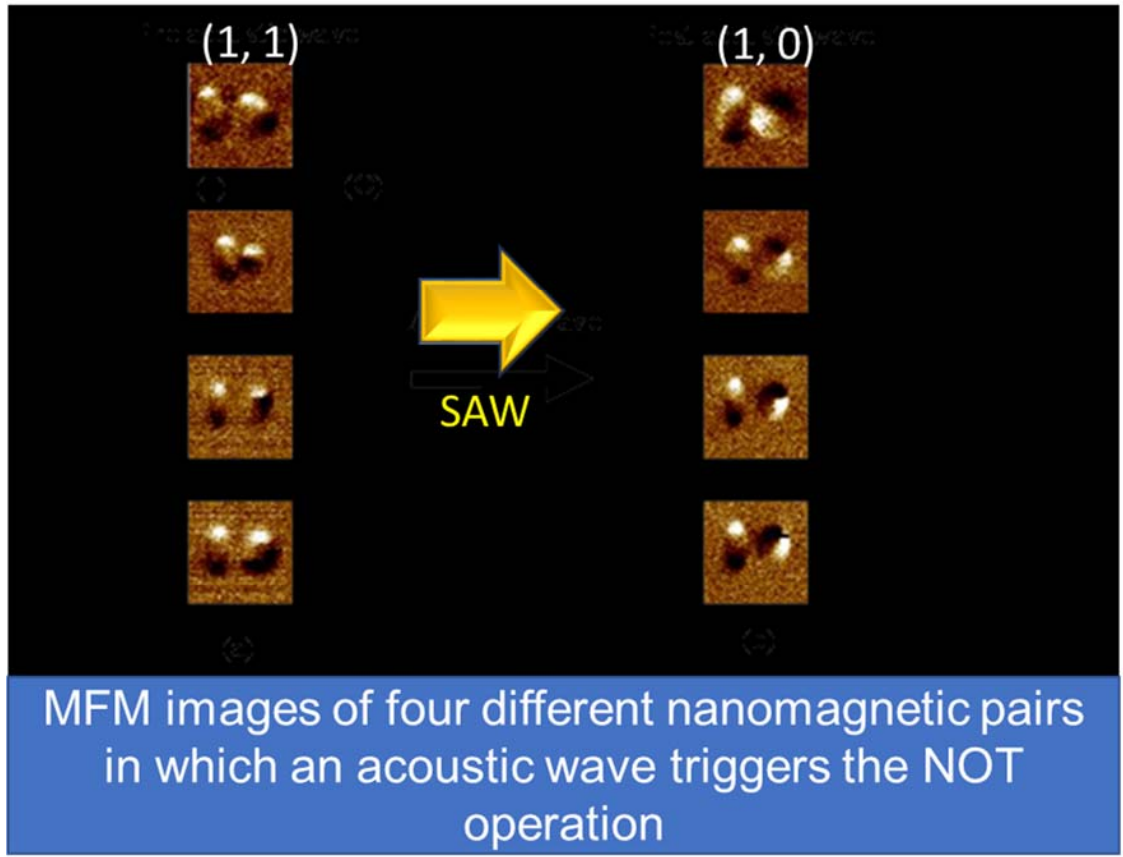

Fig. 17: (a) Magnetic force micrographs of four different Co nanomagnet pairs delineated on a LiNbO$_3$ substrate. All pairs are initially magnetized in the same direction with a global magnetic field and then subjected to a surface acoustic wave (SAW) launched with interdigitated transducers. The SAW triggers the NOT action, making the magnetizations of the left (input) and right (output) nanomagnets mutually antiparallel. Reproduced from [101] with permission of the American Institute of Physics.

## VI. STRAINTRONIC DOMAIN WALL DEVICES

Since the proposal to use domain wall (DW) devices as racetrack memory by Parkin et al [102], many schemes have been proposed to control domain wall motion in an energy efficient manner. One scheme involved the use of strain applied to a magnetostrictive racetrack to change the domain wall pinning [103]. The authors proposing this scheme showed that in the absence of voltage induced stress, the DW propagates freely in the magnetic strip (Fig 18 a). However, when a voltage is applied (Fig 18 b), the DW motion is impeded due to local pinning created by the stress, consequently the DW propagation field is doubled. The authors also showed that the paradigm can be used to create a NOR gate.

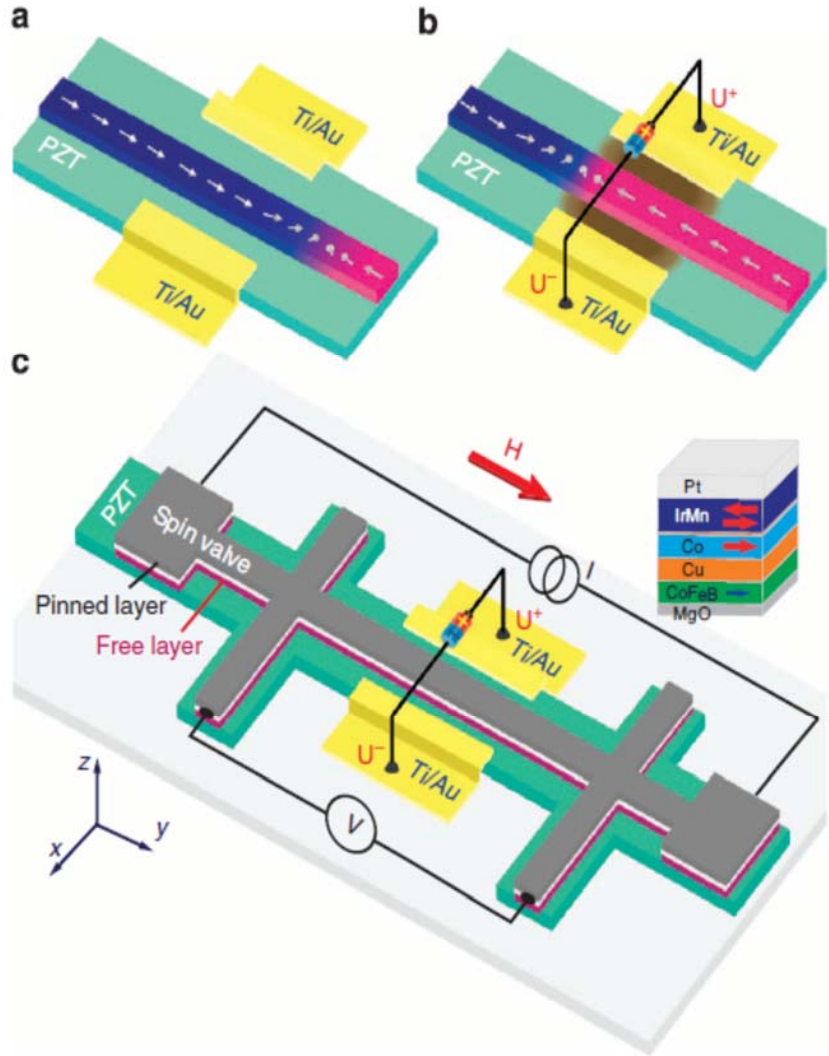

**Fig. 18**: Lateral approach used to manipulate magnetic domain wall through magnetoelectric coupling. **(a)** In the absence of applied voltages on the piezoelectric layer, the domain wall (DW) propagates freely in the magnetic stripe. **(b)** DW propagation in magnetic stripe can be controlled by voltages through lateral magnetoelectric coupling device. By applying a voltage onto the piezoelectric layer, a local stress is induced, followed by DW blockade. **(c)** Measurement configuration with hybrid PZT and spin-valve (SV) hall bar-shaped device, a single DW is injected from a large reservoir. The position of the DW is monitored by measuring the GMR between two electrodes. By applying a voltage on PZT, an induced stress results in a local modification of the domain wall dynamics. The SV multilayer structure is shown on the right. Reproduced from [103] with permission of the Nature Publishing Group.

There are several other studies on domain wall control with strain and surface acoustic waves (SAW). Recently, a method of using short strain pulses to move a DW deterministically along a nanowire was proposed and simulated [104]. Simulations have also shown that it is possible to use strain to control 360° domain wall motion in nanorings [105]. Further work has shown that resonant standing acoustic waves of frequency ~96.6 MHz and of sufficient amplitude can drive DW motion from creep to the flow regime [106]. This increases DW propagation velocity by an order of magnitude compared to field driven DW motion. In another work [107] this group also showed that high frequency SAW can help de-pin DWs and increase depinning probabilities 10-fold. An experimental study by another group also reports increase in DW propagation velocity with increasing SAW intensity [108]. While all these studies show the potential of straintronics (direct strain transferred from a piezoelectric or strain from SAW) in manipulating DWs, there is a tantalizing potential for long-term research in using strain control of DWs for neuromorphic computing applications.

In general, there are several papers, for example, references [109-111], that have studied the benefit of spintronics and DW devices for neuromorphic applications. A recent work [112] explores the use of a fixed duration-and-amplitude spin torque pulse in conjunction with voltage induced strain to control the DW position through micromagnetic simulations. Furthermore, the effect of thermal noise and edge roughness on DW dynamics and control with strain has been studied [113] for the configuration shown in Fig 19.

Figure 19 shows a schematic of a SOT driven DW device whose DW position is controlled by voltage-induced strain. Comprehensive micromagnetic modeling was performed using MUMAX of such DW dynamics driven by SOT and controlled by voltage induced strain in the presence of both thermal noise and defects. Change in the perpendicular magnetic anisotropy (PMA) due to voltage induced strain alters the DW velocity and consequently controls the final position of DWs driven by SOT. However, when the SOT is withdrawn, the DW does not immediately come to rest but its interaction with edge roughness in the presence of additional stochastic dynamics due to thermal noise leads to a stochastic distribution in the final DW position as shown in Fig 20 [113].

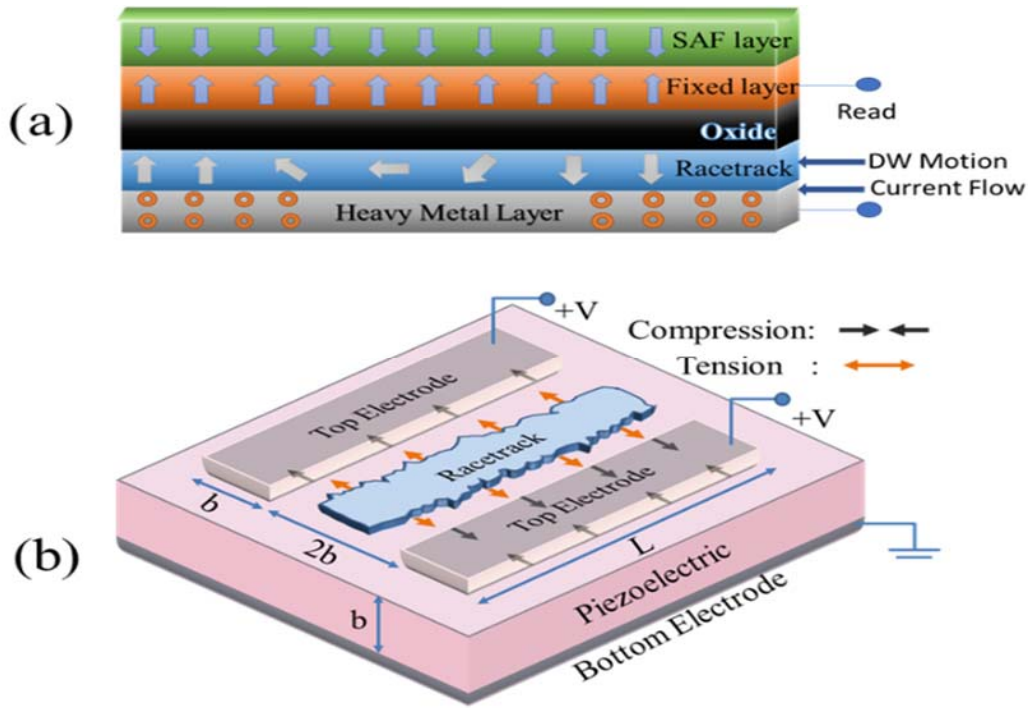

**Fig. 19:** (a) Proposed device stack where the nanoscale racetrack acts as the magnetic free layer. A DW in the racetrack moves when a current is injected into the heavy metal layer. (b) Stress generation mechanism in rough edge racetrack when a voltage is applied across the piezoelectric. Reproduced from [113].

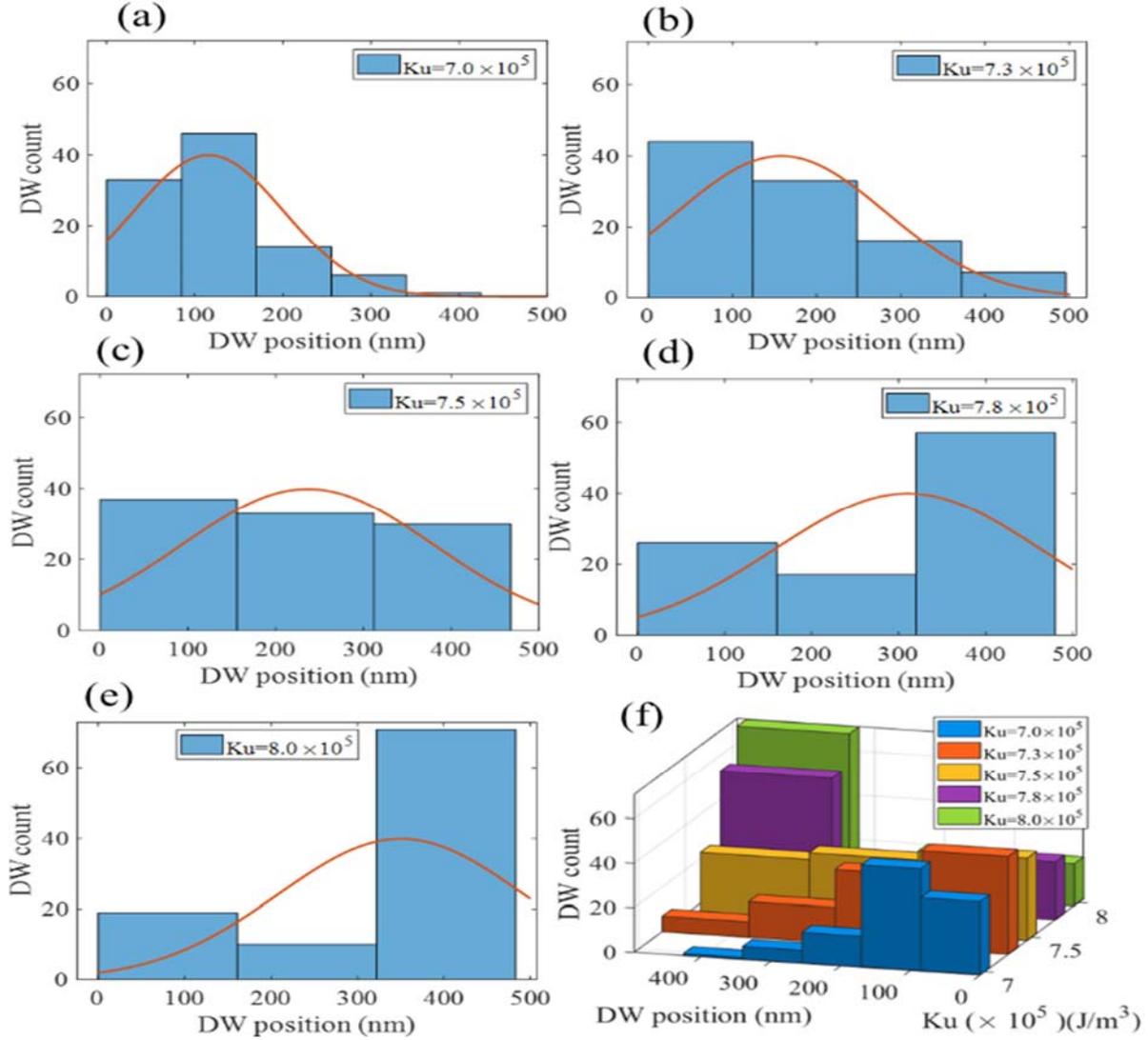

**Fig. 20:** (a)-(e) Equilibrium DW positions in one racetrack (~ 3nm rms edge roughness) at T=300K for a fixed SOT and different stresses correspond to $K_u$ values of 8.0, 7.8, 7.5, 7.3 and 7.0 ($\times 10^5$) $J/m^3$. For each figure in 18(a)-(e) a Gaussian distribution plot is overlaid having a mean and standard deviation identical to the data used to create the bins (f) 3-dimentional histogram shows combined plot of 20(a)-(e). Reproduced from [113].

In summary, strain and SAW control of DW motion has very significant potential in memory and neuromorphic computing hardware. However, nanofabrication of prototype devices that can scale competitively, switch reliably for memory devices, and use back propagation algorithms that can account for the stochastic and limited resolution synaptic weights based on DW devices, are important future areas of research to make this field have an impact.

# VII. STRAINTRONIC CONTROL OF SKYRMIONS AND THEIR DEVICE APPLICATIONS

Skyrmions are topologically protected magnetic states occupying a finite region of space in a magnetic film where the spins at the center point opposite to the spins at the periphery. One can have Bloch or Neel skyrmions depending on the way spins rotate from $z = +1$ at the center to $z = -1$ at the periphery. Skyrmions are a consequence of Dzyaloshinskii-Moriya Interaction (DMI) in a ferromagnetic materials and the DMI vector plays a role in the stabilization of the Bloch vs. Neel skyrmion.

In a Bloch skyrmion, the spin rotation is perpendicular to the line joining the center and a point on the periphery. It is stabilized if the DMI vector is parallel to the vector that joins two spin sites. On the other hand, in a Neel skyrmion, the spin rotation occurs along the direction of the line joining the center and a point on the periphery. It is stabilized if the DMI vector is perpendicular to the vector that joins two spin sites. Furthermore, the DMI vector direction in bulk systems prefers Bloch skyrmions while interface driven DMI prefers Neel skyrmions. The handedness of the chirality originates from the sign of the DMI vector, where the spins have the same sense of rotation along any diameter (in a Neel skyrmion) which distinguishes it from bubble states where no DMI is involved.

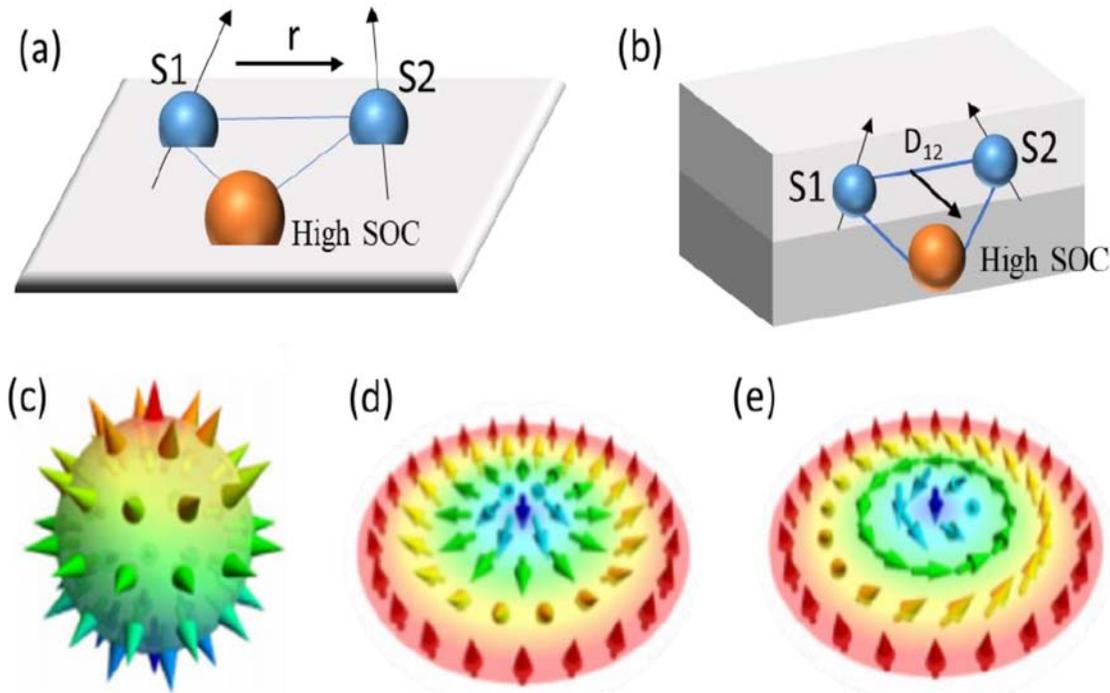

Fig. 21: (a) Bulk DMI, (b) Interfacial DMI, (c) Neel skyrmion wrapped around a sphere, (d) Stereographic projection of (c), (e) a Bloch skyrmion. Reproduced from [114] with permission of the American Physical Society.

Skyrmions [115-124] have been studied, investigated for various devices concepts given the low depinning currents, realized at room temperature so they can be adapted to practical devices, and driven at relatively high velocities. However, the use of electric field generated strain to create, manipulate and annihilate skyrmions is relatively less explored and discussed next.

There are many ways in which strain can create, destroy or manipulate a skyrmionic state. For example, a recent work on Pt/Co/Pt multi-layer showed that the DMI constant can be increased by almost one order of magnitude from 0.1 mJ/m$^2$ to 0.8 mJ/m$^2$ with change in uniaxial deformation of the film from -0.08% (or -800 microstrain) to +0.1% (or +1000 microstrain) as shown in Fig 22 [125]. Measurements of the change were made using Brillouin Light Scattering Spectroscopy (BLS) as a function of strain applied by bending a glass/Ta (2.5 nm)/Pt/Co (1.2 nm)/Pt (2nm) cantilever.

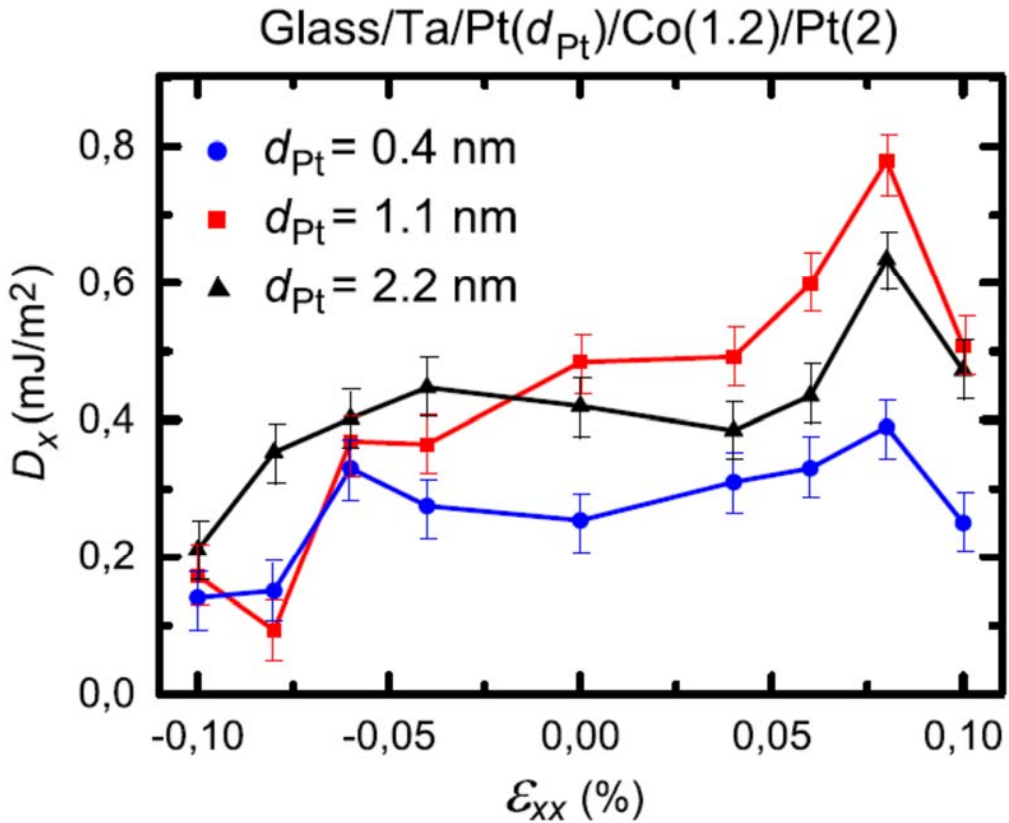

**Fig. 22:** The DMI constant $D_x$ measured along the direction of uniaxial strain (x-direction) as a function of applied strain ($\varepsilon_{xx}$). Reproduced from [125] with permission of the American Institute of Physics.

Another work [126], explored the effect of ~0.3 % (or 3000 micro-strain) deformation on skyrmions in FeGe using Lorentz Transmission Electron Microscopy (LTEM) to visualize them. Such a strain deforms the skyrmion shape from approximately circular to approximately elliptical, besides distorting the skyrmion lattice by about 20% [126]. Experimental observations, backed by complementary simulations, suggest strain induced DMI change is the predominant mechanism for deformation of skyrmions in the FeGe system studied.

Skyrmion annihilation/creation with electric field induced strain have also been demonstrated [127]. In this case, the effect of strain on DMI is characterized through Brillouin light scattering spectroscopy (BLS), while its effect on perpendicular magnetic anisotropy (PMA) that leads to a change in effective anisotropy ($K_{eff}$) is characterized through angle dependent FMR measurements and other methods. Fig 23 shows voltage induced strain-based creation and annihilation of skyrmions visualized with magnetic force microscopy (MFM) in a Ta(4.7 nm)/[Pt(4 nm)/Co(1.6 nm)/Ta (1.9 nm)] × 5 multilayers heterostructure grown on (001) cut single crystal PMN-PT substrate.

In another work involving [Pt(2.5 nm)/Co(2.2 nm)/Ta(1.9 nm)]12/Ta(5 nm)] heterostructures grown on oriented single-crystal (001)-cut PMN-PT and patterned to ~350 nm dots, the authors showed that with different electric field pulses that generate different levels of strain, one can switch between multiple states: stripe, vortex and skyrmions as shown in Fig 24 [128]. These states are non-volatile and can be imaged by MFM even after the electric field pulses have been withdrawn.

Finally, simulations [129] show that one can repeatedly create and delete a skyrmion in a nanoscale disk with voltage induced strain as shown in Fig 25. This could lead to non-volatile memory that consumes ~0.5 fJ per switching event making it ~200 times more energy efficient than Spin Transfer Torque Random Access Memory (STT-RAM).

There are also several other articles on electric field induced strain manipulation of skyrmions in films and nanostructures [130, 131], strain control of skyrmion propagation [132, 133], etc. In addition to manipulating skyrmions with electric field induced strain, there is also a proposal for driving skyrmions with acoustic waves [134] and an experimental demonstration of creation of skyrmions with acoustic waves [135].

Skyrmions also have applications in areas other than memory. Skyrmions whose magnetization dynamics are driven by strain can be used as spintronic nano oscillators [136] and resonate and fire neurons [137].

We note that while we discuss strain-mediated manipulation of skyrmions in this review, it is also possible to manipulate skyrmions with direct voltage control of magnetic anisotropy (VCMA), which is not mediated by strain and does not rely on the magnetoelastic effect. For example, creation and annihilation of isolated skyrmions in a film using direct voltage control of magnetic anisotropy (VCMA) without strain mediation has been demonstrated in a Ta (2) / IrMn (5) / CoFeB (0.52-1.21) / MgO (2.5) / $Al_2O_3$ (35)/ ITO heterostructure [138]. Such skyrmion creation and annihilation has also been demonstrated in magnetic tunnel junctions [139]. Other voltage control schemes for manipulating skyrmions have also been explored and demonstrated [140, 141].

The above experimental work provides preliminary support to device proposals [142, 143] that use simulations to show that core reversal of fixed skyrmions in nanostructure ~100 nm is possible due to the boundary conditions. One promising scheme involves using intermediate skyrmion state to switch between a ferromagnetic "up" and ferromagnetic "down" state that is robust to switching errors due to both thermal noise and inhomogeneities [144]. Such devices are potentially scalable to 20 nm lateral dimensions and beyond, but some materials and device physics challenges exist [145, 146].

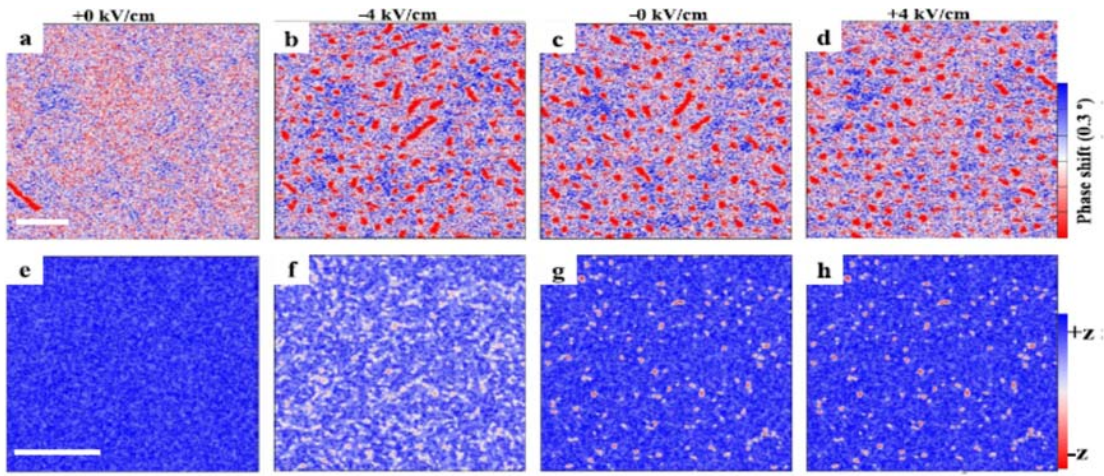

**Fig. 23:** Skyrmion creation: MFM images at electric field E=+0 kV/cm (a), −4 kV/cm (b), −0 kV/cm (c), and +4 kV/cm (d) with magnetic field $B_{bias}$ = 60 mT. Corresponding simulation results of strain-mediated skyrmion creation with $\varepsilon_{[110]} = \varepsilon_{[110]}=0$ (initial state) and DMI interaction strength D= 0.772mJ/m$^2$ (e), $\varepsilon[110]=\varepsilon[-110]=$ -0.189% and D =0.585mJ/m$^2$ (f), $\varepsilon[110]= \varepsilon[110]= $ -0.034% and D = 0.685mJ/m$^2$ and $\varepsilon[110]= \varepsilon[-110] =0.010\%$ and D = 0.727 mJ/m$^2$ (h), with the blue and red contrasts corresponding to magnetizations pointing up and down, respectively. The scale bar is 1 μm. Reproduced from [127] with permission of the Nature Publishing Group.

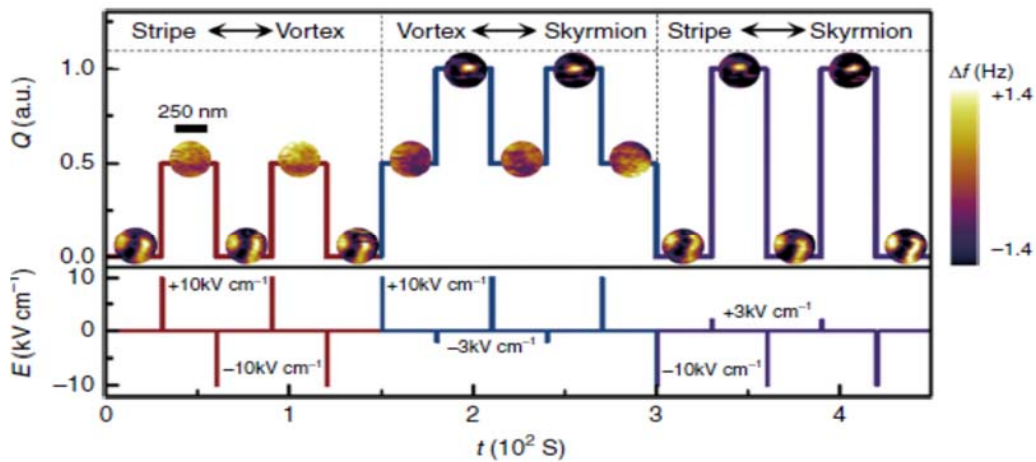

**Fig. 24:** Switching of individual skyrmions induced by pulse electric field. a Switching of topological number Q of various magnetic domains (Q = 1.0, 0.5, and 0 corresponds to skyrmion, vortex, and stripe, respectively) by applying a pulse electric field with a pulse width of 1 ms. The insets contain the corresponding MFM images for the switching. The values of E for the generation of the skyrmion, vortex, and stripe are ±3, +10, and −10 kVcm$^{-1}$, respectively. The MFM contrast represents the MFM tip resonant frequency shift (Δf). The scale bar represents 250 nm. Reproduced from [128] with permission of the Nature Publishing Group.

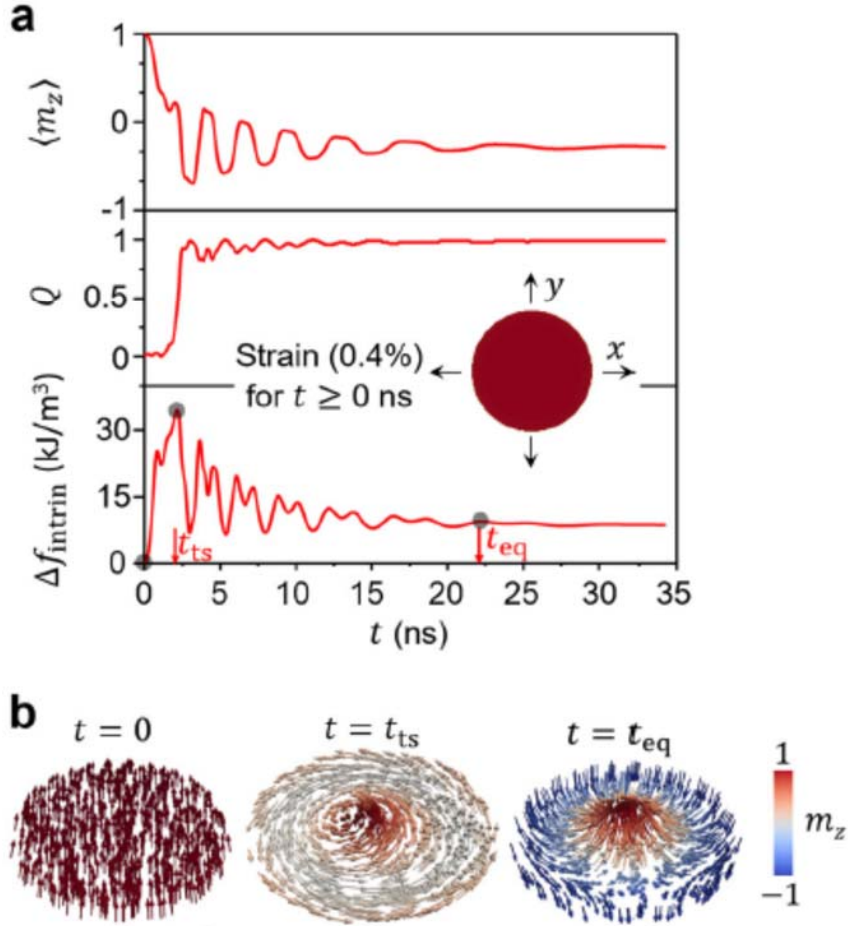

**Fig. 25:** Skyrmion creation. a, Temporal evolution of $m_z$ (volume average of the normalized perpendicular magnetization), Q (the topological charge number), and the change of the intrinsic magnetic free energy density when a biaxial in-plane isotropic tensile strain of 0.4% is applied to a 220-nm-diameter CoFeB disk and then kept on. The gray dots, as well as the downward arrows mark the initial state (t = 0), highest transitional state (t = $t_{ts}$), and the equilibrium state (t = $t_{eq}$). b, Corresponding spin structures at these three-time stages showing a strain-mediated skyrmion creation. The interfacial DMI strength D = 0.75 mJ/m². Reproduced from [129] with permission from Nature Publishing Group.

Ultimately, both strain induced or direct voltage control of skyrmions have potential for new and novel skyrmionic memory and neuromorphic computing devices provided they can scale reliably to 20 nm diameter and below.

## VIII. NON-BOOLEAN COMPUTING WITH MAGNETOSTRICTIVE NANOMAGNETS ACTUATED WITH STATIC OR TIME-VARYING STRAIN (ACOUSTIC WAVES)

Although much has been made of the potential of magnetic switches to replace transistors in Boolean logic processors, the truth is that there is not a single *magnetic* Boolean logic chip in existence today and not one is in sight. The primary reason for this is the often-neglected fact that magnetic switches are much more error-prone than transistor switches and currently their reliability is just too poor for applications in Boolean logic. While the switching error probability in transistor switches is around $10^{-15}$ in modern-day transistors, the switching error probability of a magnetic switch is larger than $10^{-9}$ as long as low-energy (e. g. voltage controlled) switching mechanisms are employed to switch. The error probability increases further by several orders of magnitude if there are extended structural defects in the nanomagnets [43]. This has doomed logic applications for magnetic switches.

There are also other reasons why magnetic switches are inappropriate for logic. Many magnetic binary logic proposals are based on the notion of using a magnetic tunnel junction (MTJ) as the switch that realizes the gate [147-149]. An important metric for a digital switch that encodes binary bits 0 and 1 is the conductance on/off ratio (for switches that encode bit information in the value of the conductance). While for transistors, this ratio exceeds $10^5$:1, for MTJs, this ratio (which is essentially the tunneling magnetoresistance ratio or TMR) is barely 7:1 [150]. This makes logic level restoration extremely challenging. Additionally, the off-current in the MTJ will be no smaller than one-seventh of the on-current, leading to unacceptable leakage and standby power dissipation. This problem is averted in MTJ based memory by placing a CMOS transistor in *series* with the MTJ (which also provides the read and write currents). It is turned off when the MTJ conductance is in the "low" state, thereby preventing a large amount of current leakage in the OFF state, but this remedy is not available in the logic proposals since the logic functionality will be impaired if such a CMOS transistor is present.

There are of course non-MTJ versions of magnetic logic as well, such as dipole coupled logic (sometimes referred to as "magnetic quantum cellular automata" although it has no connection with cellular automata) [151], but they are extremely error-prone since dipole coupling is usually too weak to withstand thermal perturbations [152-154]. In fact, the only experiment that claimed to demonstrate such dipole coupled logic [a majority logic gate] in an array of nanomagnets reported an error probability 75% [155]! Many years ago, John von Neumann had shown that the maximum tolerable error probability in a majority logic gate is 0.0073 [156], which makes a majority logic gate with such high error probability unacceptable. Because of this kind of extreme unreliability, "magnetic logic" has withered on the vine.

Fortunately, Boolean logic is not the only paradigm for building computing machinery. The human brain 'computes' with neurons, synapses, dendrites and axons which are much more error-prone and significantly slower than transistors, and yet the human brain can outshine the most powerful digital computers in many tasks such as face recognition. The brain is also very energy-frugal when it 'computes'. In the legendary chess match between Garry Kasparov and IBM's Deep Blue computer, Kasparov lost narrowly, but his brain dissipated roughly 20 Watts of power, while Deep Blue dissipated several kilowatts! Watt for watt, the brain will win handily over digital computers in solving many problems. A growing body of brain-inspired computing platforms built with magnetic devices has been proposed in the literature recently to solve NP-complete problems. In fact, there is a major effort in building artificial neural networks with magnetic devices because certain types of magnetic devices, e. g. those involving domain wall motion, are conducive to building powerful neural architectures [157-159].

The most desirable attribute of a neuron is the energy efficiency. The lower the energy it takes to operate (or 'fire') a neuron, the more desirable it is. A simple straintronic threshold firing neuron was analyzed in ref. [160] and it turned out to be orders of magnitude more energy-efficient than a comparable 'spin neuron' implemented with spin transfer torque [161]. Both used MTJs to implement the neuron. While one used strain as the activation mechanism, the other used spin transfer torque. The energy dissipated to fire the straintronic neuron in less than 1 ns was only ~ 8 aJ, while the one activated with spin transfer torque would dissipate around 50 fJ to fire with a delay of ~6 ns.

### A. Simulated annealing in an array of dipole coupled magnetostrictive nanomagnets with a surface acoustic wave for collective ground state computing

A popular approach to performing non-Boolean computation with magnetic devices is to implement a *Boltzmann machine*. This is a network of binary neurons interconnected with adjustable connection weights that are called synapses. The neurons are 'biased' in a certain way and the interconnected system is allowed to relax to the ground state which represents the optimal solution to a problem. One defines a "cost-function" for the system given by

$$E = [S]^T[x] + [S]^T[w][S]$$

where $[S]$ is $n \times 1$ column vector whose elements are 0 or 1, $[x]$ is a bias vector which is also a $n \times 1$ column and $[w]$ is a symmetric $n \times n$ matrix called the "weight matrix". The bias vector and weight matrix are picked such that the minimum value of $E$ represents the solution. The probability of relaxing to any energy state is given by $p_i = e^{-\beta E_i}/\sum_i e^{-\beta E_i}$ in accordance with Boltzmann statistics. Hence, relaxing to the ground state has the highest probability. Thus, the correct solution is found with the highest probability.

Recently, there has been significant interest in performing "ground state computing" with Boltzmann machines. Since this is a purely hardware-based approach with no software, it eliminates the need to execute instruction sets and hence will be much faster than traditional computing schemes that rely on software [162-164]. Some of these hardware systems (Boltzmann machines) are built with interacting nanomagnetic systems [165-169] because they are particularly suitable for this approach. The Achilles' heel of ground state computing, however, is the possibility that the Boltzmann machine, namely the magnetic array, may get *stuck* in a metastable state and fail to relax to the ground state. This is a catastrophic computational failure. If this happens, "simulated annealing" can release the system from the metastable state and allow it to migrate to the ground state.

Simulated annealing is an old concept that is well known in computer science. It is an algorithm to find the optimum solution of a problem (e. g. the traveling salesman problem) when the search space is discrete. It is executed either by solving kinetic equations for density functions [170, 171] or by using stochastic sampling [172, 173]. Typically, the problem to be solved will be recast as one corresponding to energy minimization. The optimum solution will correspond to the thermodynamic ground state (minimum energy state), while sub-optimal solutions will correspond to metastable states. Simulated annealing algorithms ensure that sub-optimal solutions are rejected in favor of the optimal solution by driving the system from metastable states to the ground state, following a sequence of computing steps. The term "annealing" in this context has its origin in the idea that raising the temperature of a disordered solid, and then reducing the temperature, allows the solid to approach the ordered (crystalline) state which may represent the thermodynamic ground state.

Experimental demonstration of the emulation of simulated annealing in a small interacting nanomagnetic system with the aid of time-varying strain was carried out in our group. We focused on a two-dimensional array of 3 × 3 dipole-coupled magnetostrictive nanomagnets deposited on a piezoelectric substrate [174]. The system is intentionally driven out of the ground state with an external agent and gets pinned in a metastable state. The nanomagnets are subsequently subjected to periodic time-varying strain generated with a surface acoustic wave. This returns the system to the ground state (unpinning it from the metastable state), thereby implementing a hardware emulation of simulated annealing (rejecting the sub-optimal solution represented by the metastable state magnetic order in favor of the optimal solution represented by the ground state magnetic order).

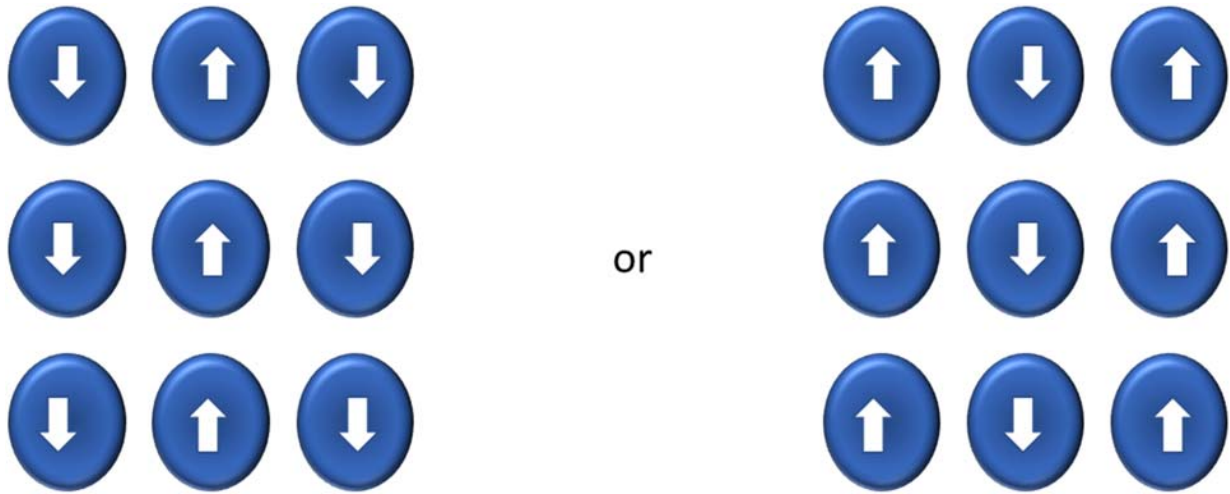

Fig. 26: Magnetization orientations in an array of dipole coupled elliptical nanomagnets with in-plane anisotropy when the system is in the ground state. Reproduced from [174] with permission of the Institute of Physics.

The ground state magnetic ordering in a system of 9 elliptical nanomagnets arranged in three rows and three columns is shown in Fig. 26. All nanomagnets along a column are magnetized in the same direction along the major axis, but alternating columns have opposite (antiparallel) magnetizations. Image processing problems can be mapped into such a system of nanomagnets where the two stable magnetization orientations can represent pixel color (black or white) and the interaction between the nanomagnets can accomplish various image processing tasks [166]. A specific problem that relates directly to the example here is representation of the color "grey" with black and white pixels. The best representation of grey is to alternate black and white pixels, and the second-best representation is to alternate black and white stripes (columns or rows). Let us suppose that magnetization pointing "up" represents the color black and that pointing "down" represents the color white. In that case, the ground state configuration shown in Fig. 26 represents "grey" with alternating vertical stripes of black and white.

Next, let us suppose that noise or other perturbations disrupt the ground state magnetizations, resulting in the system being driven out of the ground state and getting stuck in a metastable state (with a different magnetic order) that causes corruption of the grey image. In this case, the image can be restored by driving the system out of the metastable state back to the ground state by subjecting it to an appropriate external influence that performs "simulated annealing".

To demonstrate this paradigm, we fabricated a two-dimensional array of cobalt nanomagnets on a piezoelectric LiNbO$_3$ substrate (of thickness 0.5 mm) using electron beam patterning of a resist, electron beam evaporation of cobalt on to the patterned substrate and lift-off [174]. An atomic force micrograph of an arbitrarily selected 3 × 3 array is shown in Fig. 27 (a). The dimensions of the nanomagnets are: major axis = 350 nm, minor axis = 320 nm and thickness = 12 nm. The edge-to-edge separation is ∼ 30 nm.

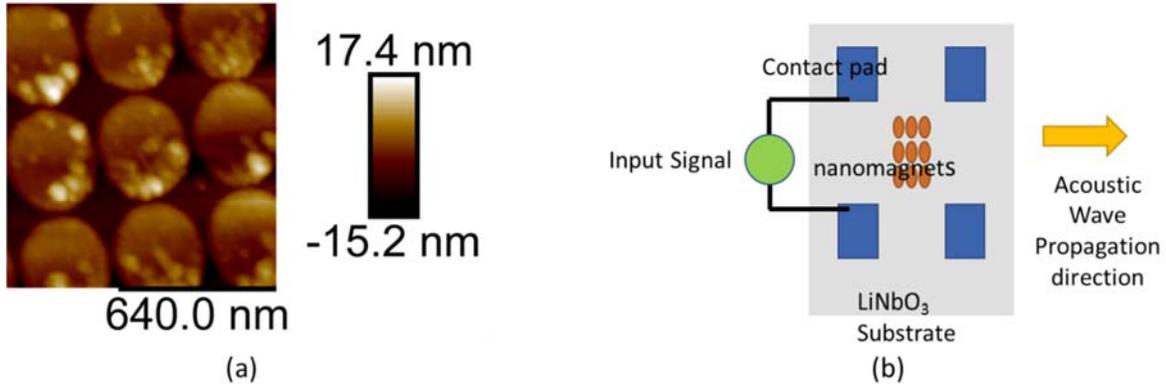

Fig. 27: (a) Atomic force micrographs of elliptical cobalt nanomagnets fabricated on a LiNbO$_3$ substrate, and (b) the device configuration. Reproduced from [174] with permission of the Institute of Physics.

Two pairs of gold contact pads (of few mm lateral dimension) are delineated on the substrate, each pair on one side of the nanomagnet assembly, to launch surface acoustic wave (SAW) in the substrate. Since we are not using interdigitated transducers, the launched SAWs are not traditional Rayleigh or Sezawa modes, but a different mode, However, this is not important for our purpose since we are only interested in time-varying strain. The distance between the edge of any pad and the nearest edge of the nanomagnet assembly is a few mm. The pads are placed such that the SAW launched by them propagates through all the nanomagnets, as shown in Fig. 27 (b). Either pair of pads can be used to launch the SAW.

Magnetic force microscopy showed that most nanomagnets do not show good magnetic contrast. This is not surprising given that the nanomagnets are large enough that they would tend to be multi-domain. Additionally, defects such as edge roughness, size and shape variation, etc. can result in significant non-uniformities. There are, however, small regions within the fabricated array, containing few nanomagnets, where one observes good magnetic contrast. We isolated one such section containing 3 × 3 nanomagnets and focused on it.

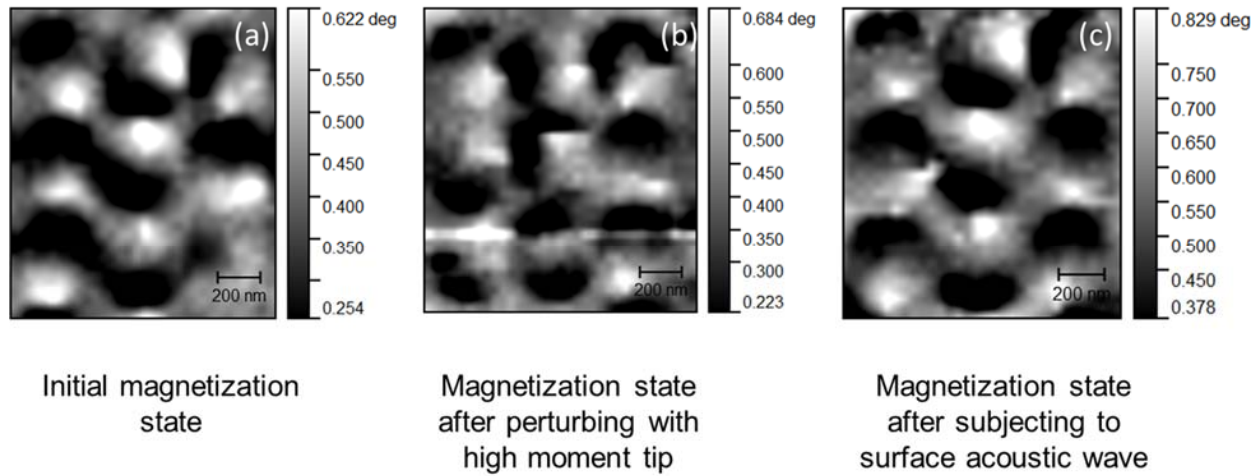

Fig. 28: Magnetic force microscopy (MFM) images showing "simulated annealing" being performed: (a) the initial magnetization state of a 3 × 3 array of elliptical Co nanomagnets, which corresponds to the ground state shown in Fig. 10. The magnetizations are mutually parallel (ferromagnetically ordered) along the columns and mutually antiparallel (anti-ferromagnetically ordered) along the rows, (b) the state of the array after perturbation with a high moment MFM tip (this is a metastable state) where the magnetizations deviate from the ground state, (c) the magnetization state after passage of the surface acoustic wave. The system gets unpinned from the metastable state and the ground state restored. If we repeat this experiment, the high moment tip may take the system to a different metastable state since the latter is not controllable, but the acoustic wave will always restore the system back to the ground state. The white and black color represent opposite magnetizations (white could be viewed as the north pole of a nanomagnet and black as the south pole, or vice versa). Reproduced from [174] with permission of the Institute of Physics.

We determined the magnetic ordering within this array with magnetic force microscopy (MFM). The MFM image of the 3 × 3 array is shown in Fig. 28(a). We clearly see the ordering depicted in Fig. 25, where the nanomagnets along a column are magnetized in the same direction and alternating columns have opposite directions of magnetization, i.e. the system is in the ground state. This image is obtained with a low-moment tip in order to carry out non-invasive imaging. Next, we intentionally perturb the magnetization in the array with a high-moment tip and we show the MFM image of the resulting configuration after the perturbation in Fig. 28(b). Clearly, the ground state ordering has been destroyed and the system has not spontaneously returned to the ground state (i.e. it is stuck in a metastable state). We then launch a surface acoustic wave (SAW) in the substrate, which periodically exerts tensile and compressive strain on the nanomagnets and rotates their magnetization via the Villari effect during one of the two cycles. The SAW is launched by applying a sinusoidal voltage of 24 V peak-to-peak and frequency 3.57 MHz to one pair of electrodes. After the SAW excitation is terminated, we image the nanomagnets again, and find that the system has returned to the ground state, indicating successful "simulated annealing" [174]. The SAW exerts time varying stress on the nanomagnets and at some time during either the positive or the negative cycle, the stress generated erodes the potential barriers that impede transition from the metastable state to the ground state and allows the system to relax to the ground state, as shown in Fig. 28(c). This is an emulation of simulated annealing. Here, the periodic strain (SAW) acted as the simulated annealing agent.

## B. Straintronic Bayesian networks

Bayesian (belief) networks are computational modules that are especially adept at computing in the presence of uncertainty (e.g. disease progression, stock market behavior, etc.). The simplest 2-node Bayesian network consists of a parent node and a child node, where the state of the "child" is influenced by the state of the "parent", but not the other way around. In the language of neural networks, the synaptic connection between the parent and child node has to be non-reciprocal, which is not the case with normal Boltzmann machine where the synapse is reciprocal. The example in Section 7.1 is that of a Boltzmann machine where the synaptic connection (dipole interaction) between two nanomagnets (binary neurons) is reciprocal because the way nanomagnet A interacts with nanomagnet B is identical to the way nanomagnet B interacts with nanomagnet A.

In order to implement a Bayesian network, one would need to make the interaction non-reciprocal. We can do this with the construct shown in Fig. 29. Here, we have two MTJs of elliptical cross-section, one of which has much more shape anisotropy (more eccentricity) than the other. The former is the parent and the latter the child. Their hard axes and easy axes are parallel. The two MTJs are placed close enough to each other to have significant dipole coupling between the soft layers. In one case, the pair is placed such that the line joining the centers of the two soft layers is parallel to the minor axes of the ellipses (hard axes), while in the other case, the line joining the centers is parallel to the major axes of the ellipses (easy axes). The former makes an anti-correlator and the latter a correlator.

In the anti-correlator, the conductance states of the two can have no correlation to perfect (100%) anti-correlation (meaning when one conductance state is high, the other is low) depending on the amount of stress applied to both MTJs simultaneously. In the correlator, the conductance states can have no correlation to perfect (100%) correlation depending on the magnitude of global stress.

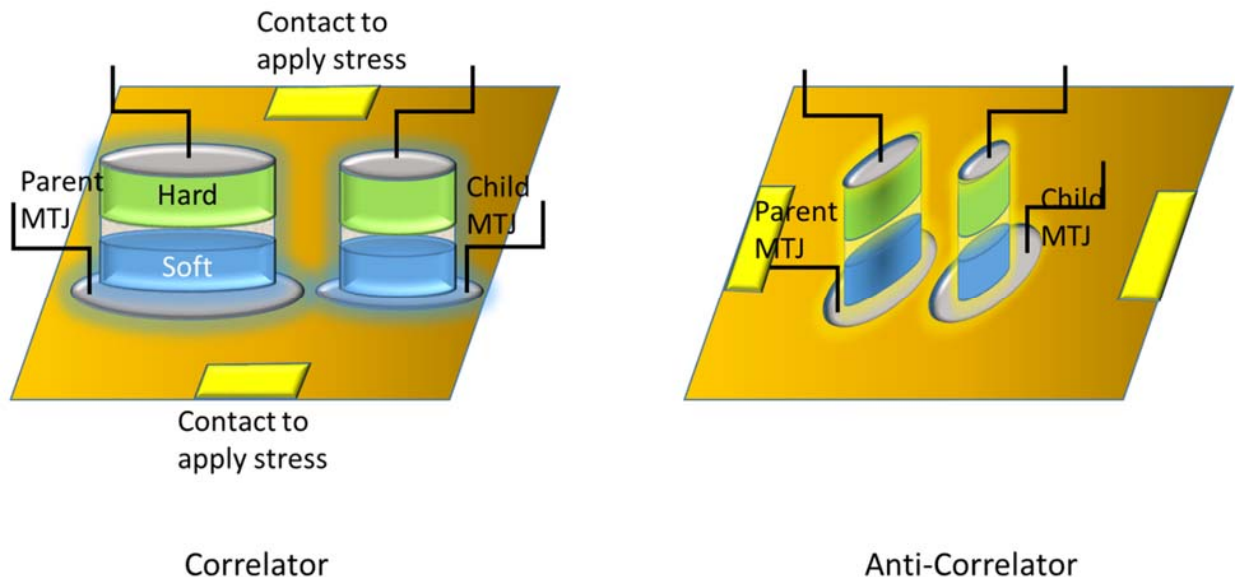

Fig. 29: A simple 2-node Bayesian network that can act as (a) a bit correlator and (b) a bit anti-correlator fabricated on a piezoelectric film.

When global stress is applied, the state of the parent MTJ is not affected since its shape anisotropy energy is too high for stress to overcome and rotate the magnetization of its soft layer. However, the child has a much lower shape anisotropy energy and hence stress can erode much or all of the energy barrier within the soft layer. When the energy barrier is lowered sufficiently, the magnetizations of the two soft layers will tend to become mutually parallel in the correlator and mutually antiparallel in the anti-correlator. In the presence of thermal noise, the probability of them becoming correlated or anti-correlated will depend on the stress and hence the probability can be varied with stress. As a result, the degree of correlation or anti-correlation can be varied with stress. We can set the *conditional probabilities* (e.g. the probability that the bit in the child node is 0 given that the bit in the parent node is 1, etc.) with stress. We have also enforced the parent-child relationship, making this a simple 2-node Bayesian network [175]. Straintronic Bayesian networks are currently an active field of research [176-178].

## IX. ACOUSTIC/ELECTROMAGNETIC SIGNAL GENERATION AND TRANSMISSION WITH THE AID OF STRAIN-SWITCHED MAGNETOSTRICTIVE NANOMAGNETS

### A. Extreme sub-wavelength *electromagnetic* antenna implemented with magnetostrictive nanomagnets actuated by time-varying strain

Antennas are used for transmitting and receiving radiated signals in the form of electromagnetic, acoustic or other types of waves. They are ubiquitous in communication systems and have other applications in range finding, geo-sensing, health monitoring, identification, etc. Recently, there has been strong interest in sub-wavelength antennas for *embedded applications* where the antenna has to be aggressively miniaturized in order to be integrated on-chip, embedded in wearable electronics, or to be medically implanted inside a patient's body. Such antennas must be much smaller than the wavelength of the radiation they emit, and consequently they usually suffer from poor radiation efficiency since the latter is typically bounded by the limit $A/\lambda^2$, where $A$ is the emitting area of the antenna and $\lambda$ is the wavelength of the radiated wave.

Magneto-elastic (or straintronic) antennas are a recent entrant in the world of antennas. They have attracted attention because they are activated (made to radiate) in a way that is very different from the way traditional electromagnetic antennas are activated. The magneto-elastic antenna consists of an array of magnetostrictive nanomagnets whose magnetizations precess when subjected to time varying strain, such as an acoustic wave, thereby exciting confined spin wave (SW) modes within the nanomagnets. Both intrinsic and extrinsic SW modes are excited [179] - the intrinsic modes depend on the array's parameters, while the extrinsic mode is excited at the same frequency as the driving acoustic wave. The latter couples to an electromagnetic (EM) mode of the same frequency that radiates out of the nanomagnets, resulting in a "spin-wave-antenna" that we have recently demonstrated experimentally [180]. A variant of this, where radio frequency electromagnetic waves are emitted by slow small angle precessions of magnetizations at radio frequencies (much smaller frequencies than the frequencies of confined spin wave modes) was recently demonstrated experimentally [181]. The radiation efficiency of such an antenna is not limited by $A/\lambda^2$ [182], but by the coupling efficiency of spin waves and electromagnetic waves. That efficiency is poor because the spin wave and the electromagnetic wave cannot be easily phase-matched (because of a large difference between their phase velocities), but even then, extreme sub-wavelength *straintronic* antennas $\left(A << \lambda^2\right)$ predicated on this principle of operation, have been shown to radiate with reasonable efficiency that exceeds the $A/\lambda^2$ limit by several orders of magnitude [180]. Interestingly, their quality factors are also not limited by Chu's limit [183, 184], which adds to their attractiveness.

Drobitch, et al. recently demonstrated an extreme sub-wavelength "straintronic" EM antenna, where the antenna elements were ~350 nm sized magnetostrictive Co nanomagnets delineated on a piezoelectric LiNbO$_3$ substrate [181]. A surface acoustic wave (or time varying strain) was launched in the substrate, which periodically strained the nanomagnets, making their magnetizations precess owing to the Villari effect. The oscillating magnetizations (spin waves excited within the nanomagnets) couple to and emit electromagnetic waves in the air. This makes the system act as an electromagnetic antenna. The measured radiation efficiency of ~ 0.1% exceeded the $A/\lambda^2$ limit by a factor exceeding 10$^5$. The antenna's emitting area was more than *eight* orders of magnitude smaller than the square of the wavelength ($A/\lambda^2$ ~ 10$^{-8}$), resulting in drastic miniaturization.

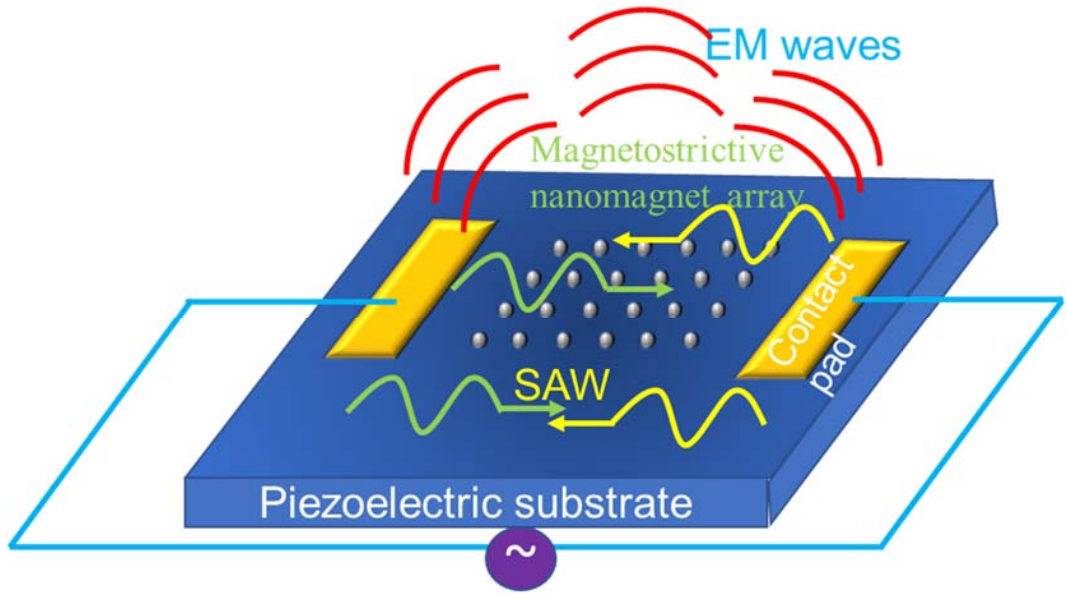

Fig. 30: Schematic of an extreme sub-wavelength straintronic electromagnetic antenna consisting of Co nanomagnets deposited on a LiNbO$_3$ substrate. A SAW is launched into the substrate by applying an ac voltage of ~144 MHz to the contact pads, which periodically strains the nanomagnets and rotates their magnetizations. The rotating magnetic field couples into electromagnetic waves that are emitted into the air at 144 MHz. The ratio of the emitting area to the square of the wavelength of emitted radiation was ~10$^{-8}$ and the measured radiation efficiency was ~144,000 times larger than that. Reproduced from [181] with permission of Wiley.

Fig. 30 shows a schematic for the antenna and Fig. 31 shows a scanning electron micrograph of the nanomagnets at low and high magnification.

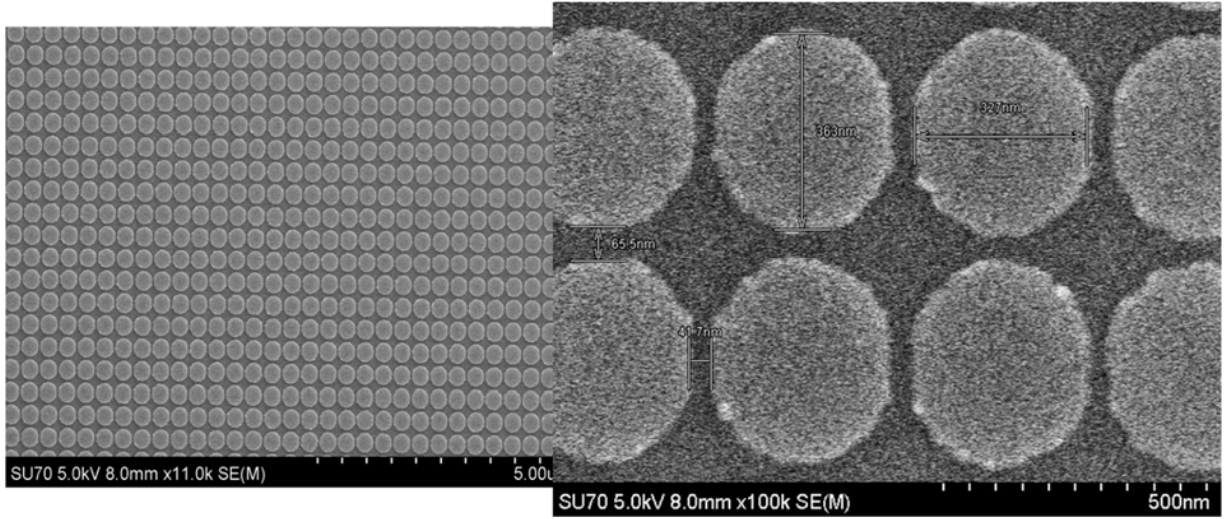

Fig. 31: Low- and high-magnification scanning electron micrographs of the nanomagnets employed in the extreme sub-wavelength electromagnetic antenna. The nanomagnets are elliptical with major axis ~360 nm, minor axis ~330 nm, vertical edge-to-edge separation 65 nm and horizontal edge-to-edge separation ~42 nm. Reproduced from [181] with permission of Wiley.

### B. Sub-wavelength *acoustic* antenna implemented with magnetostrictive nanomagnets actuated by spin-orbit torque

Straintronics can be employed to implement not just electromagnetic antennas, but also acoustic antennas. The latter radiate acoustic waves into a solid. Recently, Abeed, et al. demonstrated a sub-wavelength *acoustic antenna* that emits acoustic waves with an efficiency of ~1% [185]. The antenna linear dimension was 67 times smaller than the acoustic wavelength, so the efficiency would have been limited to $A/\lambda^2 = (1/67)^2 = 0.02\%$ if we had driven the antenna at acoustic resonance, which would have been the traditional route to actuating the antenna. In order to overcome that limit, a different principle of actuation was adopted. An alternating charge current, passed through a heavy metal (Pt) strip that is in contact with an array of magnetostrictive nanomagnets, produces alternating spin-orbit torque on the nanomagnets because of the giant spin Hall effect in Pt that injects spin current of alternating spin polarization into the nanomagnets. As long as the period of the alternating charge current exceeds the time required to rotate the magnetization of the nanomagnets through a significant angle, the magnetizations of the nanomagnets will rotate periodically with sufficient amplitude and, in the process, emit an electromagnetic wave, as described in the previous subsection. At the same time, because the nanomagnets are magnetostrictive, they will periodically expand and contract when their magnetizations are rotating, unless they are mechanically clamped by the Pt strip. If the nanomagnets are deposited on a piezoelectric substrate, then their periodic expansion/contraction will generate a periodic strain in the underlying piezoelectric, leading to the propagation of a surface acoustic wave (SAW) in the substrate that can be detected as an oscillating electrical signal with interdigitated transducers (IDT) delineated on the piezoelectric substrate. The wavelength of the acoustic wave is determined solely by the frequency of the alternating charge current (which is the frequency of the generated acoustic wave) and the velocity of acoustic wave propagation in the piezoelectric substrate. Therefore, it has no relation to the size of the nanomagnets (antenna elements) which can be much smaller than the size of the acoustic wavelength. Thus, this construct can be a sub-wavelength *acoustic* antenna with a radiation efficiency that exceeds the theoretical limit of $A/\lambda^2$. Abeed, et al. were able to exceed the theoretical limit by a factor of 50 [185].

The acoustic antenna of ref. [185] is shown schematically in Fig. 32. The nanomagnets have a protruding ledge (as shown in the top right) which is placed underneath a heavy metal (Pt) nanostrip and the entire assembly is fabricated on a $LiNbO_3$ piezoelectric substrate. Interdigitated transducers (IDTs) are delineated on the edges of the substrate to detect any surface acoustic wave (SAW) radiated in the substrate by the magnet/Pt assembly which acts as the acoustic antenna. Note that the bulk of the nanomagnet remains outside the Pt strip (only the ledge is placed underneath the strip) and hence its expansion/contraction is *not clamped* by the nanostrip.

When a charge current is passed through the Pt nanostrip, the top and bottom surfaces of the nanostrip become spin-polarized because of the giant spin Hall effect in Pt. The two surfaces have antiparallel polarizations. The polarizations of spins in either surface will depend on the direction of the current and will change sign when the current reverses direction. The accumulated spins in the *bottom* surface of the nanostrip will diffuse into the "ledges" that they are in contact with, and from there into the nanomagnets, which will exert a spin-orbit torque on the latter and rotate their magnetizations. If we reverse the direction of the injected charge current, then that will reverse the spin polarization of the bottom surface of the nanostrip and hence rotate the magnetizations of the nanomagnets in the opposite direction because the spin-orbit torque will reverse direction. Thus, if we pass an alternating current through the nanostrip, we will make the magnetizations of the nanomagnets oscillate periodically. This will emit an electromagnetic wave (as described in the preceding sub-section), and hence the system will act as an electromagnetic antenna, but because the nanomagnets are magnetostrictive, they will also periodically expand and contract, thus executing a "breathing mode" mechanical oscillation. This will generate periodic strain in the piezoelectric substrate underneath the nanomagnets and set up a SAW that can be detected with the IDTs. That makes it act as an acoustic antenna as well.

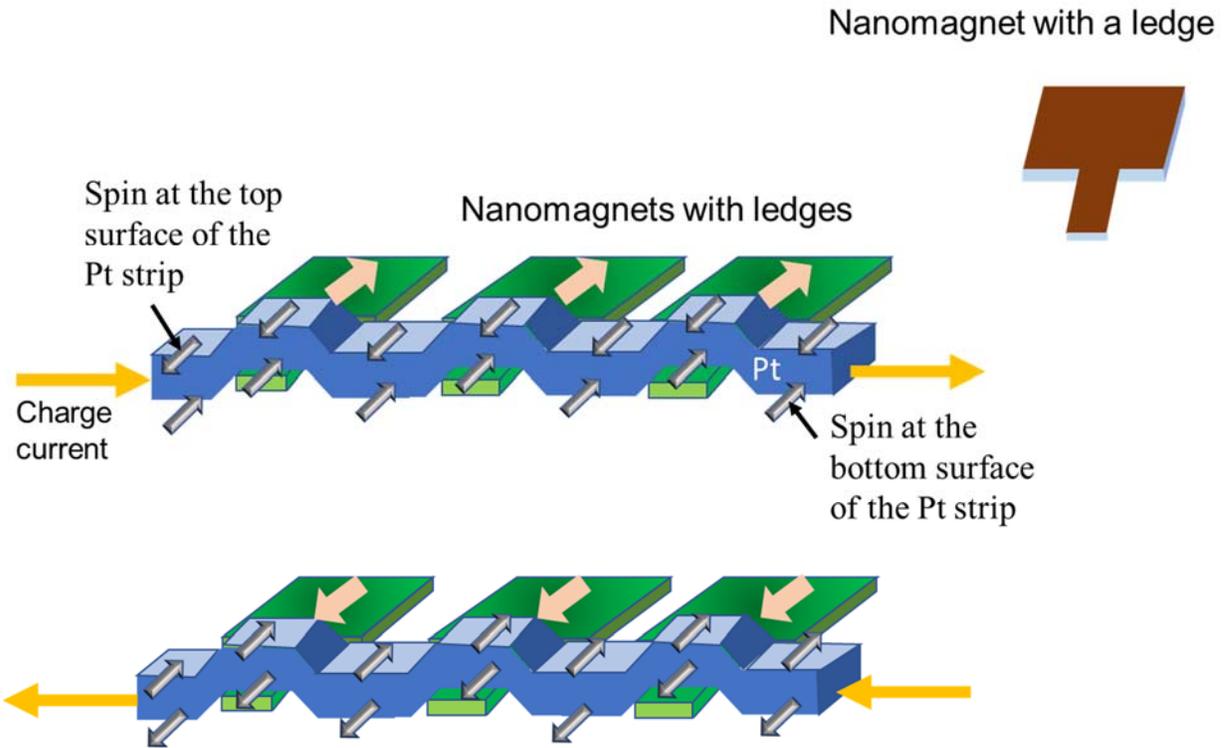

Fig. 32: Principle of actuation of the acoustic antenna by spin orbit torque generated in the nanomagnets because of spin current injection from the heavy metal (Pt) nanostrip. For one polarity of the injected charge current into the nanostrip, the torque is in one direction and for the opposite polarity, it is in the opposite direction. This makes the magnetizations oscillate periodically with the frequency of the spin (or charge) current. The nanomagnets expand and contract as their magnetizations oscillate (because they are magnetostricitve) and that generates a surface acoustic wave in the piezoelectric substrate which can be detected by interdigitated transducers delineated on the substrate. It is unlikely that the spin-orbit torque will be large enough to cause complete magnetization reversal, and what is likely is that the magnetization is rotated through an angle much smaller than $180^0$, but the accompanying expansion/contraction of the nanomagnets is sufficient to generate a detectable surface acoustic wave in the substrate. Reproduced from [185] with permission of Wiley.

Fig. 33 shows the device layout and scanning electron micrographs of various components. Figure 34 shows oscilloscope traces of the waveforms of the alternating current injected into the Pt nanostrip, as well as the oscilloscope traces of the signals picked up at the IDTs because of SAW generated by the mechanically oscillating nanomagnets. The experiment used a frequency of ~3.5 MHz for the injected current and the generated SAW was of the same frequency.

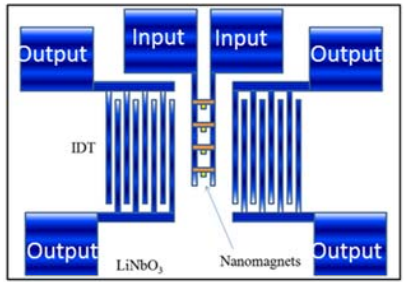
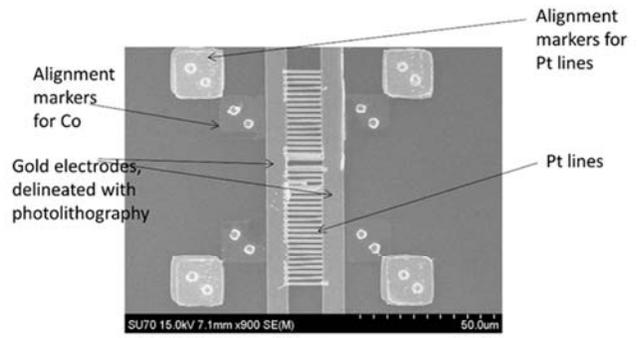
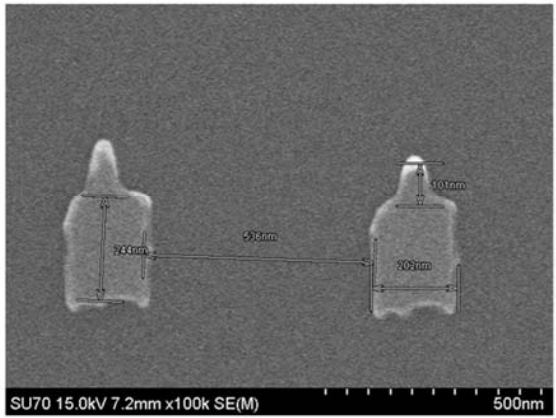
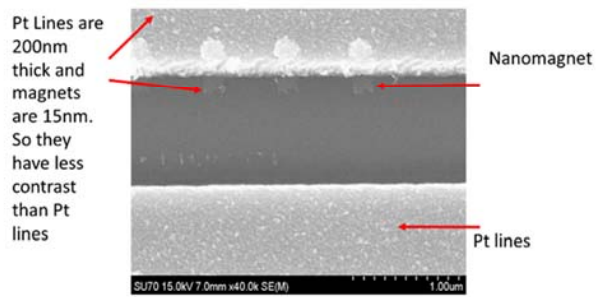

Fig. 33: (a) Pattern for the acoustic antenna. This figure is not to scale. (b) Scanning electron micrographs of the fabricated Co nanomagnets with Gaussian shaped ledges. (c) Scanning electron micrograph of the Pt lines overlying the nanomagnets. (d) Zoomed view showing the nanomagnets underneath the Pt line. Reproduced from [185] with permission of Wiley.

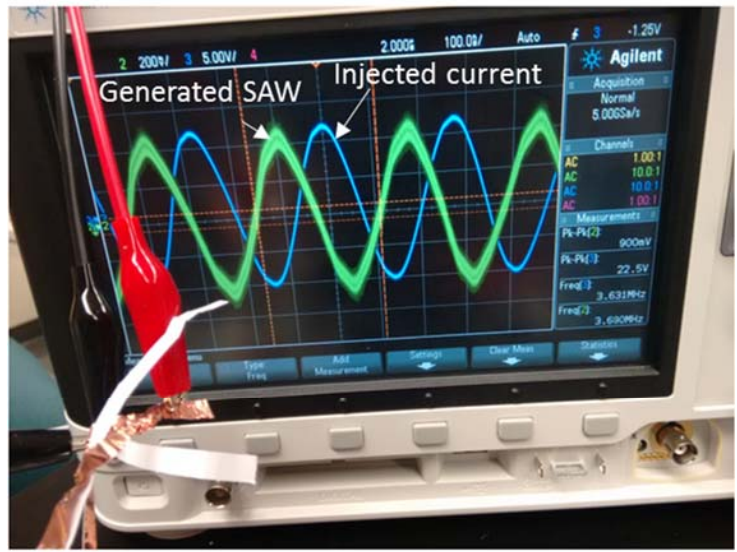

(a)

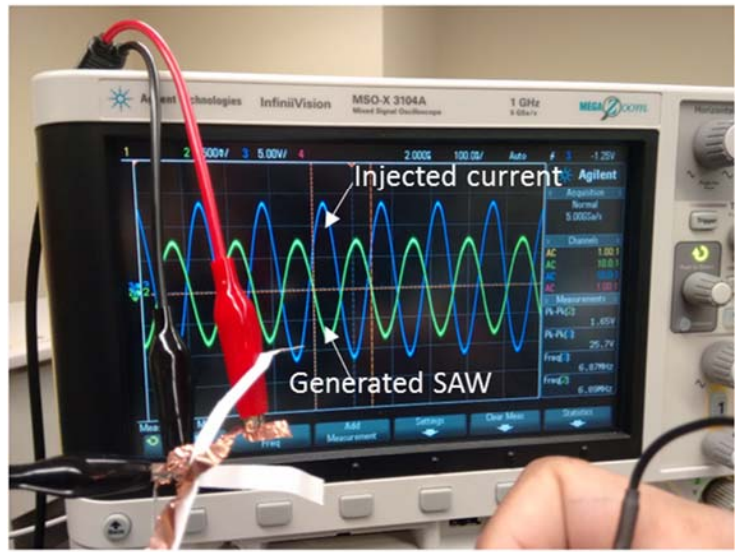

(b)

Fig. 34: Oscilloscope traces of the alternating voltage applied across the Pt lines to actuate the acoustic antenna and the alternating voltage detected at the interdigitated transducer. They are respectively the input and output signals. (a) The input voltage frequency is 3.63 MHz which is the resonant frequency of the interdigitated transducers (IDT) determined by the spacing of the IDT fingers and the velocity of surface acoustic wave in the substrate. Input voltage peak-to-peak amplitude is 22.5 V and the detected voltage peak-to-peak amplitude is 0.9 V. (b) In this case, the input voltage frequency is 6.87 MHz and the peak-to-peak amplitude is 25.7 V, while the detected voltage peak-to-peak amplitude is 1.65 V. The phase shift between the two waveforms is due to the finite velocity of the SAW that introduces a time lag between the current injection and detection of the SAW. Reproduced from [185] with permission of Wiley.

## C. An X-band microwave oscillator implemented with a straintronic magnetic tunnel junction

Ultra-small microwave oscillators, especially those operating in the X-band, have myriad applications, such as in microwave assisted writing of data in magnetic memory cells (microwave assisted magnetic recording, or MAMR) and coupled oscillator based neuromorphic computing [186]. The spintronic microwave oscillator that has been widely used is the spin-torque-nano-oscillator (STNO) [187, 188] which typically has a large linewidth $\Delta f$ at the resonant frequency $f_0$, resulting in low quality factor $Q = f_0/\Delta f$. Usually, one would observe quality factors as low as ~10 [189], although much higher quality factors, up to 10,000, have been reported after design modification [190-193].

A *single straintronic* MTJ (s-MTJ) along with a passive resistor can implement a novel microwave oscillator, very different from the STNO, based on the interplay between strain anisotropy, shape anisotropy, dipolar magnetic field, and spin transfer torque generated by the passage of spin polarized current through the soft layer of the MTJ [194]. According to results of simulations, this can have a quality factor exceeding 50. While typical STNOs output power in the few nW range, this oscillator can, in principle, output much higher power in the mW range. Its disadvantages are the relatively low quality factor (not an issue for MAMR) and the inability to tune the frequency of the oscillation easily with the power supply voltage, which would make it unsuitable for coupled oscillator based neuromorphic computing.

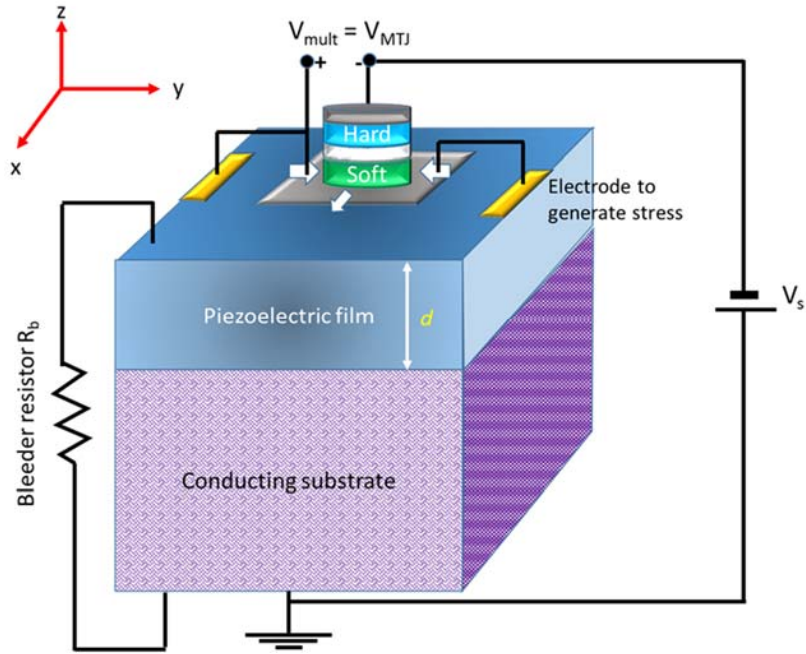

Fig. 35: A microwave oscillator implemented with a single straintronic magneto-tunneling junction (MTJ) and a passive resistor. The output voltage of the device is $V_{mult}$ which is the voltage dropped over the MTJ. The strain generated in the elliptical MTJ soft layer due to the voltage dropped over the piezoelectric is biaxial (compressive along the major axis and tensile along the minor axis). The white arrows show the strain directions. The piezoelectric layer is poled in the vertically down direction. The lateral dimension of the soft layer, the spacing between the edge of the soft layer and the nearest electrode, and the piezoelectric film thickness are all approximately the same and that generates biaxial strain in the soft layer. Reproduced from [194] with permission of the American Physical Society.

The device structure is shown in Fig. 35. To understand how this can act as a microwave source, consider the situation when there is some small residual dipole coupling between the hard layer (made of a synthetic anti-ferromagnet) and the soft layer made of a magnetostrictive material of the MTJ sitting on top of the piezoelectric film.

When no current passes through the MTJ (i. e. the voltage source is absent), the magnetizations of the hard and soft layers will be mutually antiparallel owing to the residual dipole coupling field and the MTJ will be in the high resistance state. If we now turn on the voltage supply $V_s$ (with the polarity shown in Fig. 35), spin-polarized electrons will be injected from the hard layer into the soft layer and they will gradually turn the soft layer's magnetization in the direction of the hard layer's magnetization because of the generated spin transfer torque. This will take the MTJ toward the low resistance state.

Note that the voltage dropped over the piezoelectric film is

$$V_{piezo} = \frac{R_{piezo} \parallel R_b}{R_{piezo} \parallel R_b + R_{MTJ}} V_s$$

where $R_{piezo}$ is the resistance of the piezoelectric film between the s-MTJ soft layer and the conducting substrate, $R_b$ is the resistance of a current bleeder resistor in parallel with the piezoelectric film as shown in Fig. 35 and $R_{MTJ}$ is the resistance of the s-MTJ. The resistance of the conducting substrate is assumed to be negligible.

The last equation shows that when the s-MTJ goes into the low resistance state $(R_{MTJ} \to \text{low})$, the voltage dropped over the piezoelectric film $V_{piezo}$ increases and that generates sufficient strain in that film which is partially transferred to the soft layer. Since the soft layer is magnetostrictive, this strain will rotate its magnetization away from the major axis toward the minor axis because of the Villari effect, as long as the product of the strain and magnetostriction coefficient is negative (i.e. strain will have to be compressive if the magnetostriction coefficient of the soft layer is positive and tensile if the magnetostriction coefficient is negative). This rotation, which increases the angle between the magnetizations of the hard and soft layer, will increase the resistance of the MTJ and reduce the spin polarized current flowing through it (for a constant supply voltage $V_s$). At that point, the dipole coupling effect can overcome the reduced spin transfer torque associated with the reduced current and swing the soft layer's magnetization toward an orientation antiparallel to that of the hard layer, causing the MTJ resistance to approach the high resistance state. Once that happens, the voltage dropped over the piezoelectric film falls (see the last equation) and the strain in the soft layer subsides. However, the spin polarized current still flows through the soft magnet, and over sufficient time, will transfer enough torque to the soft layer's magnetization to make it once again attempt to align parallel to the hard layer's magnetization. This will take the MTJ back to the low resistance state and the process repeats itself. The MTJ resistance $R_{MTJ}$ therefore continuously oscillates between the high and low states. This will make the voltage $V_{MTJ}$ dropped over the MTJ $\left[ V_{MTJ} = V_s R_{MTJ} / \left( R_{piezo} \parallel R_b + R_{MTJ} \right) \right]$ also continuously oscillate between two values, resulting in an oscillator.

Fig. 36 shows the simulated oscillation waveform. The simulation was based on the stochastic Landau-Lifshitz-Gilbert equation, which takes the effect of thermal noise into account. The inset shows the Fourier transform of the oscillations. This can be an oscillator with fairly high power output and for the parameters used in ref. [193], the output power was calculated to be tens of mW. This is another example of a magnetostrictive nanomagnet's interaction with time-varying strain resulting in a useful device.

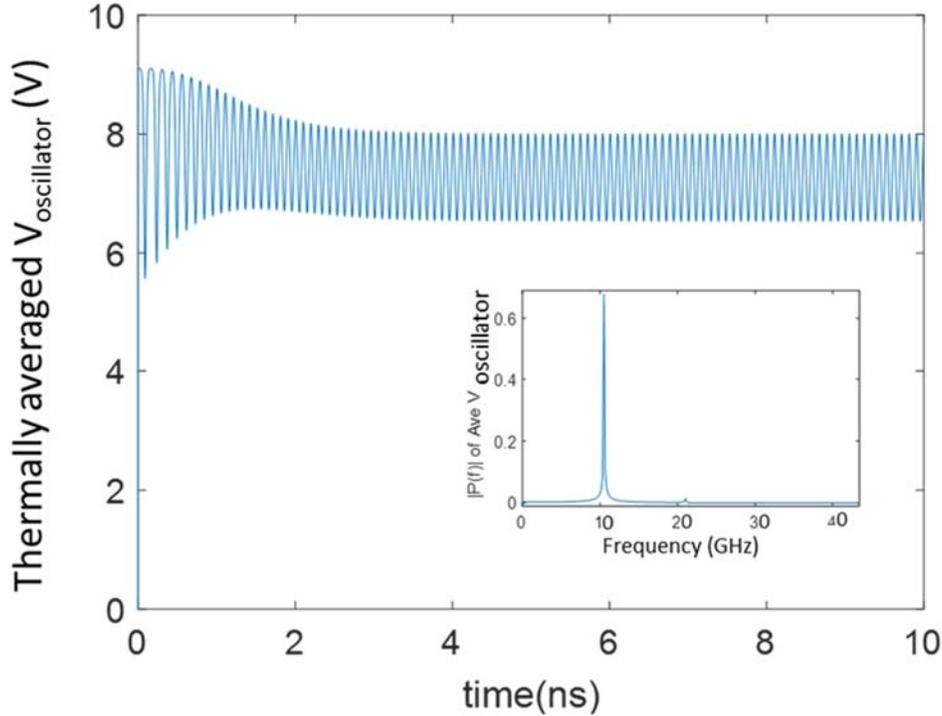

Fig. 36: Thermally averaged variation in the oscillator voltage as a function of time at 300 K in the presence of thermal noise. This plot was obtained by averaging 1000 trajectories (results of simulations) which are all slightly different from each other because of the randomness of the noise field. It takes about 3 ns to reach steady state amplitude. The steady state period is ~100 ps (frequency = 10.52 GHz, wavelength = 3 cm). The dc offset is about 7.3 V and the steady state peak-to-peak amplitude is 1.5 V. The inset shows the Fourier spectra of the oscillations after suppressing the dc component. The fundamental frequency is 10.52 GHz and there is a second harmonic at ~21 GHz whose amplitude is ~60 times less than that of the fundamental. Surprisingly the output is spectrally pure and this is almost a monochromatic (ideal) oscillator. The resonant frequency is 10.52 GHz and the bandwidth (full width at half maximum) is ~200 MHz, leading to a quality factor of 52.6. Reproduced from [194] with permission of the American Physical Society.

## X. HYBRID STRAINTRONICS-MAGNONICS

The excitation, propagation, control and detection of spin waves (or magnons) using *periodic* magnetic media led to a fascinating research field known as "magnonics". It entails several advantages over its photonic or phononic counterparts, namely shorter wavelength (at the same frequency) leading to smaller device size, anisotropic properties, negative group velocity, non-reciprocity, lower energy consumption, easier integration and compatibility with complementary metal-oxide semiconductor (CMOS) platforms, re-programmability, and efficient tuning by various external stimuli [195]. Recently, hybrid quantum systems based on collective spin excitations (or magnons) in artificial magnetic crystals (magnonic crystals), synthesized with a periodic array of nanomagnets, became important components for novel quantum technologies [196]. The magnons in these arrays can interact coherently with microwave and optical photons, phonons, magnons and superconducting qubits via magnetic dipole, magneto-optical, magnetostrictive and electric dipole interactions. This portends the development of new quantum technologies, e.g. microwave-to-optical quantum transducers for quantum information processing [197] and quantum-enhanced detection of magnons for applications in magnon spintronics [198] and in the search for dark matter.

To this end, mechanical or elastic degrees of freedom in ferromagnetic crystals is a natural and exciting avenue for hybrid quantum systems based on magnonics, and this is where straintronics can assume a significant role. The deformation modes in a ferromagnetic crystal are intrinsic elastic modes, which can couple to magnetostatic modes through magnetostrictive forces [199]. When a mechanical mode corresponding to a deformation mode is present in a ferromagnetic material, magnetostrictive forces lead to a radiation pressure-like interaction between a magnetostatic mode and the mechanical mode. This enables phenomena such as sideband cooling of the mechanical mode and parametric enhancement of the coupling strength. The experimental demonstration of this phenomenon at room temperature was made with a millimeter-sized YIG sphere in a three-dimensional microwave cavity [200]. For a sphere with a diameter of approximately 250 μm, the frequencies of the low-order mechanical modes reach ~10 MHz.

Optical generation and characterization of picosecond acoustic pulse is well-known [201-203], but excitation of spin waves in multiferroric nanomagnets by strain pulse or acoustic waves is a more recent trend. The initial challenge was to establish that the strain pulse can indeed excite and control spin precession in a ferromagnetic thin film integrated with a piezoelectric layer to form a two-phase multiferroic. In 2010, Scherbakov et al. demonstrated that a picosecond strain pulse can drive magnetization dynamics in a 200-nm-thick GaMnAs layer due to changes in magnetocrystalline anisotropy induced by the strain pulse [204]. Picosecond strain pulses were generated in a 100-nm thick Al film deposited on the back side of the GaAs substrate. The latter was excited by an optical pump pulses from an amplified femtosecond laser. The strain pulse resulted in a tilt of the magnetization vector **M**, followed by a coherent precession of **M** around its equilibrium orientation. At bias fields above 2 kOe, the magnetization became strongly aligned with the bias field resulting in a negligible tilt by the strain pulse and the ensuing precession. Kim et al. studied the room temperature magnetoacoustic dynamics in a 200-nm-thick Ni film excited by femtosecond laser pulses. The propagation of the acoustic pulse to the back side of the film modified the magnetoelastic energy of the Ni film due to the large lattice deformation, inducing precession motion. The latter could be controlled by matching the round-trip duration of the strain pulse echoes with the precessional period [205]. In another study, magnetization dynamics in a 140 nm-thick $Bi_2Y_1Fe_3O_{15}$ (Bi-YIG) film was excited by acoustic pulses generated from the Pt surface in Pt/Cu/Bi-YIG trilayers. The generated strain pulse propagated through the Cu layer and launched on to the Bi-YIG layer to induce a coherent magnetization precession at the ferromagnetic resonance (FMR) frequency, which was subsequently detected by a probe laser from the Bi-YIG surface. The observed phenomena were interpreted as strain-induced changes of magnetocrystalline anisotropy via the inverse magnetostriction effect [206].

## A. Excitation of spin waves in multiferroic nanomagnets by static and time-varying strain

Manipulation of magnetization by electric field in strain coupled artificial multiferroic nanostructures has become an active field of research with the advent of straintronics. Using x-ray photoemission electron microscopy, 90º electric field-induced uniform magnetization rotation in single domain ferromagnetic sub-micrometer islands grown on a ferroelectric single crystal was demonstrated [207]. Sadovnikov et al. developed strained controlled 4 mm long and 500 µm wide channels on yttrium iron garnet (YIG) in two-phase multiferroics made of YIG and lead zirconate titanate (PZT) with Cr electrodes. Strain coupling between the PZT and the YIG stripes occurred via the heat-cured two-part epoxy strain gage adhesive. Using Brillouin light scattering (BLS) and microwave spectroscopy, these authors observed voltage-controlled spin-wave transport and spin wave routing between the strain-reconfigurable magnetic channels [208].

In 2018, Yang et al. reported an experimental study of magnetization dynamics in magnetic tunnel junction (MTJ) nanopillars of 100 nm × 550 nm lateral dimensions driven by femtosecond-laser-induced surface acoustic waves (SAWs) [209]. On top of the MTJ, a 30-nm-thick $(Ti_{10}W_{90})_{100-x}N_x$ layer and a 300-nm-thick Al layer were deposited and patterned, serving as the top contact and transducer to convert ultrafast laser pulses into acoustic phonon pulses. The acoustic pulses induced a magnetization precession in the free layer of the MTJ through magneto-elastic coupling. Comparison of the acoustic-wave-induced precession frequencies with those by charge currents and with micromagnetic simulations, revealed edge modes to be responsible for this acoustically driven magnetization dynamics. These authors also achieved coherent control of the magnetization precession using double acoustic pulses, showing promise for future applications requiring ultrafast spin manipulation (Fig. 37).

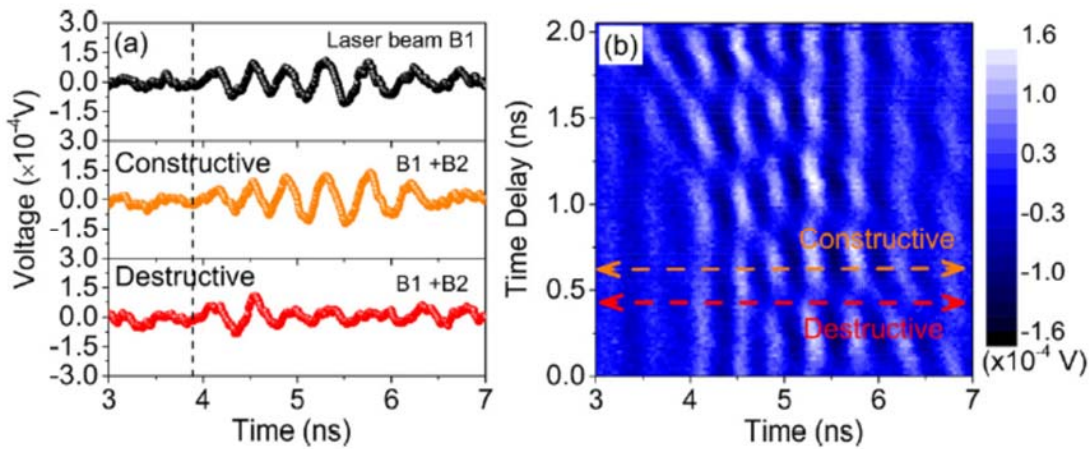

Fig. 37: (a) Coherent control of magnetization dynamics using two time-delayed laser pulses B1 and B2 being focused onto the same position of the upper contact approximately 10 µm away from the MTJ stack. Upper plot: only pulse B1, middle plot: constructive interference between B1 and B2, and lower plot: destructive interference between B1 and B2. (b) Contour plot of the magnetization dynamics versus time delay between B1 and B2. The horizontal lines correspond to the two lower plots of (a). Reproduced from [209] with permission of the American Institute of Physics.

Spin wave modes excited by time-varying strain in a *single* magnetostrictive nanomagnet were first studied by Mondal et al. [210]. They investigated the hybrid magneto-elastic modes generated by surface acoustic waves in a single quasi-elliptical magnetostrictive Co nanomagnet deposited on a poled piezoelectric $Pb(Mg_{1/3}Nb_{2/3})O_3$-$PbTiO_3$ (PMN-PT) substrate by employing all-optical time-resolved magneto-optical Kerr effect (TR-MOKE) measurements. The femtosecond laser pulse of the pump beam in the TR-MOKE played a dual role: it triggered precessional motion of the nanomagnet's magnetization about an applied bias magnetic field and it also generated surface acoustic waves (SAWs) in the PMN-PT substrate. The latter produced periodic strain in the magnetostrictive nanomagnet and modulated its precessional dynamics. Two different mechanisms generated the SAWs in the substrate. First, the alternating electric field of the pump laser generated periodic compressive and tensile strain in the PMN-PT substrate from $d_{33}$ and/or $d_{31}$ coupling. The strain was tensile when the electric field in the substrate was in the same direction as the poling and compressive when the electric field was opposite to the direction of the poling. Second, the differential thermal expansions of the nanomagnet and the substrate due to periodic heating by the pulsed pump beam also produced some periodic strain. Figure 38 shows the time-resolved Kerr oscillations from a single Co nanomagnet on the SAW-carrying PMN-PT substrate as a function of bias magnetic field. These oscillations have multiple frequency components as observed in the power spectra. All peaks in the power spectra shift to lower frequencies with decreasing bias magnetic field.

This bias-magnetic field dependence confirms that these modes have mixed magnetic components in them. The experimental results were in excellent agreement with theoretical results obtained by introducing a periodic strain anisotropy field due to the SAW in micromagnetic simulations. Simulated spin-wave mode profiles in the absence and the presence of the strain field revealed the spatial nature of the hybrid magneto-dynamical modes as shown in Fig. 39. Instead of the characteristic center and edge mode behavior of a single nanomagnet excited optically or by pulsed magnetic field, the hybrid magneto-dynamical modes of frequencies $F_L$, F and $F_H$ exhibit complex profiles with their unique characteristics, besides displaying rich variation with bias magnetic field. This study demonstrated that strain can affect the magnetization state of even a weakly magnetostrictive nanomagnet in time scales far shorter than 1 ns leading towards their possible applications in energy efficient high frequency spintronics applications.

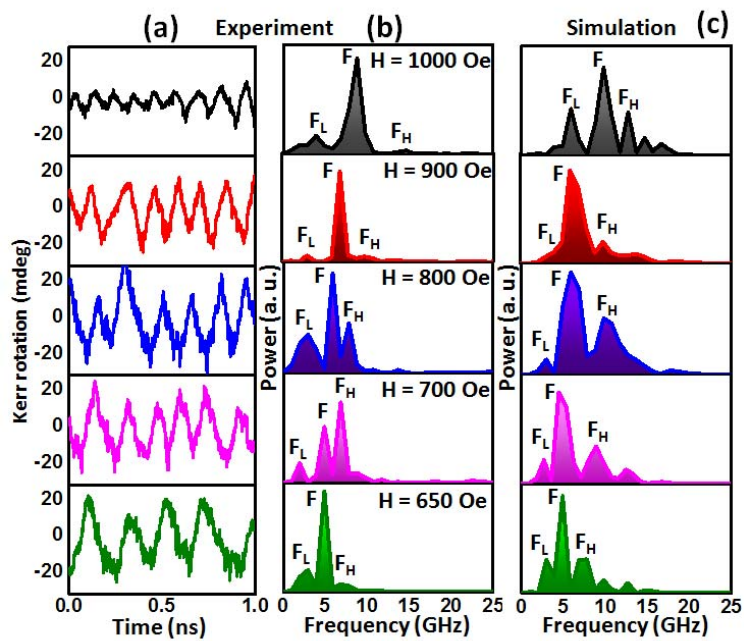

Fig. 38. Bias magnetic field dependence of the (background-subtracted) time-resolved Kerr oscillations from a single Co nanomagnet on a PMN-PT substrate. The pump fluence is 15 mJ/cm$^2$. (a) The measured Kerr oscillations in time and (b) the fast Fourier transforms of the oscillations. The Fourier transform peaks shift to lower frequencies with decreasing bias magnetic field strength. There are multiple oscillation modes of various Fourier amplitudes. The dominant mode is denoted by F and its nearest modes are denoted by $F_H$ and $F_L$ (at all bias fields except 700 Oe). (c) Fourier transforms of the temporal evolution of the out-of-plane magnetization component at various bias magnetic fields simulated with the micromagnetic simulator MuMax3 where the amplitude of the periodically varying strain anisotropy energy density $K_0$ is assumed to be 22,500 J/m$^3$. The simulation has additional (weak) higher frequency peaks not observed in the experiment. The spectra in the two right panels are used to compare simulation with experiment. Reproduced from [210] with permission of the American Chemical Society.

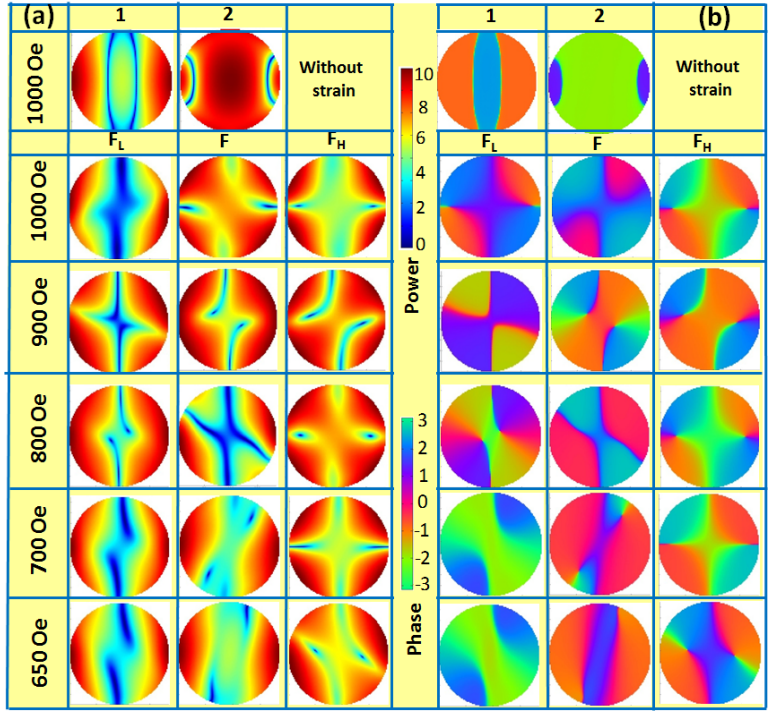

Fig. 39. (a) Simulated power and (b) phase profiles of the spin waves associated with the three dominant frequencies $F_L$, F and $F_H$ in the Kerr oscillations at any given bias field. The top most row shows edge and center modes at the two dominant frequencies in the Kerr oscillations in the absence of strain at 1000 Oe bias field. The units of power and phase are dB and radians, respectively. Reproduced from [210] with permission of the American Chemical Society.

## XI. INTERACTION BETWEEN ACOUSTIC AND SPIN WAVES

Magnons are generally difficult to manipulate but recent studies have shown that they could be controlled through coupling with acoustic vibration or phonons. A strong mutual interaction can cause back and forth energy transfer between them leading to the formation of hybrid quasiparticle magnon-polaron [211], which is neither a magnon nor a phonon. Strong mutual interaction can also lead to energy transfer to spin waves from acoustic waves, leading to *parametric amplification* of the spin wave. Significant amplification would require significant energy transfer and hence a high degree of coupling. This is usually challenging since efficient coupling requires phase matching, namely at the given frequency, the wavevectors of the two waves will be equal. In effect, that would require the phase velocities of the spin wave and the acoustic wave to be the same, which is usually not the case. However, the acoustic wave velocity can be much closer to the spin wave velocity than an electromagnetic wave. Hence, parametric amplification of spin waves with acoustic waves would usually yield higher amplification compared to electromagnetic waves.

Parametric amplification of spin waves coupled to electromagnetic waves via voltage-controlled-magnetic-anisotropy was proposed in the past [212] but not experimentally demonstrated, while parametric amplification of spin waves in ferromagnetic thins films [213] and in magnonic crystals [179], via coupling to acoustic waves, have been demonstrated. In ref. [179], the spin wave power could be amplified by a factor of 7-8, even though the spin wave and acoustic wave were not phase matched. Spin wave amplification has important applications in spin wave logic and other spin wave devices since spin waves decay rapidly in ferromagnetic materials and will have to be amplified for logic level restoration.

The interaction between spin waves and SAW was studied in nanomagnet arrays by Yahagi et al. in 30-nm-thick nickel elliptic disks with varying array pitch (*p*) fabricated on a (100) silicon substrate with a 110-nm-thick hafnium oxide antireflection (AR) coating. The optical pump-probe experiment using a TR-MOKE setup showed the simultaneous excitation of both acoustic and magnetic resonances. A strong coupling between these excitations was observed when they were brought near degeneracy by varying an external magnetic field [214]. Three important manifestations of the magneto-elastic coupling were: (a) pinning of the spin-wave frequency over an extended range of the bias magnetic fields, (b) generation of new mode with frequency differing by more than 100% from the intrinsic element response, and (c) sudden enhancement of the Fourier amplitude of the spin-wave mode at crossovers with acoustic modes since coupling becomes most efficient at these cross-over points (see Fig. 40 for details). These features were explained by invoking an additional effective field component created by magneto-elastic coupling, which were also accurately reproduced by simulations.

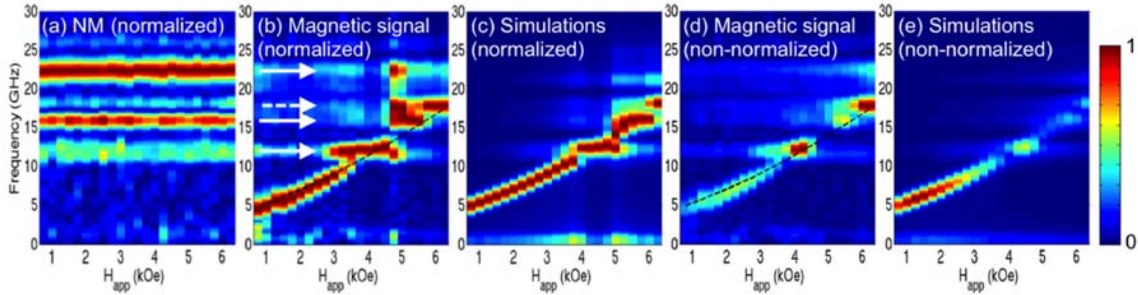

Fig. 40: Fourier spectra for nickel elliptic disks for *p* = 212 nm. (a) Measured nonmagnetic signal. (b), (d) Measured magnetic signal. Dashed line is the simulation result for nickel elliptic disks without magnetoelastic contribution. The solid and dashed arrows in (b) indicate the RW and SSLW frequencies. (c), (e) Simulated magnetization dynamics including magnetoelastic coupling. The Fourier amplitudes in (a), (b), and (c) are normalized for better visualization of oscillation modes. The nonnormalized Fourier spectra in (d) and (e) illustrate the enhanced Fourier amplitude at the crossover points at 12.2, 15.8, 17.7, and 22.3 GHz. Reproduced from [213] with permission of American Physical Society.

Further work by this group showed that the nanomagnet geometry plays an important role in the magneto-elastic coupling. By changing the nanomagnet array pattern from periodic to aperiodic, the magneto-elastic effect of SAWs on the magnetization dynamics was effectively removed. The efficiency of this method was found to depend on the amount of residual spatial correlations, which was quantified by spatial Fourier analysis of the two structures [215]. A combined experimental and theoretical study showed that the pinning magnetic field range of the spin-wave resonance to SAW resonance is determined by the effective damping coefficient $\alpha_{\text{eff}}$ of the nanomagnets, instead of the magneto-elastic coupling coefficient. Using this insight, the field dependent $\alpha_{\text{eff}}$ was directly extracted from the pinning linewidth, and the intrinsic Gilbert damping was recovered at large applied fields [216].

## A. Coupling between magnon and phonon and formation of magnon-polaron

As mentioned earlier, when magnons and phonons are strongly coupled, they form a hybridized magnon-phonon quasiparticle. In this state, the modes do not possess specifically either a magnon or a phonon character, but they co-exist in both states. Berk et al. reported the direct observation of coupled magnon-phonon dynamics in a single rectangular shaped Ni nanomagnet of dimensions 330 × 330 × 30 nm grown on a (100) Si substrate capped by a 110-nm-thick hafnium oxide layer. They utilized the vibrational modes of the single Ni nanostructure to stimulate phonon dynamics optically in the frequency range of 5 GHz–25 GHz along with the intrinsic precessional modes of the nanomagnet. The confined geometry of the Ni nanostructure created a confined cavity for the phonon and magnon modes. By varying an external magnetic field in the appropriate geometries, the magnon mode was tuned through the phonon resonances leading towards the observation of avoided crossings characteristic of coupled systems [217]. Two different anti-crossings between (1,1) and (2,0) modes were observed, and from the loss rate and coupling rate of those anti-crossings, the cooperativity was found to be about 1.14 and 0.74, i.e. in the intermediate coupling regime.

A magnon polaron, i.e. the hybridized state of phonon and magnon in a magnetically ordered material, can be formed at the intersection of the magnon and phonon dispersions, where their frequencies meet. However, the observation of this entity in the time domain remained elusive due to the weak interaction of phonons and magnons and short lifetime of the polaron, which prohibit the strong coupling required for the formation of a hybridized state. In a recent seminal work, Godejohann et al. managed to overcome these difficulties by imposing spatial overlapping of magnons and phonons in a Galfenol ($Fe_{0.81}Ga_{0.19}$) thin film nanograting (NG) grooved on an epitaxially grown Galfenol film on a (001)-GaAs substrate [211]. Galfenol, being a highly magnetostrictive alloy, possesses both enhanced magnon-phonon interaction and well-defined magnon resonances. Parallel grooves on the 105-nm Galfenol film were milled using a focused beam of Ga ions along the [010] crystallographic axis of the GaAs substrate, having depth $a = 7$ nm and width $w = 100$ nm, which equals their separation; the NG lateral period was $d = 200$ nm. The spatial overlap of the desired phonon and magnon modes occurred in the NG structure, resulting in a high coupling strength. This, in combination with their long lifetimes, allowed clear evidence of an optically excited magnon polaron. The authors showed that the symmetries of the localized magnon and phonon states play a decisive role in the magnon polaron formation and its ensuing manifestation in the optically excited magnetic transients (time-resolved Kerr rotation (KR)) measured by conventional time-resolved magneto-optical pump-probe spectroscopy. By changing the external magnetic field, they were able to realize resonance conditions for a magnon mode with two localized phonon modes with different polarizations. These were: (a) a lower-frequency Rayleigh-like standing wave with dominant displacement perpendicular to the NG plane, referred to as a quasi-transverse acoustic (QTA) mode and (b) a higher-frequency second-order Rayleigh-like mode (Sezawa mode) with predominant in-plane displacement, referred to as a quasi-longitudinal acoustic (QLA) mode. At the resonance with the lower phonon mode (QTA), a frequency splitting was observed indicating excitation of a hybridized state, i.e., a magnon polaron. At resonance with the higher phonon mode (QLA), a strong driving of the magnon mode by the phonon mode was observed without any detectable frequency splitting. Figure 41 displays the hybridization of magnon and phonon modes.

Vidal-Silva, et al. explored the theoretical possibility of generating magnon-polaron excitations via a spatiallyvarying magnetic field [218]. In the presence of a magnetic field gradient, the spatial dependence of the magnetic field in the Zeeman interaction resulted in a magnon-phonon coupling. Such a coupling depends directly on the strength of the magnetic field gradient. The theory also predicted that control over coupling of specific phonon polarization to the magnons in the material can be obtained by tuning the direction of the magnetic field gradient [218]. The magnetoelastic interaction mechanism is complex and nontrivial. For propagating waves, the coupling is known to require the matching of both the frequencies and the wavelengths [219]. This condition can be fulfilled for spin waves and SAWs existing in the same range of frequencies and wave vectors. Recently Babu et al. showed using Brillouin light scattering (BLS) measurements and theoretical calculation that even for a favorable orientation of the field for the coupling, the magnetoelastic interaction can be significantly reduced for SAWs with a particular profile in the direction normal to the surface at distances much smaller than the wavelength [220].

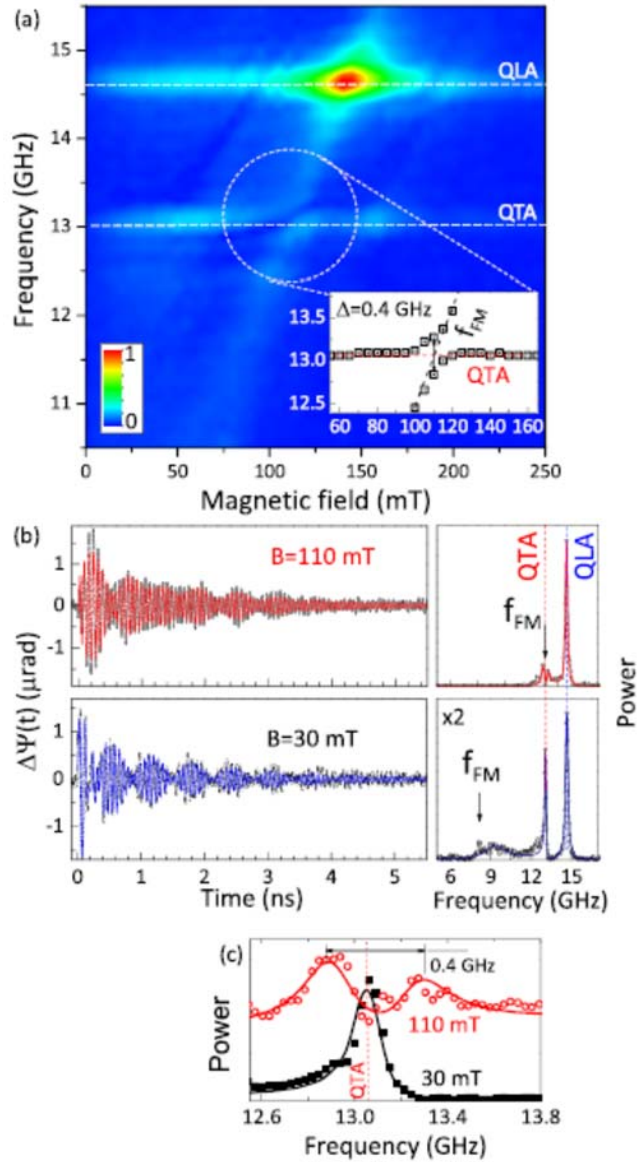

Fig. 41: Hybridization of magnon and phonon modes. (a) Color map which shows the spectral density of the measured KR signal as a function of the external magnetic field applied along the NG diagonal when the interaction between the magnon and phonon modes of NG has maximal strength. The anticrossing is observed at $f$ = 13 GHz and $B$ = 110 mT. The inset shows the magnetic field dependence of the spectral peaks in the magnon spectrum around the intersection of the QTA and FM modes. (b) Transient KR signals (left panels) and their FFTs (right panels) at nonresonant ($B$ = 30 mT) and resonant ($B$ = 110 mT) conditions. Symbols show the measured signals and their FFTs; solid lines show respective fits and their FFTs. (c) Zoomed fragments of the FFT spectra shown in (b) around the resonance frequency. The splitting of the line in the resonance at $B$ = 110 mT is clearly seen. Reproduced from [211] with permission of the American Physical Society.

### B. Investigation of single nanomagnet dynamics using magneto-elastic coupling

The concept of magnet-elastic excitation of nanomagnets was further probed to extract the damping of a single Ni cylindrical nanomagnet of 200 nm diameter and 30 nm thickness by exciting the dynamics with SAW. SAWs with specific frequencies were optically generated by two sets of Al bars with varying pitch deposited on a Si (100) substrate coated with hafnium oxide. The counter propagating SAWs form a standing wave effectively increasing the magneto-elastic field amplitude and driving a narrowband "cold" excitation of the magnetization precession of the nanomagnet. It was shown that the intrinsic Gilbert damping of the single nanomagnet can be directly extracted, which is in contrast to optically excited precessions of single nanomagnets that show larger effective damping presumably due to the thermal excitation [221]. This cold excitation was exploited to excite two identical elliptical Ni nanomagnets (316 × 160 × 30 nm$^3$) with orthogonal orientations between two sets of identical Al bars as described above. TR-MOKE measurements showed that one can preferentially excite one of the nanomagnets by controlling the applied field due to the shape anisotropy. The damping behavior also showed that the magneto-elastic coupling efficiency for these two nanomagnets depends on the relative orientation between the SAW and the sample geometry [222].

Recent study on nanomagnet size and SAW wavelength dependence of magnetization dynamics showed that the efficiency of magneto-elastic resonance depends on the Gilbert damping in addition to the relative nanomagnet size and acoustic wavelength. Simulations showed that inhomogeneous broadening of the elastically driven spin dynamics results in enhanced damping for nanomagnets larger than the SAW wavelength [223]. The authors of [223] claimed that the losses associated with acoustically driven spin dynamics scale favorably with nanomagnet dimensions.

In another interesting development, a 10-fold enhancement of the precessional dynamical excitation in single 30-nm-thick Ni nanomagnets with diameters ranging between 75 nm and 200 nm was observed using focused SAW (FSAW). In this experiment the same batch of single Ni nanomagnets was defined between two sets of straight (pitch ~400 nm) and arc-shaped Ni gratings (pitch ~400 nm, focal distance ~2 μm, and arc angle 120º) for optical excitation of the SAW, which, in turn, excited the magnetization dynamics. Using this FSAW excitation, magneto-elastically controlled magnetization dynamics in a sub-100 nm single nanomagnet was excited. Subsequently, using multiple phononic gratings, selective activation of magnetization dynamics of single nanomagnets by varying an external field was demonstrated (Fig. 42), which may lead to the development of SAW-driven nano-oscillators [224]. A recent review of the interaction of SAW with magneto-dynamics is ref. [225].

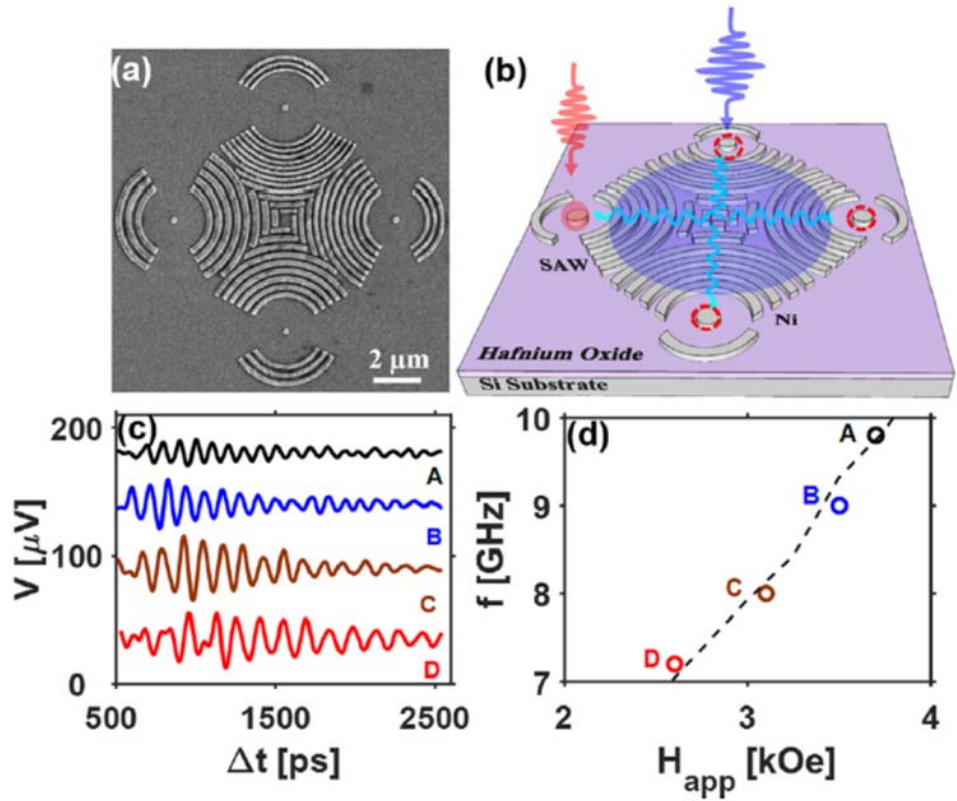

Fig. 42. (a) SEM image of four identical Ni nanomagnets (A, B, C, and D) surrounded by the corresponding four different pitches of gratings with pitches of 250 nm, 300 nm, 350 nm, and 400 nm, respectively. (b) Schematic plot of FSAW driven magnetization excitation in four identical Ni nanomagnets. (c) Time traces of the FSAW-driven four nanomagnets. (d) The precession frequency of the four nanomagnets at the corresponding resonant fields. The dashed line is the magnetic frequency of a single nanomagnet measured using the all-optical TRMOKE. Reproduced from [222] with permission of the American Institute of Physics.

## XII. CONCLUSION

Straintronics is an emerging field with vast promise. It is exceptionally energy-efficient and hence a very important field of investigation as cloud computing, data mining, machine learning, artificial intelligence, communication, information storage and related activities become increasingly demanding of energy. Much of the demand on energy can be relieved by the use of energy-efficient devices, which may also have other attributes (e.g. non-volatility) that can enable superior hardware platforms that are faster and more reliable. Devices with lower dissipation can also prolong the celebrated Moore's law [226] that has guided the electronics industry for the last six decades.

In addition to offering better energy efficiency, straintronic devices can also reduce circuit complexity and device footprint dramatically. The correlator/anti-correlator in Section VIII.B would normally have required microcontrollers, shift registers and/or other circuits, if implemented with conventional electronics [227], but requires only stress-modulated dipole interaction which is wireless and hence consumes no area on a chip. This further reducing energy consumption and cost.

The Achilles' heel of straintronics is the high error rate associated with switching. This is an unfortunate trait, which seems to be ordained by an unavoidable trade-off between energy-efficiency and error resilience [228]. There are, however, some computing architectures that are forgiving of errors, such as collective computational models where the circuit functionality is derived from the cooperative actions of many devices acting in unison (e. g. neuromorphic networks, Bayesian inference engines, Boltzmann machines). There, the failure of a single device (or even a significant fraction of devices) does not inhibit circuit operation [162]. It is in these areas, which are increasingly attracting attention [229], where straintronics might make serious inroads.

In addition to digital information processing, straintronics also plays a very valuable role in analog applications such as high frequency signal generation and antennas. Antennas based on straintronic principles can overcome the theoretical limits on radiation efficiencies of traditional antennas actuated with an electromagnetic source. This allows ultra-miniaturization of antennas for embedded applications.

Finally, straintronics is playing an increasingly important role in the understanding of fundamental physical phenomena such as hybrid straintronics-magnonics, magnon-phonon coupling and magnon-polaron formation. Studies of these phenomena reveal a plethora of rich physics and can even enable useful devices and systems such as spin wave amplifiers based on parametric amplification. The blending of straintronics and magnonics has spawned an extremely fertile ground for seeding new ideas and revealing new physics.

**ACKNOWLEDGEMENTS**

Numerous students contributed to the work reviewed here and they are too numerous to mention individually. J. A. acknowledges Dr. Dhritiman Bhattacharya, currently a post-doctoral associate at Georgetown University, for the literature on strain-based manipulation of skyrmions. Researchers who collaborated with us on some of the work reported here are Prof. Amit Ranjan Trivedi from the University of Illinois at Chicago, Prof. Jian-ping Wang from the University of Minnesota, Prof. Csaba Andras Moritz from the University of Massachusetts at Amherst, Prof. Avik Ghosh from the University of Virginia and Prof. Kang Wang from the University of California at Los Angeles. The US authors were supported to perform some of the work reported here by the US National Science Foundation (NSF) grants ECCS 1124714, CCF 1216614, CMMI 1301013, ECCS 1609303, DMR 1726617 and CCF 1815033. J. A. received a NSF CAREER grant CCF 1253370. Additionally, S. B. is currently supported by the NSF grants CCF 2006843 and CCF 2001255, while J. A. is supported by the NSF grants CCF 1909030 and ECCS 1954589, for their work on straintronics. Additional funding sources for the US authors were the Semiconductor Research Corporation, the Center for Innovative Technology at Virginia, the Virginia Microelectronics Consortium and the Virginia Commonwealth University Commercialization Fund. All authors acknowledge the Indo-US Science and Technology Fund grant "Center for Nanomagnetics for Energy-Efficient Computing, Communications and Data Storage" (IUSSTF/JC-030/2018).

**DATA AVAILABILITY**

Data sharing is not applicable to this work since no new data were generated or analyzed.